\documentclass[aps,prd,twocolumn,floatfix,nofootinbib,preprintnumbers,superscriptaddress,notitlepage]{revtex4-1}

\usepackage{times}
\usepackage[usenames]{color}
\usepackage{amsfonts}
\usepackage{amsmath}
\usepackage{amssymb}
\usepackage{xcolor}
\usepackage{bm}
\usepackage{booktabs}
\usepackage{dcolumn}
\usepackage{enumerate}
\usepackage{epsfig}
\usepackage{graphicx}
\usepackage{graphics}
\usepackage[latin1]{inputenc}
\usepackage{latexsym}
\usepackage{rotating}
\usepackage{hyperref}
\usepackage[caption=false]{subfig}
\usepackage{multirow}

\newcommand{\<}{\begin{equation}}
\newcommand{\?}{\end{equation}}

\newcommand{\cQ}{\mathcal{Q}}

\newcommand{\hh}{\mathbf{h}}

\newcommand{\fc}{f_\text{cut}}

\newcommand{\Mf}{M_\text{f}}
\newcommand{\Mfin}{M_{\text{f}}^\text{insp}}
\newcommand{\Mfmr}{M_{\text{f}}^\text{postinsp}}

\newcommand{\af}{\chi _\text{f}}
\newcommand{\afin}{\chi _{\text{f}}^\text{insp}}
\newcommand{\afmr}{\chi _{\text{f}}^\text{postinsp}}

\def\gw#1{Gravitational wave#1 (GW#1)\gdef\gw{GW}}
\def\psd#1{power spectral density#1 (PSD#1)\gdef\psd{PSD}}
\def\gr#1{General Relativity#1 (GR#1)\gdef\gr{GR}}
\def\snr#1{signal-to-noise ratio#1 (SNR#1)\gdef\snr{SNR}}

\DeclareMathOperator{\Real}{Re}

\begin{document}

\title{Investigating the relation between gravitational wave tests of general relativity}

\author{Nathan~K.~Johnson-McDaniel}
\affiliation{Department of Applied Mathematics and Theoretical Physics, Centre for Mathematical Sciences, University of Cambridge, Wilberforce Road, Cambridge, CB3 0WA, UK}
\affiliation{Department of Physics and Astronomy, University of Mississippi, University, Mississippi 38677, USA}
\author{Abhirup Ghosh}
\affiliation{Max Planck Institute for Gravitational Physics (Albert Einstein Institute), D-14476 Potsdam-Golm, Germany}
\author{Sudarshan Ghonge}
\affiliation{Center for Relativistic Astrophysics, Georgia Institute of Technology, Atlanta, Georgia 30332, USA}
\author{Muhammed Saleem}
\affiliation{School of Physics and Astronomy, University of Minnesota, Minneapolis, Minnesota 55455, USA}
\affiliation{Chennai Mathematical Institute, Siruseri 603103, Tamil Nadu, India}
\author{N.~V.~Krishnendu}
\affiliation{Max Planck Institute for Gravitational Physics (Albert Einstein Institute), Callinstr.~38, D-30167 Hannover, Germany}
\affiliation{Leibniz Universit{\"a}t Hannover, D-30167 Hannover, Germany}
\author{James~A.~Clark}
\affiliation{LIGO, California Institute of Technology, Pasadena, California 91125, USA}
\affiliation{Center for Relativistic Astrophysics, Georgia Institute of Technology, Atlanta, Georgia 30332, USA}

\date{\today}

\begin{abstract}
Gravitational wave observations of compact binary coalescences provide precision probes of strong-field gravity. There is thus now a standard set of null tests of general relativity (GR) applied to LIGO-Virgo detections and many more such tests proposed. However, the relation between all these tests is not yet well understood. We start to investigate this by applying a set of standard tests to simulated observations of binary black holes in GR and with phenomenological deviations from GR.
We consider four types of tests: residuals, inspiral-merger-ringdown consistency, parameterized phasing (two varieties), and modified dispersion relation. We also check the consistency of the unmodeled reconstruction of the waveforms with the waveform recovered using GR templates. These tests are applied to simulated observations similar to GW150914 with both large and small deviations from GR and similar to GW170608 just with small deviations from GR. We find that while very large deviations from GR are picked up with high significance by almost all tests, more moderate deviations are picked up by only a few tests, and some deviations are not recognized as GR violations by any test at the moderate signal-to-noise ratios we consider. Moreover, the tests that identify various deviations with high significance are not necessarily the expected ones. In particular, the parameterized tests recover PN test parameters much closer to zero than their true values in some cases. Additionally, we find that of the GR deviations we consider, the residuals test is only able to detect extreme deviations from GR that no longer look like binary black hole coalescences in GR. The reconstruction comparison shows more promise for detecting relatively small GR deviations in an unmodeled framework, at least for high-mass systems.
\end{abstract}

\maketitle

\section{Introduction}
\label{sec:intro}

\gw{} observations have provided our first probes of the dynamics of general relativity (GR) in the strong field, highly dynamical regime. A variety of tests were applied to the first detection~\cite{GW150914_TGR} and other detections in the first observing run of the advanced GW detectors (O1)~\cite{O1_BBH} and more tests have been added as subsequent detections have been analyzed by the LIGO and Virgo collaborations~\cite{GW170104, GW170814, GW170817_GW_GRB, GW170817_TGR, O2_TGR, GW190521_discovery, GW190521_implications, O3a_TGR,O3b_TGR}. These observations have so far revealed no inconsistencies with the predictions of GR and future observations with upgraded and new detectors~\cite{Aasi:2013wya,Hild:2010id,Reitze:2019iox,Hall:2020dps,Maggiore:2019uih,Baker:2019nia,LISA} will place even more stringent bounds on any possible deviations from GR's predictions, or potentially reveal more subtle deviations that have eluded detection to date. These improved results will come both from more sensitive observations of individual signals, as well as by combining together many detections. See, e.g.,~\cite{Perkins:2020tra} for predictions of the sensitivity of tests of GR with future detectors.

However, the tests so far applied by the LIGO and Virgo collaborations are all null tests of one sort or another---none of them is testing a specific alternative theory. There are ways of mapping the results of some of these tests to constrain specific alternative theories, e.g., using the results of the parameterized tests that vary post-Newtonian (PN) coefficients~\cite{Yunes:2016jcc,Tahura:2018zuq,Nair:2019iur,Tahura:2019dgr,Wang:2021yll}. However, these mappings require significant assumptions that may not be valid in practice. This is due in part to the relatively low PN order and linearization in the coupling constant of the calculations used in the mapping. However, perhaps even more importantly, the LIGO-Virgo constraints on deviations in individual PN coefficients from their GR values that are mapped onto the constraints on modified theories are not necessarily valid constraints on those PN coefficients in a theory where multiple PN coefficients vary, and/or when the merger-ringdown phase is also modified, as we shall see here. The LIGO-Virgo analyses are designed to detect deviations from GR, not to measure individual PN coefficients. As discussed in~\cite{Chua:2020oxn}, tests against specific alternative theories are preferable, and there is preliminary work on such constraints in~\cite{Sennett:2019bpc,Perkins:2021mhb}, though still with simplified waveform models.

In fact, numerical simulations in alternative theories are not yet quite advanced enough to provide even single waveforms that could be used to check the performance of the tests carried out to date, though such simulations are progressing (see, e.g.,~\cite{Hirschmann:2017psw,Witek:2018dmd,Okounkova:2019zjf,Okounkova:2020rqw,East:2020hgw,Silva:2020omi,East:2021bqk} for binary black hole simulations and~\cite{Barausse:2012da,Shibata:2013pra,Bezares:2021dma} for simulations of binary neutron stars), as are analytical calculations (e.g.,~\cite{Sennett:2016klh,Bernard:2018hta,Bernard:2018ivi,Bernard:2019yfz,Tahura:2018zuq,Khalil:2018aaj,Julie:2018lfp,Julie:2019sab,Sennett:2019bpc,Huber:2020xny,Shiralilou:2020gah,Shiralilou:2021mfl,Brax:2021qqo,Battista:2021rlh}). Additionally, while there is progress in simulations that could be used as proxies for non-GR effects, e.g., of charged binary black holes~\cite{Bozzola:2020mjx,Bozzola:2021elc}, and of binaries of black hole mimickers, such as boson stars~\cite{Liebling:2017fv, Cardoso:2016oxy, Bezares:2017mzk, Palenzuela:2017kcg, Helfer:2018vtq, Sanchis-Gual:2018oui, Bezares:2018qwa, Helfer:2021brt}, these are also not quite advanced enough to provide waveforms of the quality required for data analysis.

Given this, as well as the continued proposals for new null tests of GR or of binary black hole nature (e.g., \cite{Krishnendu:2017shb, Krishnendu:2019tjp, Dhanpal:2018ufk, Islam:2019dmk, Johnson-McDaniel:2018uvs, Kastha:2018bcr, Kastha:2019brk, Carullo:2018gah, Carullo:2019flw, Maselli:2019mjd, Carullo:2021dui, Ghosh:2021mrv, Asali:2020wup, Haster:2020nxf, Capano:2020dix, Edelman:2020aqj, Psaltis:2020ctj, Bhagwat:2021kfa}), 
it is important to understand the relation between the tests being applied to the data, to decide which set of tests is most efficacious at detecting a range of plausible deviations from GR and how a deviation from GR would show up in the various tests. For the present study, we restrict ourselves to the four waveform-based tests applied to the binary black hole signals through O2~\cite{O2_TGR} (with updated results for more events in~\cite{O3a_TGR,O3b_TGR}), namely the residuals test; the inspiral-merger-ringdown (IMR) consistency test; the parameterized test of GW generation, both the Test Infrastructure for GEneral Relativity (TIGER) and Flexible Theory-Agnostic (FTA) varieties; and the parameterized test of propagation. We also compare the unmodeled waveform reconstructions with the waveforms inferred from the modeled GR analysis, as carried out in~\cite{GWTC-1_paper, GWTC-2_paper, GWTC-3_paper}.

To do this, we apply the tests to simulated observations of waveforms with parameters similar to GW150914~\cite{GW150914} and GW170608~\cite{GW170608}, as a paradigmatic high- and low-mass event, respectively.
We also apply the tests to simulated observations of non-GR waveforms of the type used in the parameterized tests (purely phenomenological in the case of the tests of GW generation, and coming from the dispersion due to a massive graviton for the GW propagation test), as well as the self-consistently modified effective-one-body (EOB) waveforms used to check the performance of the IMR consistency test in~\cite{Ghosh:2016qgn,Ghosh:2017gfp}.

We give an overview of the tests considered in Sec.~\ref{sec:test_overview}, describe the specifics of the simulated observations in Sec.~\ref{sec:inj}, present the results of the tests in Sec.~\ref{sec:results}, and conclude in Sec.~\ref{sec:concl}. We give the two-dimensional IMR consistency test plots in the Appendix.

\section{Tests of GR}
\label{sec:test_overview}

Here we give an overview of the tests of GR we consider in this study. All these tests rely on accurate models for binary black hole waveforms in GR, for which we primarily use the IMRPhenomPv2 model~\cite{Hannam:2013oca,Khan:2015jqa,Bohe:PPv2} as in all but the latest LIGO-Virgo tests of GR with binary black holes (e.g.,~\cite{GW150914_TGR, O2_TGR, O3a_TGR}). IMRPhenomPv2 models gravitational waves from black hole binaries on quasicircular orbits including the leading effects of spin precession. The LIGO-Virgo analyses also use the SEOBNRv4\_ROM model~\cite{Bohe:2016gbl} to give a check of the effects of waveform systematics, though it only allows for nonprecessing spins. Here we only use SEOBNRv4\_ROM for the FTA parameterized test, since this is the only test for which that model is used to obtain the primary results in the LIGO-Virgo analyses.

The analysis of GW data is often carried out in the framework of Bayesian inference and this framework is also used for all these tests in some form. In particular, to sample the likelihood, we use the implementation of nested sampling~\cite{Skilling2004a} in the LALInference code~\cite{Veitch:2014wba}, which is part of the LIGO Scientific Collaboration Algorithm Library Suite (LALSuite)~\cite{LALSuite}. We compute the likelihood integral from a low frequency of $20$~Hz to the Nyquist frequency of $1024$~Hz, as used in the LIGO-Virgo tests of GR for GW150914 and GW170608 (the two events we use as models for our simulated observations) in~\cite{O2_TGR}.\footnote{The analysis of GW170608 required a larger minimum frequency of $30$~Hz in the LIGO Hanford detector, due to the detector state at the time~\cite{GW170608}. We have used a low frequency of $20$~Hz in all detectors.} Additionally, two tests also use the BayesWave code~\cite{Cornish:2014kda,Cornish:2020dwh}, which uses Morlet-Gabor wavelets to model the gravitational wave signal, as opposed to a waveform model based on GR.

The first two tests described below use BayesWave to check the consistency of the waveforms inferred from the data using a GR model and LALInference. In one test we subtract the best-fit GR waveform inferred by LALInference from the data and use BayesWave to compute the residual \snr{}. For the second test, we use BayesWave to reconstruct the waveform directly from the data and compare the overlap of the reconstructed waveform with the GR waveforms that the LALInference analysis finds are good fits to the data. The third test we consider also tests the consistency of the signal, this time of the low- and high-frequency portions. Both of these portions are used to infer the final mass and spin and these two inferences are then checked for consistency. The next pair of tests checks that various parameterized deviations from the GR waveform model are consistent with their GR value of zero---the two tests differ in how these parameterized deviations are introduced. The final test we consider introduces a parameterized dispersion relation and constrains deviations from the nondispersive propagation of GWs in GR. Henceforth, we use the following abbreviations to refer to the various tests: IMR: Inspiral-Merger-Ringdown consistency test; TIGER: Test Infrastructure for GEneral Relativity parametrized test; FTA: Flexible Theory-Agnostic parameterized test; MDR: test of the Modified Dispersion Relation. We now describe these tests in full detail.

\subsection{Residuals}

Ground based \gw{} detectors are characterized by noise from various sources in different parts of the frequency spectrum (see, e.g.,~\cite{LIGOScientific:2014pky,LIGOScientific:2016gtq} for a discussion of noise sources in the Advanced LIGO detectors). The noise in the detector in any given time is generally assumed to be stationary Gaussian noise colored by the detector \psd{}. To analyze a potential \gw{} candidate event,  we model the detector time series as a summation of signal and noise (see, e.g.,~\cite{LIGOScientific:2019hgc}),
\begin{equation}
\mathbf{d}(t) = \mathbf{h}(t) + \mathbf{n}(t).
\end{equation}
Here $\mathbf{d}(t)$ is the detector data, $\mathbf{h}(t)$ is the \gw{} model waveform, and $\mathbf{n}(t)$ is Gaussian noise. The boldface notation here is used to specify that the quantities are vectors with one component each for every detector in the ground-based \gw{} detector network. We infer the best-fit (maximum likelihood) waveform, $\mathbf{h}_\text{maxL}(t)$, using LALInference and a GR model waveform, here IMRPhenomPv2. If $\mathbf{h}_\text{maxL}(t)$ is an accurate estimate of the true signal, then the residual, defined as $\mathbf{r}(t) = \mathbf{d}(t) - \mathbf{h}_\text{maxL}(t)$, should be consistent with noise. We test this consistency by analyzing $\mathbf{r}(t)$ with BayesWave. Since BayesWave relies on wavelets to model the signal waveform, any loud multi-detector coherent features in $\mathbf{d}(t)$ not accounted for by the signal model $\mathbf{h}(t)$ are potentially reconstructed as parts of the signal reconstruction. In the case of a faithful reconstruction of the true underlying signal by $\mathbf{h}_\text{maxL}(t)$, the BayesWave signal model will produce waveform samples whose median is consistent with noise.

Similar to \cite{O2_TGR, O3a_TGR, O3b_TGR}, we constrain the loudness of the residual by calculating the $90\%$ credible upper limit on the network \snr{} $\rho_\text{res}$ of the waveform samples. For the case of Gaussian noise, this tends to be $\lesssim 5$ for the LIGO-Hanford, LIGO-Livingston, and Virgo network. Specifically, we generated $200$ sets of simulated Gaussian noise timeseries in each detector colored with the same ``O3low'' detector \psd{}s~\cite{Aasi:2013wya} used in this analysis, and analyzed them with BayesWave. We then computed the $90\%$ credible upper limit on the network \snr{} on each of these observations and found that $90^\mathrm{th}$ percentile of this distribution is $\sim 5$.  

\subsection{Reconstructions}
\label{ssec:rec}

The residuals test is used to place constraints on the quality of the signal reconstruction by characterizing the residual and studying its consistency with background noise. The waveform reconstructions test on the other hand, approaches the question of signal consistency by studying the waveform itself. LALInference and BayesWave both offer signal reconstructions, $\hh_{\mathrm{LI}}$ and $\hh_{\mathrm{BW}}$, respectively. Both these algorithms rely on fundamentally different waveform models, i.e., GR-based and wavelet-based respectively. An agreement between their signal reconstructions gives support to the GR model used in the LALInference reconstruction. We quantify this agreement by computing the overlap between the two waveforms. The overlap is defined as the noise weighted inner product of two normalized signals (discussed further in, e.g.,~\cite{LIGOScientific:2019hgc}). In our case, following~\cite{Ghonge:2020suv}, we define
\begin{equation}
{\cal{O}}_\textrm{B,L} := \frac{\langle \hh_\textrm{LI} | \hh_\textrm{BW} \rangle}{\sqrt{\langle \hh_\textrm{LI}|\hh_\textrm{LI} \rangle \langle \hh_\textrm{BW}|\hh_\textrm{BW} \rangle}},
\label{eq:network_overlap}
\end{equation}
where $\langle \cdot | \cdot \rangle$ applied to boldface quantities indicates an inner product taken over the network, defined by
\begin{equation}
    \langle \mathbf{a}|\mathbf{b} \rangle :=  \sum_{i = 1}^{n} 4 \Real \int_{0}^{\infty} \frac{\tilde{a}^i(f)\tilde{b}^{i*}(f)}{S_n^i(f)}df.
    \label{eqn:overlap}
\end{equation}
Here $n$ is the number of detectors ($3$ in the cases we consider) and the superscript $i$ is used to denote the signal in the $i^{\mathrm{th}}$ detector, whose \psd{} is $S_n^i(f)$. 
$\tilde{a}^{i}(f)$ is the Fourier transform of the time series $a^{i}$, and the superscript $*$ denotes the complex conjugate. Dividing by the \psd{} makes the overlap most sensitive to differences in the waveforms at frequencies where the detectors are most sensitive. The absolute value of the overlap is bounded between $1$ (complete agreement) and $0$ (complete disagreement). The overlaps for GW150914 and GW170608 are $\sim 0.98$ and $\sim 0.58$, respectively (see Fig.~2 in~\cite{Ghonge:2020suv}). The much smaller overlap for GW170608 is due to its smaller \snr{} and particularly its lower mass, which spreads out the power in the signal over a longer time and makes it more difficult for BayesWave to reconstruct the signal accurately. We expect (and find) larger overlaps for the simulated GR cases with no noise considered here, since both reconstructions are made less precise by noise, particularly the non-Gaussian noise that is actually present in gravitational wave detectors.

Since the values of the overlaps we find are generally quite close to $1$, we present results in terms of $\bar{\mathcal{O}} := 1 - \cal{O}_\textrm{B,L}$. We compute $\bar{\mathcal{O}}$  for the entire distribution of $\hh_{\mathrm{LI}}$ with the median $\hh_{\mathrm{BW}}$ waveform and obtain a distribution on $\bar{\mathcal{O}}$. We use a point estimate for $\hh_{\mathrm{BW}}$ (i.e., the median waveform) since individual BayesWave waveforms samples do not represent a physical waveform, but their median does represent a physically stable estimate of the true waveform. More details about this choice can be found in~\cite{Ghonge:2020suv}.

\subsection{IMR consistency test}
\label{ssec:imr}

A binary black hole coalescence goes through three distinct phases: an initial \emph{inspiral} where the two black holes spiral in due to the backreaction from GW emission, a \emph{merger} where the two black holes coalesce to form a single remnant object, and a final stage of \emph{ringdown} where the remnant black hole settles into a stable Kerr configuration through the emission of a quasinormal-mode spectrum of gravitational waves. Within the stationary phase approximation, the low- (high-)frequency portion of the frequency domain gravitational-wave signal frequencies comes from the early (late) portion of the time domain signal (see, e.g., the illustration in Fig.~10 of~\cite{Ghosh:2017gfp}). Thus, one can test the consistency of the inspiral and merger-ringdown portions of the signal by checking the agreement of the low- and high-frequency portions of the signal. Specifically, we choose to split the analysis at a frequency $\fc$ given by the (redshifted) frequency of the innermost stable circular orbit corresponding to the remnant black hole~\cite{Bardeen:1972fi}. One can then use these mutually exclusive parts of the signal to obtain two independent measurements of the initial masses and spins and then apply analytical fits to numerical relativity simulations~\cite{Hofmann:2016yih,Healy:2016lce,Jimenez-Forteza:2016oae} to these quantities to infer independent estimates of the mass and spin of the final black hole, $(\Mf,\af)$.\footnote{As in~\cite{GW170104}, we average the fits and augment the aligned-spin final spin fits with the contribution from in-plane spins~\cite{spinfit-T1600168}. However, as in~\cite{O2_TGR, O3a_TGR, O3b_TGR}, we do not evolve the spins before applying the fits, for technical reasons.}

The inspiral-merger-ringdown (IMR) consistency test checks that these two independent estimates of the final mass and spin are consistent with each other, as they must be if the data is well described by the waveform model used to perform this inference. We thus define fractional deviations in the estimates of the final mass and spin,
\begin{subequations}
\begin{align}
\frac{\Delta \Mf}{\bar{M}_\text{f}} &:= 2 \frac{\Mfin - \Mfmr}{\Mfin + \Mfmr}, \\
\frac{\Delta \af}{\bar{\chi}_\text{f}} &:= 2 \frac{\afin - \afmr}{\afin + \afmr},
\end{align}
\end{subequations}
where the ``insp'' and ``postinsp'' superscripts denote the estimates obtained from the inspiral and postinspiral portions of the signal. These fractional deviations should be consistent with zero if the waveform model is a good description of the observed signal. As in~\cite{O3a_TGR, O3b_TGR}, we present results with a flat prior on $\Delta \Mf/\bar{M}_\text{f}$ and $\Delta \af/\bar{\chi}_\text{f}$.

We now describe how we obtain $\fc$. As in the applications of the test to real gravitational wave data, e.g.,~\cite{O2_TGR, O3a_TGR, O3b_TGR}, we use the median values from the analysis of the simulated observation using GR waveform models.
Additionally, for comparison, we also apply the test using the same $\fc$ one obtains from the GR simulated observations corresponding to the modified GR cases (which are close to
the values one would obtain from the simulated waveforms themselves, as one would expect). We show these comparisons of $\fc$ and the resulting IMR consistency results in the Appendix.

For comparison with the \snr{}s recovered by the other tests, we compute a combined \snr{} from the inspiral and postinspiral analyses. To do this, we note that the \snr{}s from the two portions add in quadrature, so we can compute a probability density for them as is done for the fractional deviations in the Appendix of~\cite{Ghosh:2017gfp}, giving
\<
P(\rho_\text{tot}) = \int_0^{\rho_\text{tot}}\frac{P_\text{insp}(\rho_\text{insp})P_\text{postinsp}(\sqrt{\rho_\text{tot}^2 - \rho_\text{insp}^2})}{\sqrt{\rho_\text{tot}^2 - \rho_\text{insp}^2}}\rho_\text{tot}d\rho_\text{insp},
\?
where $P(\rho_\text{tot})$, $P_\text{insp}(\rho_\text{insp})$, and $P_\text{postinsp}(\rho_\text{postinsp})$ denote the probability densities for the total \snr{} and the \snr{}s in the inspiral and postinspiral, respectively. Specifically, we change variables from $\{\rho_\text{insp}, \rho_\text{postinsp}\}$ to $\{\rho_\text{insp}, \rho_\text{tot}\}$, where $\rho_\text{tot}  = \sqrt{\rho_\text{insp}^2 + \rho_\text{postinsp}^2}$, and marginalize over $\rho_\text{insp}$, noting that it is nonnegative and at most $\rho_\text{tot}$.

\subsection{Parameterized tests of gravitational-wave generation}
\label{ssec:gen}

In an alternative theory of gravity, the equations of motion describing the orbital evolution of a coalescing compact binary will in general be different from those in GR. Thus, the frequency evolution of the GW emission will in general be different from the one predicted by GR. The parametrized tests aim to detect GR violations by allowing for deviations in the frequency-domain phase coefficients of the GR waveform models, as initially proposed in~\cite{Arun:2006yw,Arun:2006hn,Yunes:2009ke,TIGER2012} 

The early inspiral dynamics of the compact binary is described with good accuracy using the well known PN approximation to GR (see, e.g.,~\cite{Blanchet:2013haa}). The frequency-domain phase in the PN approximation is obtained as an expansion in powers of the velocity parameter $v$, which is defined as a function of frequency $f$, i.e., $v = (\pi M_z f)^{1/3}$, where $M_z$ is the total redshifted mass of the binary. In the frequency domain, the phasing for a nonprecessing (i.e., aligned-spin) binary can be schematically written as (omitting additive constants and the effect of a time shift)
\begin{equation}
\Phi(f)=\frac{3}{128\eta\,v^5}\sum_k (\varphi_k v^k + \varphi_{k\text{l}}v^k \ln v),
\end{equation}
where  $\eta := m_1 m_2/(m_1+m_2)^2$ is the symmetric mass ratio ($m_{1,2}$ are the individual masses of the components of the binary) and the summation is taken over all the PN orders for which we know the phase evolution. 
The terms $\varphi_{k}$ and $\varphi_{k\text{l}}$ are PN coefficients which encode various physical effects in the dynamics of the binary and hence are functions of binary parameters such as masses and spins. 

In the IMRPhenomPv2 waveform model, the inspiral portion of the GW phasing is described using the PN phasing augmented by phenomenological coefficients obtained by fitting to numerical relativity waveforms. Similarly, the late inspiral and merger-ringdown portions of the GW phasing are described by phenomenological expressions in powers of frequency, which include 
the late inspiral ({\it a.k.a.}\ intermediate) coefficients $\beta_i$ and merger-ringdown coefficients $\alpha_i$. 
The dependence of $\alpha_i$ and  $\beta_i$ on binary parameters is also obtained by fitting to numerical relativity waveforms. 
For convenience, we denote the phasing coefficients in all three regimes collectively by $p_i$.

One version of the parametrized test is the Test Infrastructure for GEneral Relativity (TIGER) approach~\cite{TIGER2012,TIGER2014,Meidam:2017dgf}, whose implementation applied to LIGO-Virgo detections to date uses IMRPhenomPv2 as the underlying GR model. In this method, one introduces dimensionless fractional deviations $\delta {\hat p}_k$ in each phasing coefficient $p_k$ as fractional deviations from GR such that $p_k \rightarrow p_k^\text{NS} (1+ \delta {\hat p}_k) + p_k^\text{S}$. Here the superscripts NS and S denote the nonspinning and spinning parts of the GR phasing coefficients. The fractional deviations are only scaled by the nonspinning part of the coefficients to avoid cases where the spin contributions cause the GR coefficient to vanish. Additionally, these coefficients only modify the phasing of the underlying aligned-spin waveform model (IMRPhenomD), not the precessing dynamics used to twist up those waveforms to obtain the final IMRPhenomPv2 precessing waveform model.

A second approach is the Flexible Theory-Agnostic (FTA) approach, introduced in~\cite{GW170817_TGR}, which is not tied to a specific waveform model, but only considers the inspiral PN coefficients. Here it is applied using the SEOBNRv4\_ROM waveform model as the GR baseline, as in the LIGO-Virgo analyses~\cite{O2_TGR,O3a_TGR,O3b_TGR}. Unlike the TIGER approach, where the deviations in the early-inspiral and late-inspiral coefficients affect all the higher-frequency portions of the waveform through the $C^2$ matching used to stitch together the different parts of the IMRPhenomD model~\cite{Khan:2015jqa}, in the FTA approach, the deviations are tapered to zero at some frequency. Here, as in the LIGO-Virgo analyses~\cite{O2_TGR,O3a_TGR,O3b_TGR}, this frequency is chosen to be $0.35$ times the peak frequency of the SEOBNRv4 model. This choice is designed to be consistent with the end of the early-inspiral portion of IMRPhenomD ($G M_z f / c^3 = 0.018$)~\cite{Khan:2015jqa} used in TIGER.
The FTA deviations are parameterized in terms of the spinning PN coefficients, unlike TIGER's use of just the nonspinning part. They are then mapped to the TIGER parameterization to obtain the final results, as discussed in~\cite{O2_TGR,O3a_TGR,O3b_TGR}. Since both tests only modify the nonprecessing phasing, their results are then directly comparable.

If the waveform model used (e.g., IMRPhenomPv2 or SEOBNRv4\_ROM) is a good description of the signal, all of the fractional deviations introduced in the test should be consistent with zero. Thus, one ideally would constrain all the $\delta \hat{p}_k$ simultaneously, as one would in general expect all of them (at least above some PN order) to simultaneously deviate from their GR values if there is a deviation from GR. 
However, the fractional deviations are highly correlated, so in practice it is very difficult to measure all of them with current SNRs---see~\cite{GW150914_TGR} for an explicit illustration of this with GW150914. Nevertheless, recent works have shown that the ability to constrain multiple PN coefficients together can be improved with multiband observations by LISA and third-generation ground-based detectors~\cite{Gupta:2020lxa,Datta:2020vcj}, and/or with the use of principal component analysis~\cite{Shoom:2021mdj,Saleem:2021nsb}.  
However, here we follow the procedure in the most recent LIGO-Virgo testing GR papers~\cite{O2_TGR, O3a_TGR,O3b_TGR} and only vary one deviation parameter at a time. This has been shown to be sufficient to detect at least some deviations from GR~\cite{Meidam:2017dgf}.
 
For this first study, we consider the $2$PN early-inspiral coefficient $\delta\hat{\varphi}_4$, since it corresponds to the leading order of the deviation from GR in the inspiral phasing of our modified EOB waveforms. We also consider the $1$PN coefficient $\delta\hat{\varphi}_2$ in a few cases, since it corresponds to the PN order of the massive graviton dephasing we use for simulated observations and the modified dispersion relation test. We also consider the TIGER late-inspiral and merger-ringdown parameters $\delta\hat{\beta}_2$ and $\delta\hat{\alpha}_2$ since they are somewhat better constrained than the other late-inspiral and merger-ringdown parameters, respectively (see the Appendix of~\cite{O2_TGR}).

\begin{table*}
\caption{\label{tab:inj_pars}
Parameters of the waveforms considered in this study (most given to three significant digits): $M_z$ and $D_\text{L}$ are the binary's redshifted total mass and luminosity distance, while ``RA'' and ``dec.''\ denote the right ascension and declination, respectively. The mass ratio is $1.22$ for the (modified) EOB waveforms and $1$ for all others and the polarization angle is $3.9$~rad in all cases. Each non-GR parameter corresponds to a separate waveform (in the TIGER/FTA case, two different waveforms).
}
\begin{tabular}{ccccccccccc}
\hline\hline
\multirow{3}{*}{Name} & \multicolumn{6}{c}{GR parameters} & \hphantom{X} & \multicolumn{3}{c}{non-GR parameters}\\
\cline{2-7}
\cline{9-11}
& $M_z$ & $D_\text{L}$ & inclination & RA & dec.\ & GPS time & & mod.~EOB & MDR & TIGER/FTA\\
& [$M_\odot$] & [Mpc] & [rad] & [rad] & [rad] & [s] & & $a_2$ & $\tilde{A}_0$ & $\delta\hat{\varphi}_4$\\
\hline
\multirow{2}{*}{GW150914-like ($\text{M}_{72}$)} & \multirow{2}{*}{$72.2$} & \multirow{2}{*}{$452$} & \multirow{2}{*}{$2.83$} & \multirow{2}{*}{$1.68$} & \multirow{2}{*}{$-1.27$} & \multirow{2}{*}{$1126259462$} & & $400$ & $5$ & $-13$\\
& & & & & & & & $40$ & $1$ & $-2$\\
\hline
GW170608-like ($\text{M}_{20}$) & $19.9$ & $364$ & $2.15$ & $3.64$ & $\hphantom{-}0.89$ & $1180922494$ & & $40$ & $10$ & $-2$\\
\hline\hline
\end{tabular}
\end{table*}

\begin{figure*}[htb]
\centering
\subfloat{
\includegraphics[width=0.42\textwidth]{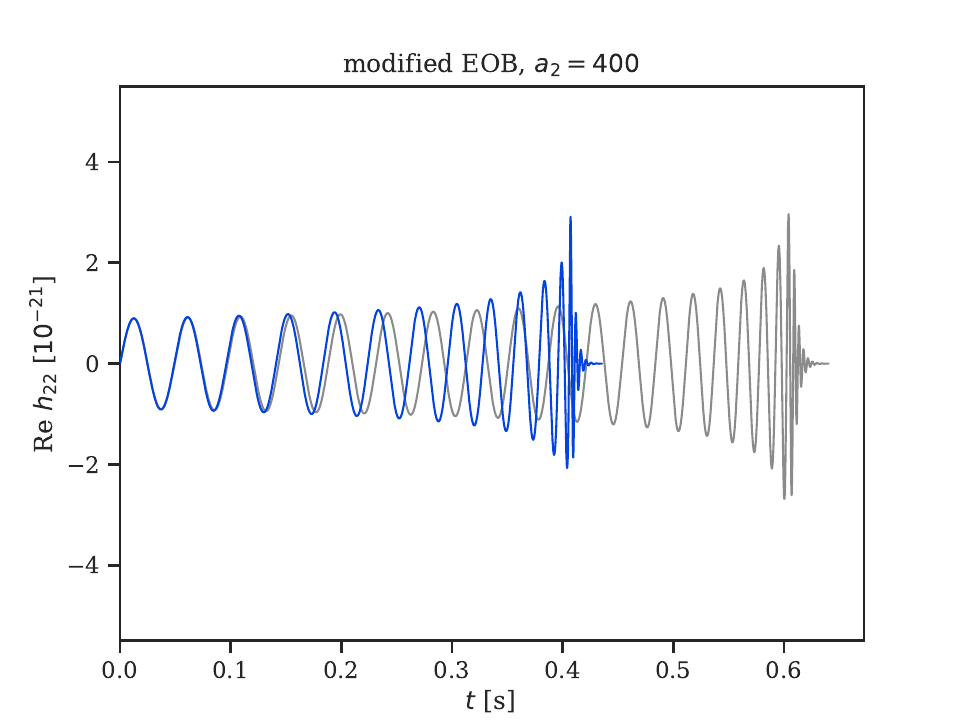}
}
\quad
\subfloat{
\includegraphics[width=0.42\textwidth]{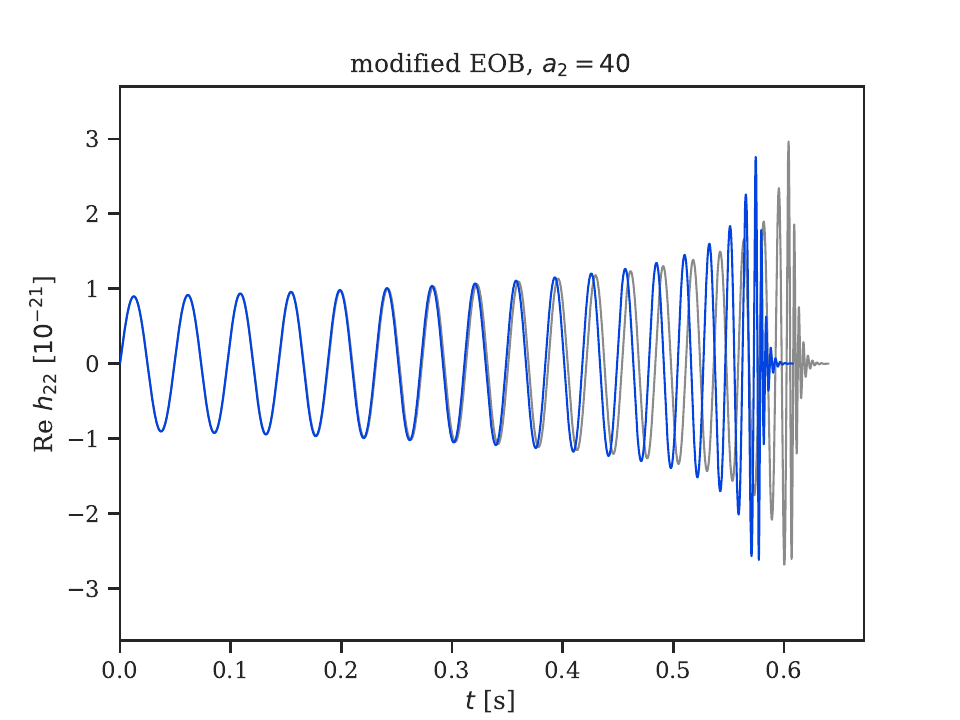}
}\\
\subfloat{
\includegraphics[width=0.42\textwidth]{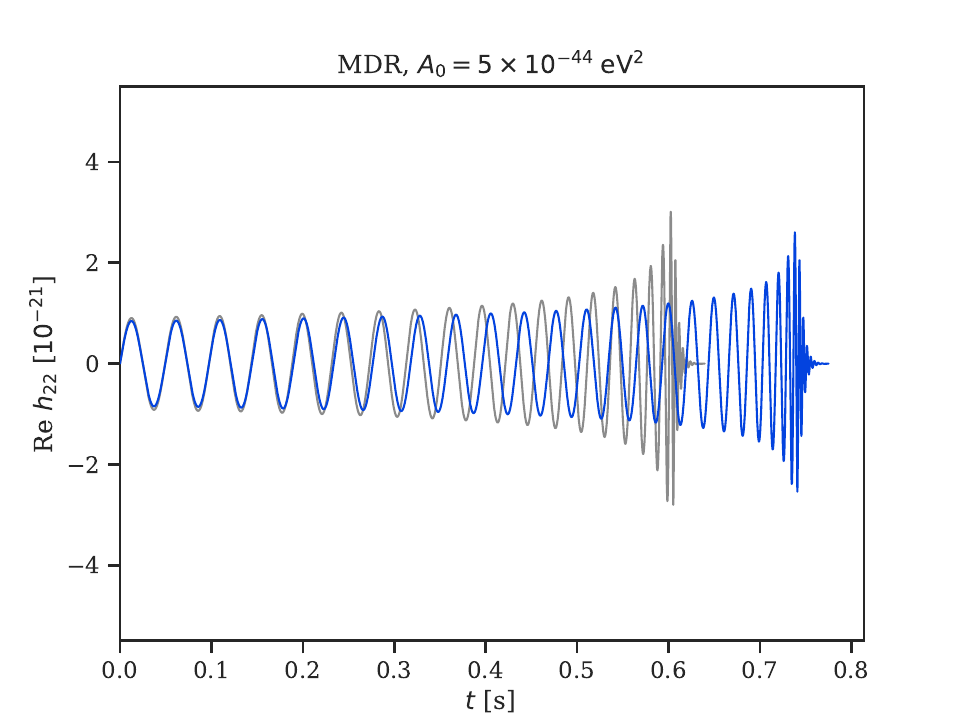}
}
\quad
\subfloat{
\includegraphics[width=0.42\textwidth]{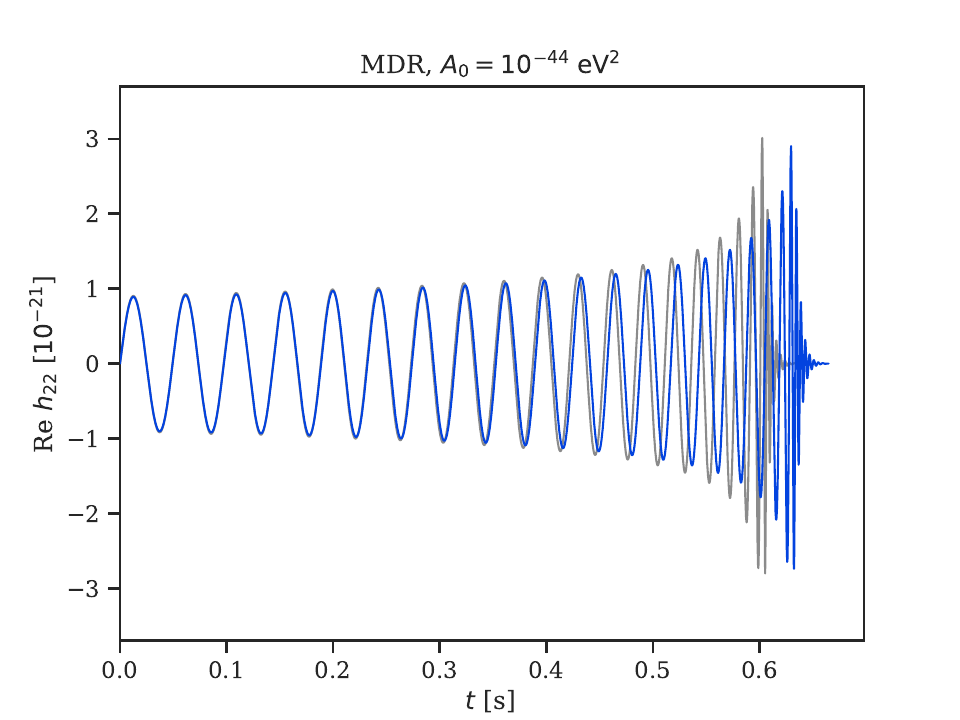}
}\\
\subfloat{
\includegraphics[width=0.42\textwidth]{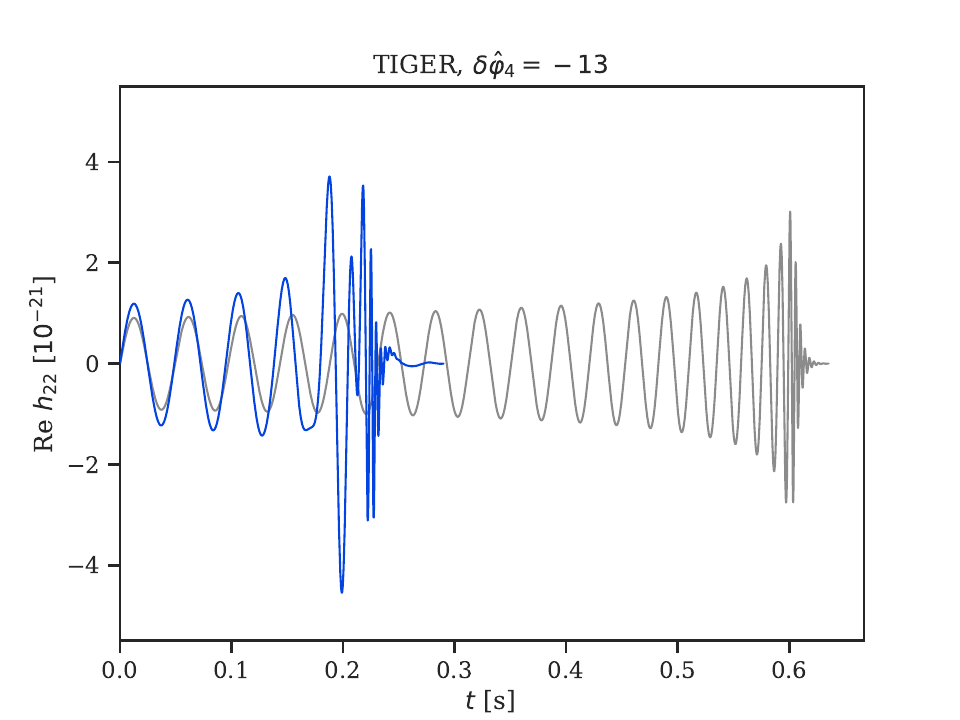}
}
\quad
\subfloat{
\includegraphics[width=0.42\textwidth]{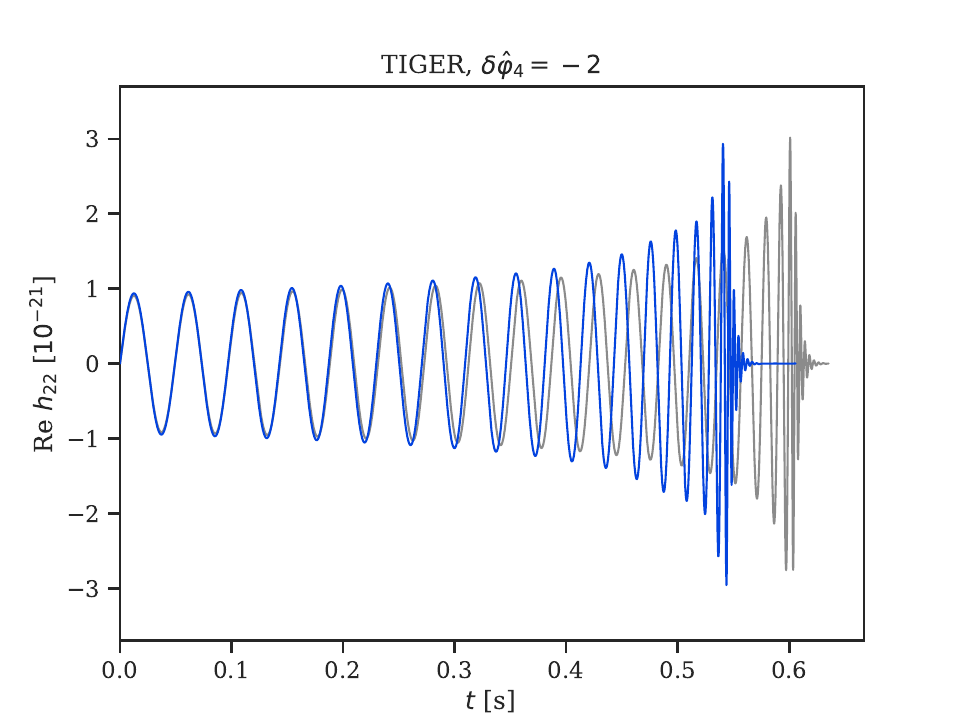}
}\\
\subfloat{
\includegraphics[width=0.42\textwidth]{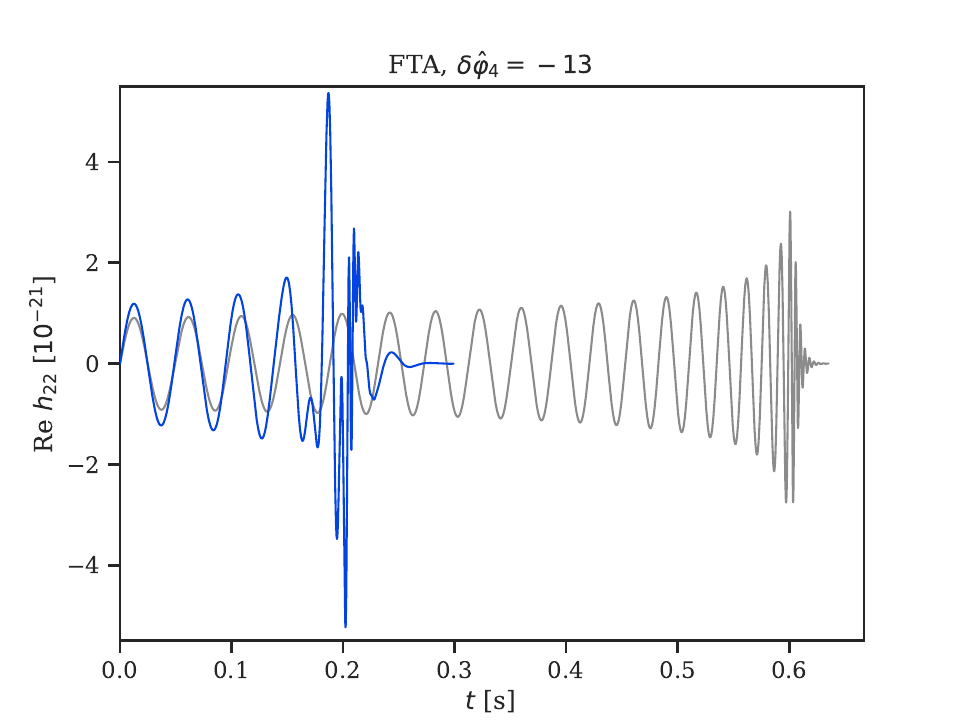}
}
\quad
\subfloat{
\includegraphics[width=0.415\textwidth]{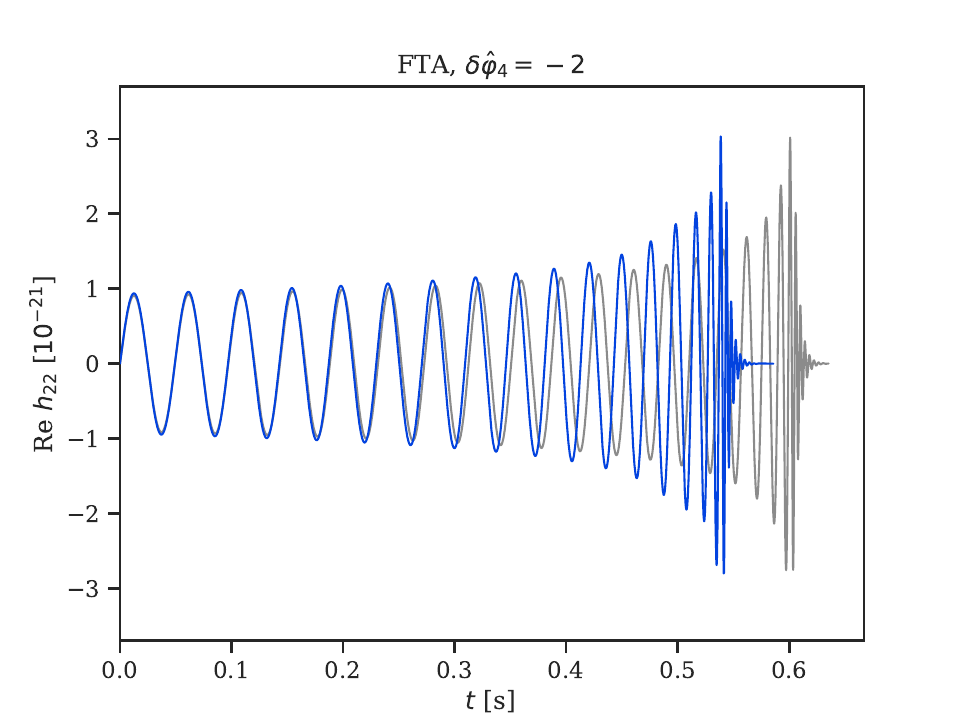}
}
\caption{\label{fig:inj_TD_GW150914-like} The GW150914-like waveforms in the time domain. We show the real part of the $\ell = m = 2$ spin-$(-2)$-weighted spherical harmonic mode of the strain, aligning the non-GR waveforms (blue) with the corresponding GR waveforms (grey) at $20$~Hz, which is also the frequency at which the plots start. The larger GR deviation is on the left, and thus those plots have a larger vertical axis range than those on the right.
}
\end{figure*}

\begin{figure*}[htb]
\centering
\subfloat{
\includegraphics[width=0.42\textwidth]{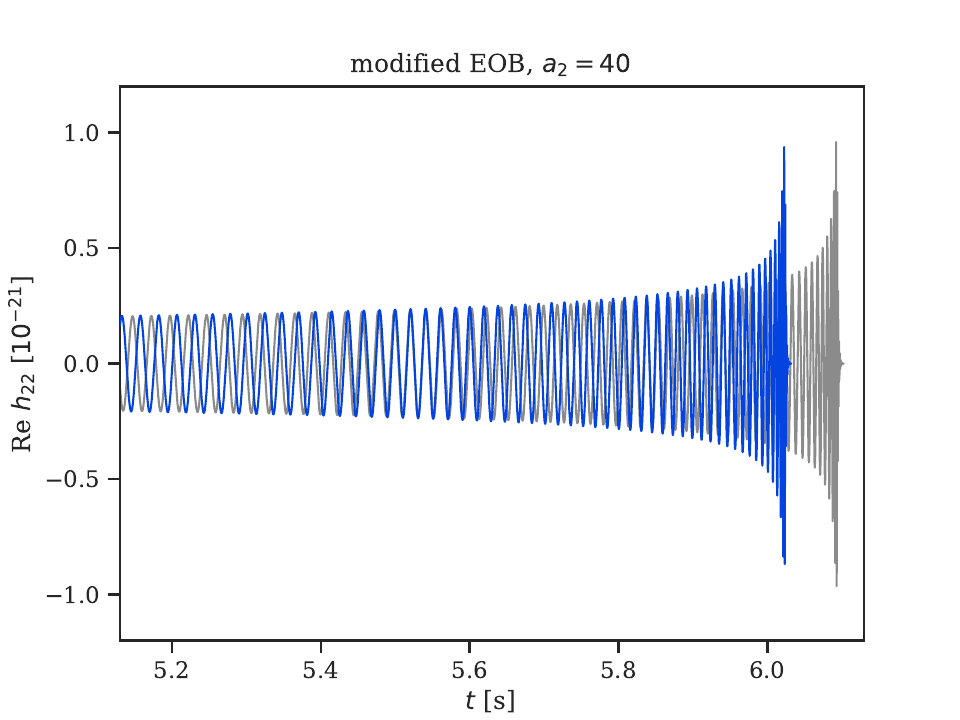}
}
\quad
\subfloat{
\includegraphics[width=0.42\textwidth]{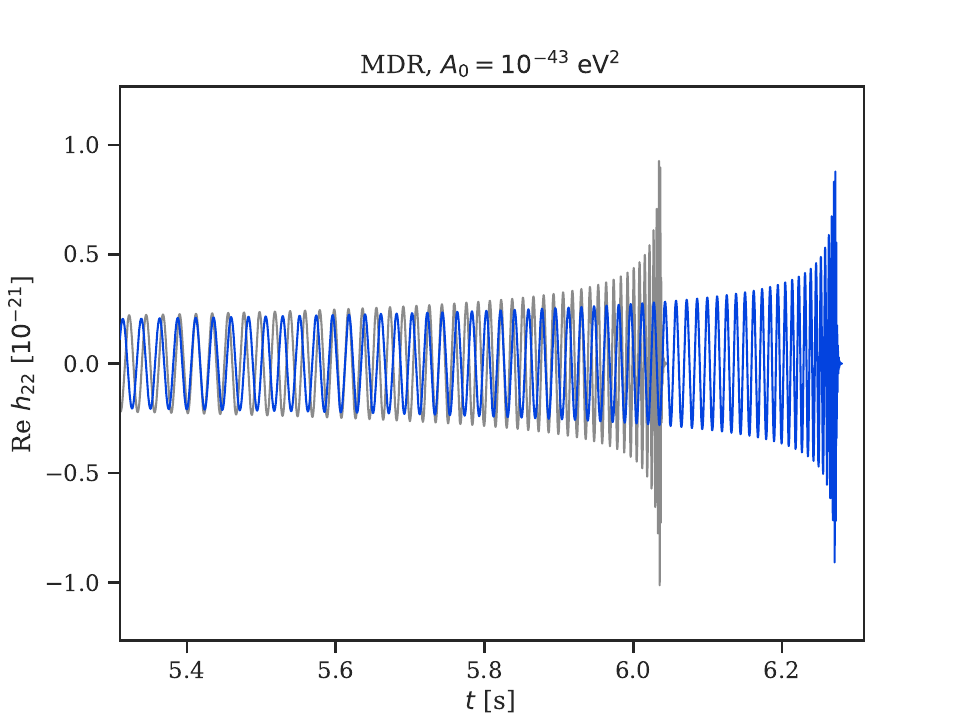}
}\\
\subfloat{
\includegraphics[width=0.42\textwidth]{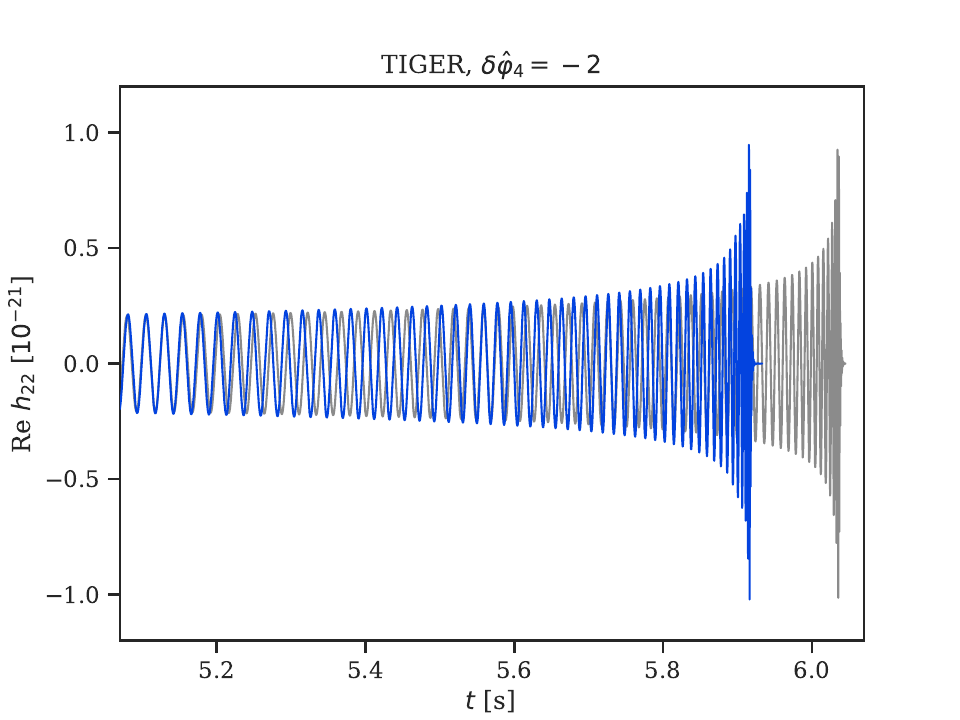}
}
\quad
\subfloat{
\includegraphics[width=0.42\textwidth]{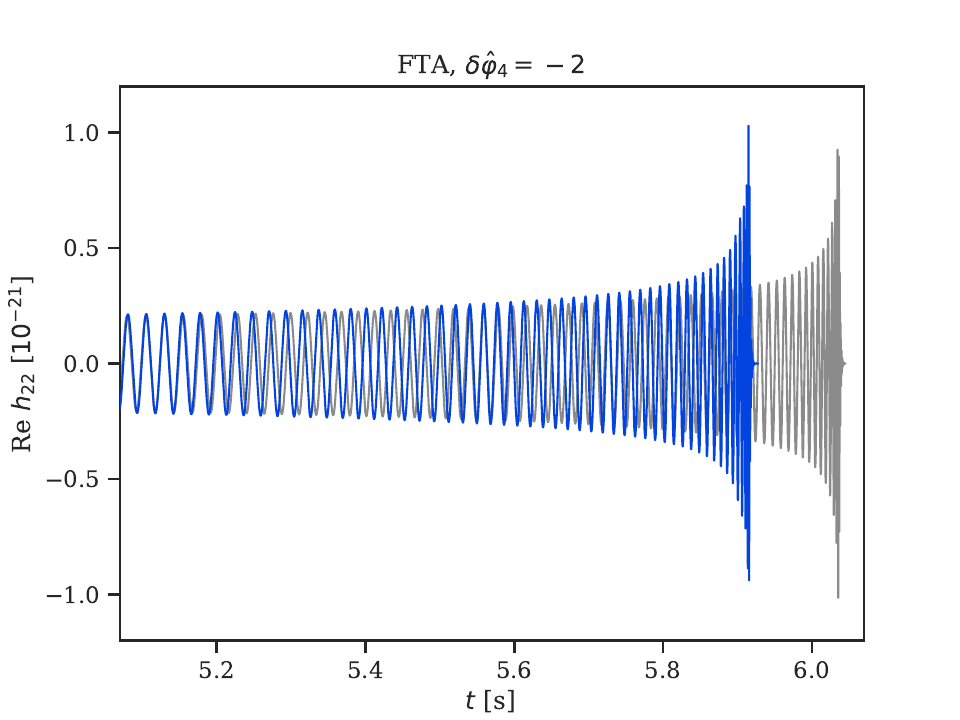}
}
\caption{\label{fig:inj_TD_GW170608-like} The analog of Fig.~\ref{fig:inj_TD_GW150914-like} for the GW170608-like waveforms. We still align the waveforms at $20$~Hz, but only show the final $\sim 1$~s of the signal.
}
\end{figure*}

\begin{figure*}[htb]
\centering
\subfloat{
\includegraphics[width=0.75\textwidth]{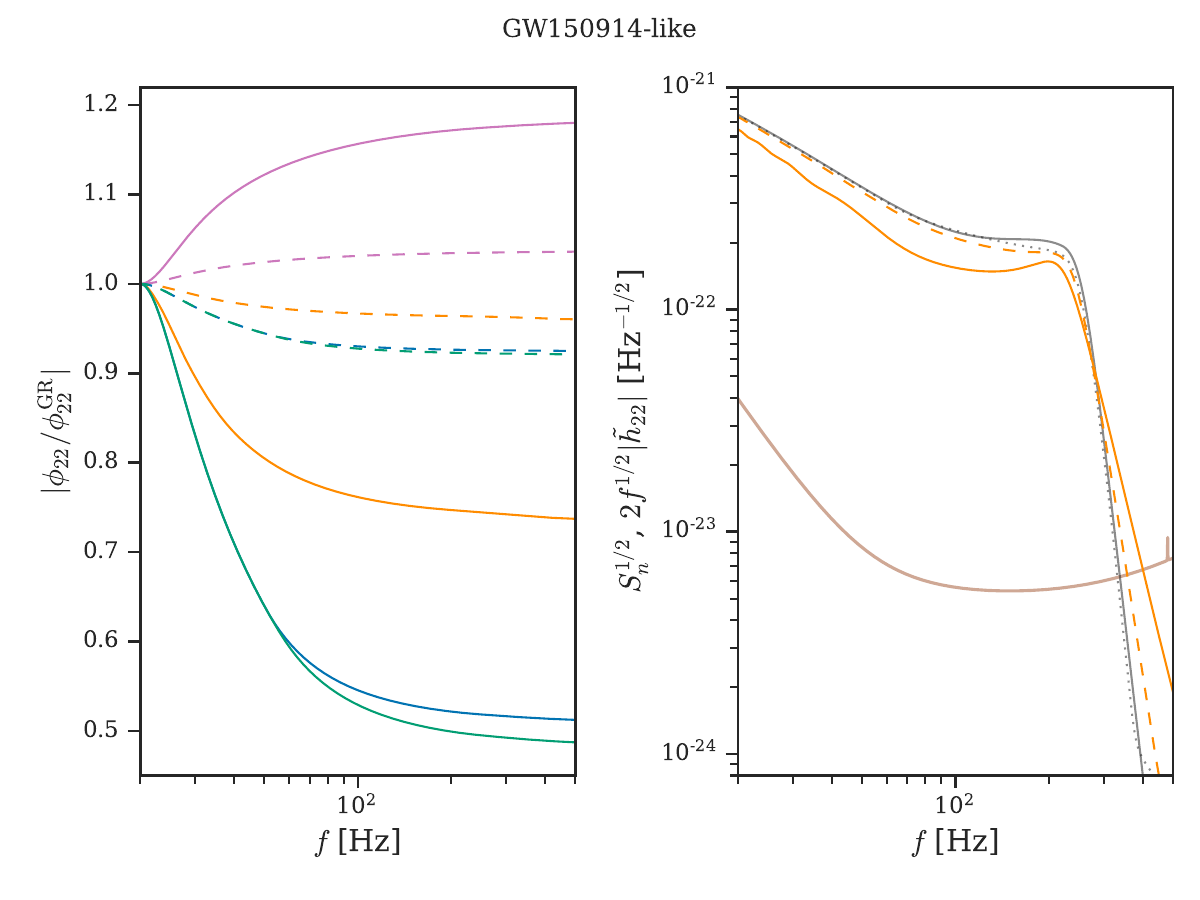}
}\\
\subfloat{
\includegraphics[width=0.75\textwidth]{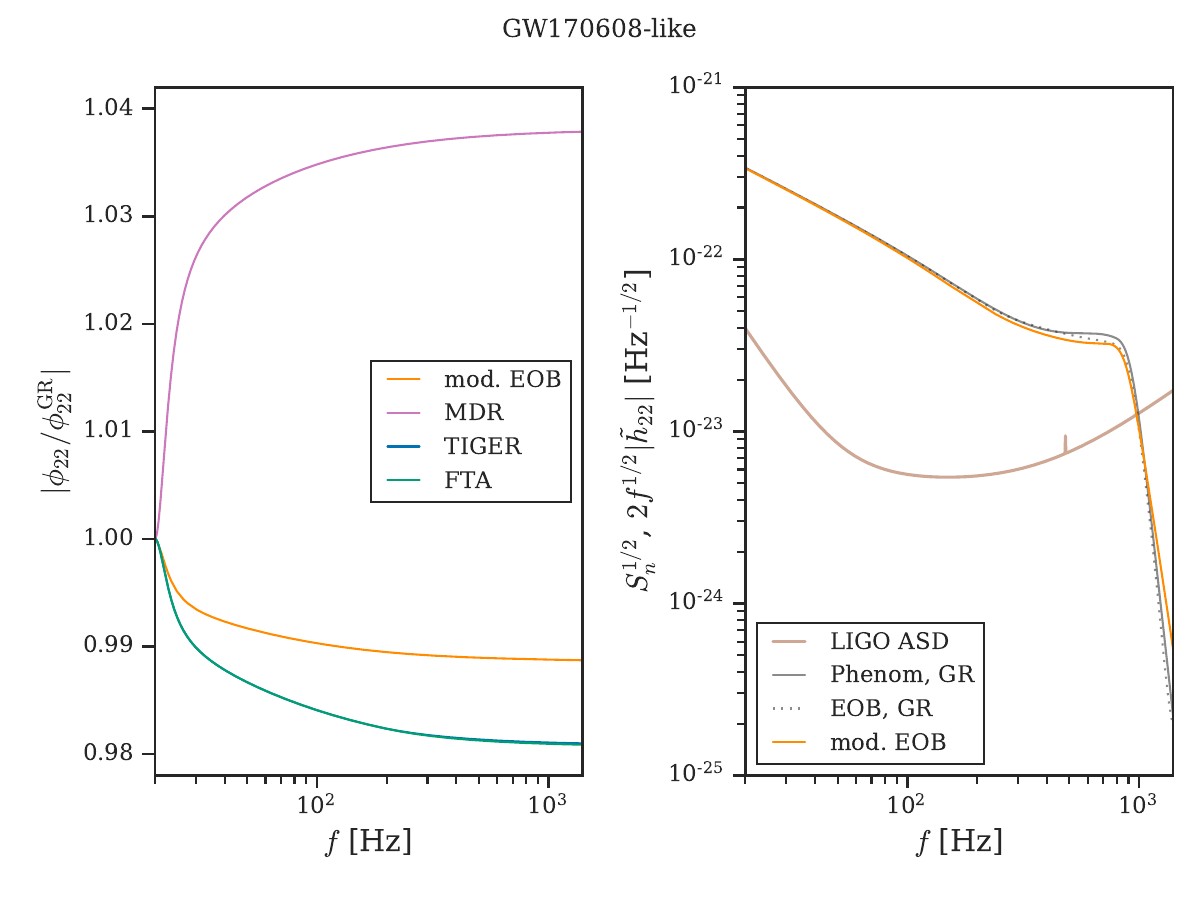}
}
\caption{\label{fig:inj_FD} The GW150914-like and GW170608-like waveforms in the frequency domain. Each of the pairs of plots shows the (frequency-domain) phase (left) and amplitude (right) of the real part of the $\ell = m = 2$ spin-$(-2)$-weighted spherical harmonic mode of the strain. We align the phases at $20$~Hz and set the time shift so that the phase derivative at $20$~Hz is $1$~s (a somewhat arbitrary choice selected to reduce sharp gradients in the plot near $20$~Hz). We then plot the ratio of the phase of the non-GR waveforms to the corresponding GR phase. The dashed lines correspond to the waveforms with smaller GR violations. In the GW170608-like case, the FTA and TIGER phases are almost identical, but the FTA dephasing is slightly greater, as it is in the other cases. For the non-EOB waveforms, we only plot the amplitude of the GR waveform, since these non-GR waveforms only modify the frequency-domain phase. The amplitude is scaled to show the contributions to the SNR integrand, compared to the amplitude spectral density of the noise (cf., e.g., Fig.~1 in~\cite{O1_BBH}); we also show the LIGO noise curve we use in the analysis, for comparison (the O3low one from~\cite{Aasi:2013wya}).
}
\end{figure*}

\subsection{Modified dispersion test}

General relativity predicts that GW propagation is nondispersive.  That is, the velocity of propagation is independent of the waves' frequency.  This property is equivalent to a massless graviton (using quantum terminology here and in the following for convenience, though we are only concerned with classical effects here). On the other hand, certain alternative theories of gravity predict a massive graviton or other dispersion effects as the waves travel from the source to the observer (see, e.g.,~\cite{Mirshekari:2011yq,GW170104}). We thus consider a parameterized dispersion relation, following~\cite{Mirshekari:2011yq}, which encompasses the leading predictions of a number of different alternative theories,
\begin{equation}
E^2=p^{2}\,c^2+A_\alpha\,p^{\alpha}\,c^{\alpha}.
\label{eq:liv}
\end{equation}
Here $A_\alpha$ is the amplitude of the modified dispersion (zero in GR) and has dimension of $\rm{[Energy]^{2-\alpha}}$; $\alpha$ is a dimensionless constant. In particular, $\alpha = 0$ and $A_0 > 0$ corresponds to a massive graviton with mass $m_\text{g} = A_0^{1/2}/c^2$. We will frequently use the dimensionless quantity $\tilde{A}_0 := A_0/(10^{-44} \text{ eV}^2)$, since this is a convenient scale for the amplitudes we are considering.

As discussed in~\cite{O2_TGR}, one can reasonably take  gravitational waves near the source to be those predicted by GR to a very good approximation. The only modification to the waveform is the dephasing that builds up over the waves' propagation to Earth [see~\cite{O2_TGR} for the explicit expressions, though the exponent in that paper's Eq.~(4) should be $1/(2 - \alpha)$, as noted in~\cite{O3a_TGR}].\footnote{We use the same TT + lowP + lensing + ext cosmological parameters from~\cite{Ade:2015xua} as in~\cite{O2_TGR,O3a_TGR,O3b_TGR}.} For instance, in the massive graviton case, the length scale of the Yukawa modification to the Newtonian potential is constrained to be much larger than the size of the binary, so this modification's effect on the binary's dynamics is negligible.
For this first analysis, we restrict to the case $\alpha = 0$, thus including the massive graviton case, though we also allow for $A_0 < 0$ along with $A_0 > 0$. 

As in~\cite{GW170104,O2_TGR,O3a_TGR,O3b_TGR}, we sample separately for $A_0 > 0$ and $A_0 < 0$ (the sampling is carried out in the logarithm of an effective wavelength and the results are then reweighted to a flat prior in $A_0$). We also combine together the $A_0 > 0$ and $A_0 < 0$ probability distributions (weighted by their evidences) to allow us to quote the quantile of the distribution corresponding to the GR value of $A_0 = 0$, again as in~\cite{O2_TGR,O3a_TGR,O3b_TGR}, as well as the SNR distribution.

The dephasing from the modified dispersion relation maps onto a modified PN coefficient in the inspiral for certain values of $\alpha$, notably corresponding to a modified $1$PN coefficient for the $\alpha = 0$ case we consider. However, the modified dispersion dephasing affects the entire waveform, and thus is quite different from the TIGER and FTA modifications to the PN coefficients described in Sec.~\ref{ssec:gen}, which are only applied to the inspiral portion of the waveform.

\section{Simulated observations}
\label{sec:inj}

We consider a variety of simulated observations, both with and without deviations from GR. Specifically, we consider waveforms with the modifications used in both the parameterized tests of GW generation and in the modified dispersion tests and their GR analogs, given by the IMRPhenomPv2 waveform model.
We still use IMRPhenomPv2 as the base waveform in the FTA case, instead of SEOBNRv4\_ROM, which is used in the application of the test, to avoid introducing another GR model.
We also consider the EOB waveforms used to check the performance of the IMR consistency test in~\cite{Ghosh:2016qgn,Ghosh:2017gfp}, described in more detail below, as well as their GR analogs. The modified EOB waveforms are only available for nonspinning systems, since they are obtained by modifying an EOB code for nonspinning binaries. Thus, we only consider simulated observations of nonspinning binaries in this paper. Additionally, since we use waveform models that only contain the dominant quadrupolar modes of the GW signal to analyze these waveforms, we just include these modes of the modified EOB waveforms, to avoid any systematics from omitted higher modes, though these would be expected to be minor for the nonspinning close-to-equal-mass cases we consider.

\subsection{EOB waveforms with modified energy flux}

As detailed in~\cite{Ghosh:2017gfp}, the modified EOB waveforms modify the energy flux in the IHES EOB model~\cite{Damour:2012ky, IHES_EOB_url} by multiplying the $(\ell, m)$ modes of the waveform that first enter the energy flux at $2$PN, viz., the $(3, \pm 2)$, $(4, \pm 4)$, and $(4, \pm 2)$ modes, by a factor $a_2^{1/2}$, so the modes' contribution to the energy flux is multiplied by $a_2$.\footnote{We call this factor $a_2$ instead of $\alpha_2$ (as in~\cite{Ghosh:2017gfp}) to avoid confusion with the $\delta\hat{\alpha}_2$ parameterized test parameter.} We then iteratively adjust the mass and spin of the final black hole used to calculate the quasinormal mode (QNM) frequencies for the ringdown model so that the waveforms satisfy energy and angular momentum balance (using modes through $\ell = 7$, to match the highest $\ell$ in the tabulated QNM results). We still use this older EOB model instead of a more recent one because it is implemented in Matlab and thus easy to modify and has a ringdown model given purely in terms of Kerr QNMs, without any further fits. Updating this modified EOB waveform construction to more recent EOB models that include spin, such as~\cite{Bohe:2016gbl,Nagar:2018zoe} (with extensions to higher modes in~\cite{Cotesta:2018fcv,Nagar:2020pcj} and precession in~\cite{Ossokine:2020kjp,Akcay:2020qrj,Gamba:2021ydi}), will be the subject of future work.

To give an idea of the effect of these large deviations in the energy flux on the binary's dynamics, we consider the mass of the final black hole as a fraction of the total mass $M_\text{f}/M$ and the dimensionless spin of the final black hole $\chi_\text{f}$. For the $a_2 = 400$ case, these are $M_\text{f}/M = 0.86$ and $\chi_\text{f} = 0.30$, while for $a_2 = 40$ they are $M_\text{f}/M = 0.92$ and $\chi_\text{f} = 0.57$. Both of these pairs are outside the region of pairs obtainable in GR, shown in Fig.~6 of~\cite{Ghosh:2017gfp}. For comparison, the final mass and spin in the GR case obtained using the fit in the IHES EOB code are $M_\text{f}/M = 0.95$ and $\chi_\text{f} = 0.68$, which agree with those obtained from the self-consistent calculation with the GR value of $a_2 = 1$ to the number of digits shown, only differing from them by one in the next decimal place (i.e., by $\sim 10^{-3}$). For all the other non-GR waveforms considered, the radiated energy and angular momentum are unchanged from their GR values, since only the frequency domain phase is altered, and these quantities just depend on the frequency domain amplitude, as one can see by converting the expressions in, e.g.,~\cite{RANT} to the frequency domain using the Parseval-Plancherel identity (i.e., the unitarity of the Fourier transform) and noting that the time derivatives of the strain become multiplication by the frequency in the frequency domain.

\subsection{Parameter choices}

For the binary's GR parameters, we consider a case like GW150914~\cite{GW150914} as well as a lower-mass case like GW170608~\cite{GW170608} (as both of these are consistent with being nonspinning)..\footnote{We expect that the different total masses of the two cases will lead to a difference in the accuracy to which the parameters can be inferred, due to, e.g., the number of cycles in band, as discussed in, e.g.,~\cite{Cutler:1994ys}, in addition to the difference in accuracy from the different SNR.} For the GW150914-like cases, we consider both large and moderate GR deviations, while for the GW170608-like cases, we only consider moderate GR deviations (for the massive graviton [MDR] case, this corresponds to a larger value of $A_0$ than in the GW150914-like case, since the test is less sensitive for this lower-mass system at a somewhat smaller distance). The mass ratio of the (modified) EOB waveforms was chosen to be the same as a numerical relativity simulation, SXS:BBH:310~\cite{SXS:catalog,Boyle:2019kee,Abbott:2016apu}, in case it proved necessary to compare with these waveforms, though this did not end up being the case. The other simulated observations are equal mass, due to a bug in their creation. The parameters of all the simulated observations are given in Table~\ref{tab:inj_pars}. The other GR parameters were obtained from the sample from the GWTC-1~\cite{GWTC-1_paper} release~\cite{GWOSC:GWTC-1} that corresponds to the median of the marginalized total mass distribution. There is a slight difference in the right ascension of the GW150914-like case from what this procedure gives, due to a transcription error---its value for the closest sample is $1.59$~rad. Additionally, there was a transcription error in the inclination angle, right ascension, and declination for the GW170608-like case and their values were permuted, while they should have been $2.46$, $2.15$, and $0.50$. The right ascension and declination values given in Table~\ref{tab:inj_pars} are wrapped so the declination has a magnitude less than $\pi/2$.

We chose the magnitude of the larger GR deviation for the modified EOB waveforms to be the same as the one used in~\cite{Ghosh:2016qgn} to illustrate that the IMR consistency test could pick up a self-consistent deviation from GR that is significant but passes the binary pulsar tests and is close enough to GR waveforms that it would likely be detected by a matched filtering pipeline using GR waveforms, since it gives a fairly high SNR in such a pipeline. In the inspiral, this deviation corresponds to a parameterized test modification of $\delta\hat{\varphi}_4 \simeq -14$.\footnote{We compute this using the TaylorF2 stationary phase approximation expression for the frequency domain phase in terms of the binary's PN binding energy and (modified) energy flux, as in~\cite{Buonanno:2009zt}.} We thus chose $\delta\hat{\varphi}_4 = -13$ for the parameterized test simulated observations with the larger GR deviation. (We used $-13$ instead of $-14$ by mistake, but found that the resulting TIGER and FTA waveforms are already very different from the GR waveforms, so we did not want to use a larger magnitude deviation.) For the massive graviton simulated observation with the larger deviation, we chose a value of $A_0$ that is well outside of the posterior probability distribution for GW150914 shown in Fig.~8 of~\cite{O2_TGR}.

For the smaller GR deviation, we chose a ten times smaller deviation in the modified EOB waveforms (still twice as large as the deviation used to check the IMR consistency test for a population of detections in~\cite{Ghosh:2017gfp}) and the corresponding parameterized test deviation rounded to the nearest integer, $\delta\hat{\varphi}_4 = -2$ (so $1 + \delta\hat{\varphi}_4$ is about ten times smaller than in the larger case). In the massive graviton case, we chose the $A_0$ values to be around the $90\%$ bounds for GW150914 and GW170608 given in Table~IV of~\cite{O2_TGR}. Even the smallest massive graviton deviation is still much larger than we would expect given the latest bound of $1.27\times 10^{-23}\text{ eV}/c^2$ on the graviton mass $m_\text{g}$ from the analysis of confident GWTC-2 binary black hole events~\cite{O3a_TGR}, which corresponds to $\tilde{A}_0 \leq 1.61\times 10^{-2}$, since $m_\text{g} = A_0^{1/2}/c^2$. We choose these relatively large values of the graviton mass because we are interested in deviations from GR that could potentially be detected with a single event at the moderate SNRs we consider, while the observational bound on the mass of the graviton is obtained by combining together the constraints from many events.

We plot all these waveforms in the time domain in Figs.~\ref{fig:inj_TD_GW150914-like} and \ref{fig:inj_TD_GW170608-like} and in the frequency domain in Fig.~\ref{fig:inj_FD}. The frame files for all these simulated observations are available at~\cite{frame_files_Zenodo}.

\section{Results}
\label{sec:results}

We now analyze all the simulated observations from Sec.~\ref{sec:inj} with the tests described in Sec.~\ref{sec:test_overview}. In all cases we use a three-detector LIGO-Virgo network with the O3low noise curves from~\cite{Aasi:2013wya} used to construct the likelihood. However, the simulated observations themselves do not contain any noise, so we are effectively averaging over noise realizations~\cite{Nissanke:2009kt}. We use a low-frequency cutoff of $20$~Hz, so the GW150914-like simulated observations have network optimal signal-to-noise ratios (SNRs) of $54$ ($53$ for the GR EOB simulated observation), except for the modified EOB simulated observation with the larger (smaller) GR deviation, which has a network optimal SNR of $40$ ($50$). (The GR deviations in the non-EOB cases do not affect the SNR, since they do not affect the frequency-domain amplitude.) For the GW170608-like simulated observations, the network optimal SNRs are all $21$ (though there are small differences between the Phenom and EOB GR waveforms that are hidden by the rounding to the nearest integer), except for the IHES modified GR simulated observation, which has a network optimal SNR of $20$. We use the same priors here as in the LIGO and Virgo collaborations' application of these tests to GW150914 and GW170608 in~\cite{O2_TGR}, though we had to increase some prior ranges to account for the wider posteriors due to the GR deviations. In particular, we choose flat priors on the IMR consistency test, TIGER, FTA, and MDR deviation parameters.

\begin{figure*}[htb]
\centering
\subfloat{
\includegraphics[width=0.42\textwidth]{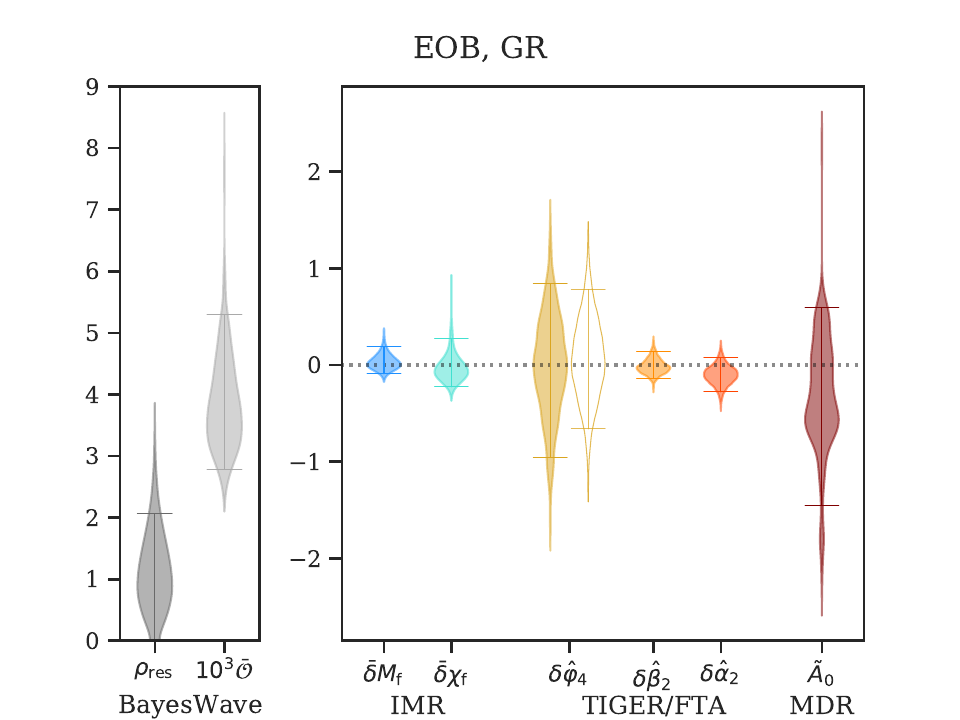}
}
\quad
\subfloat{
\includegraphics[width=0.42\textwidth]{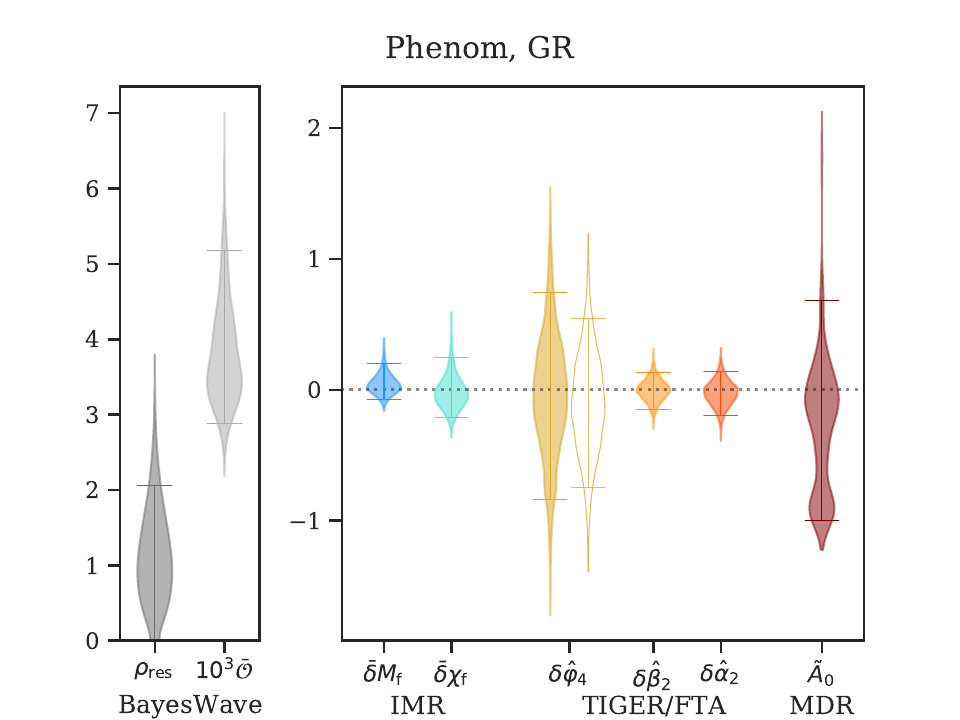}
}\\
\subfloat{
\includegraphics[width=0.42\textwidth]{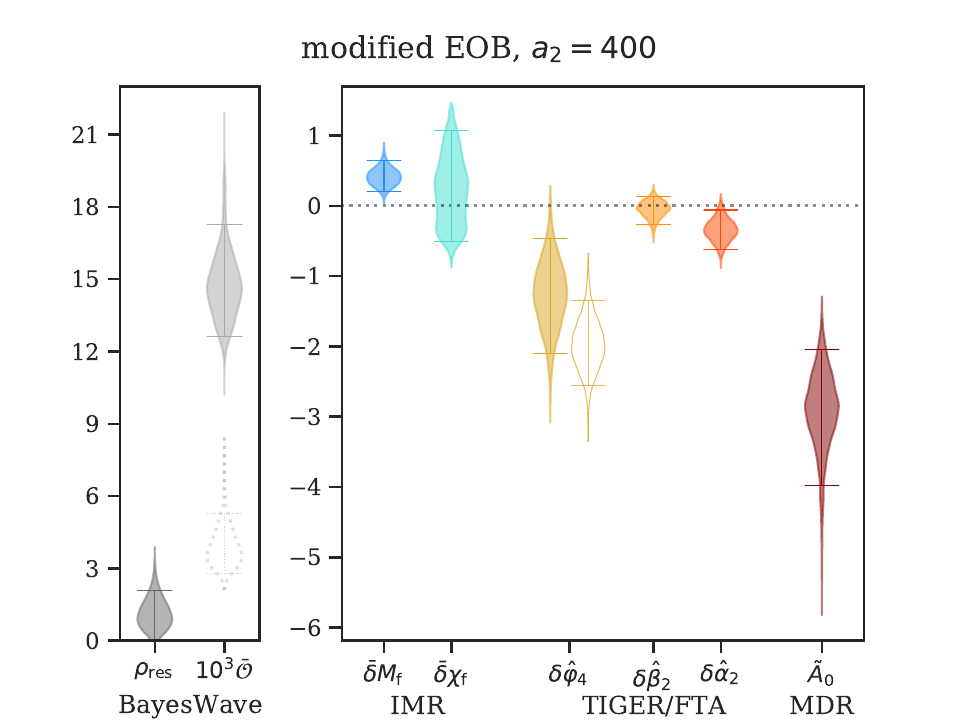}
}
\quad
\subfloat{
\includegraphics[width=0.42\textwidth]{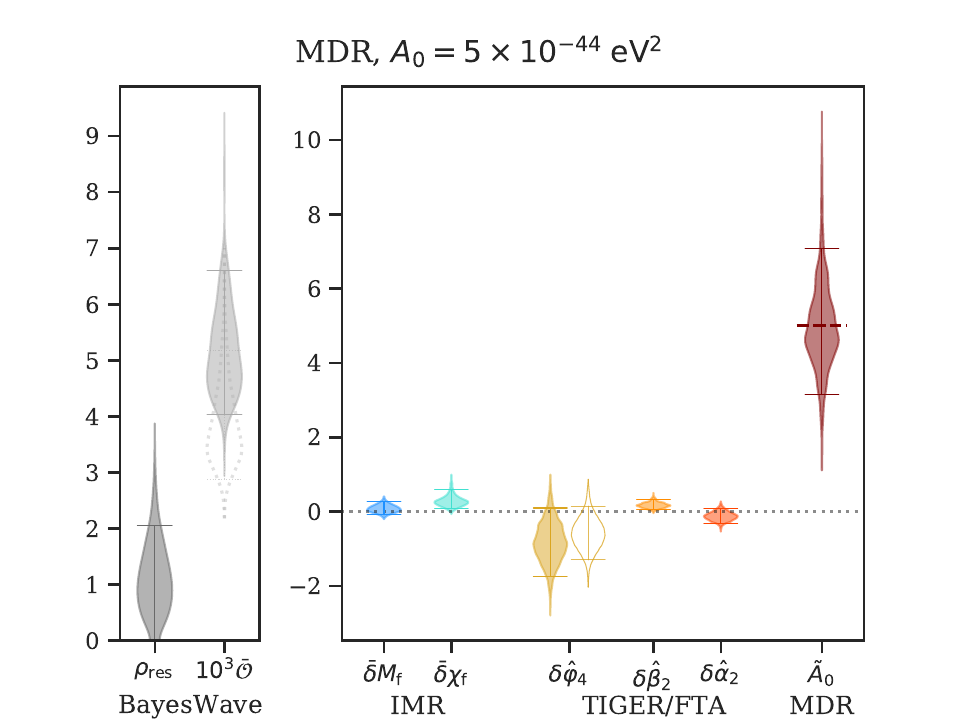}
}\\
\subfloat{
\includegraphics[width=0.42\textwidth]{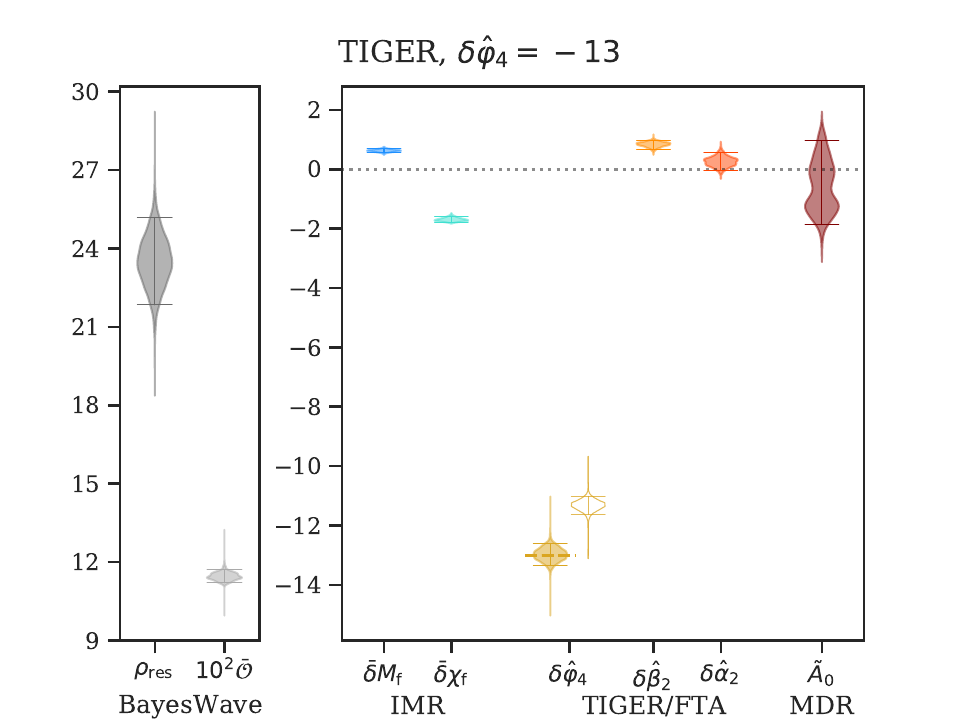}
}
\quad
\subfloat{
\includegraphics[width=0.42\textwidth]{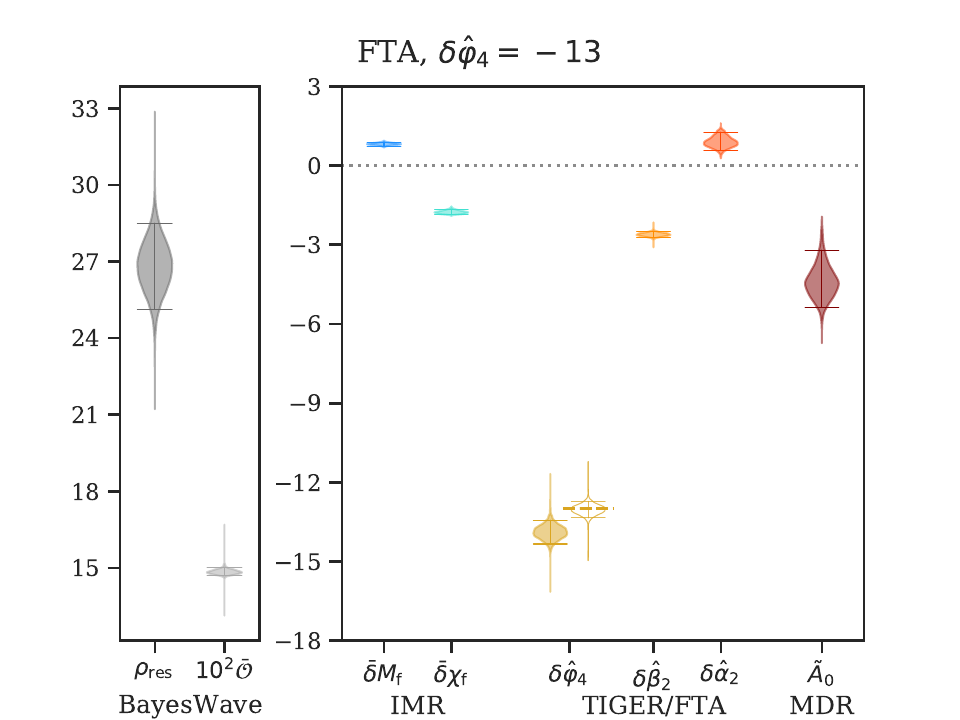}
}
\caption{\label{fig:violin_plots_GW150914-like_larger} The results of the various tests on the GW150914-like GR simulated observations and those with the larger GR deviation presented as violin plots of the posteriors on the deviation parameters and the associated $90\%$ credible intervals. For the residual SNR, we give $90\%$ upper bounds instead of the $90\%$ credible interval around the median, except for the TIGER and FTA cases where the distribution peaks well away from zero. We write $\bar{\delta}M_\mathrm{f} := \Delta M_\mathrm{f}/\bar{M}_\mathrm{f}$ and $\bar{\delta}\chi_\mathrm{f} := \Delta \chi_\mathrm{f}/\bar{\chi}_\mathrm{f}$ to save space. We scale $\bar{\mathcal{O}} := 1 - \mathcal{O}_\mathrm{B,L}$ differently for the TIGER and FTA simulated observations, which give much larger values. We mark the GR value of zero with a dotted line for the non-BayesWave tests. Additionally, for the TIGER, FTA, and massive graviton (MDR) simulated observations, we mark the true value of the corresponding test's parameter with a dashed line. For the modified EOB and massive graviton cases, we show the distribution of $\bar{\mathcal{O}}$ for the corresponding GR case as a dotted, unfilled violin, for comparison, since the distributions overlap or are relatively close to doing so.}
\end{figure*}

\begin{figure*}[htb]
\centering
\subfloat{
\includegraphics[width=0.42\textwidth]{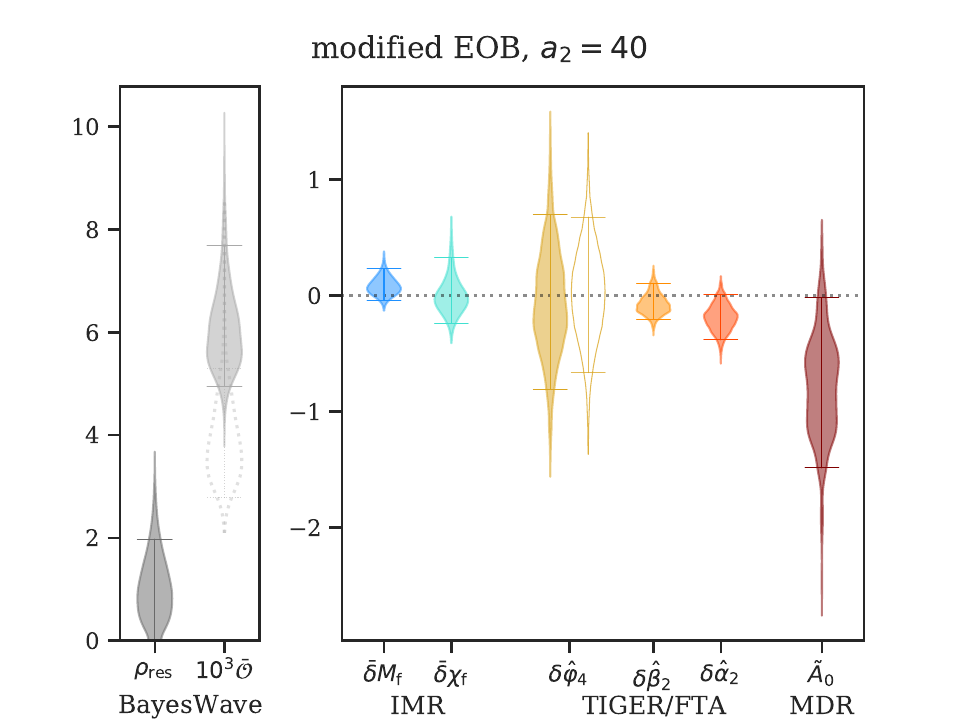}
}
\quad
\subfloat{
\includegraphics[width=0.42\textwidth]{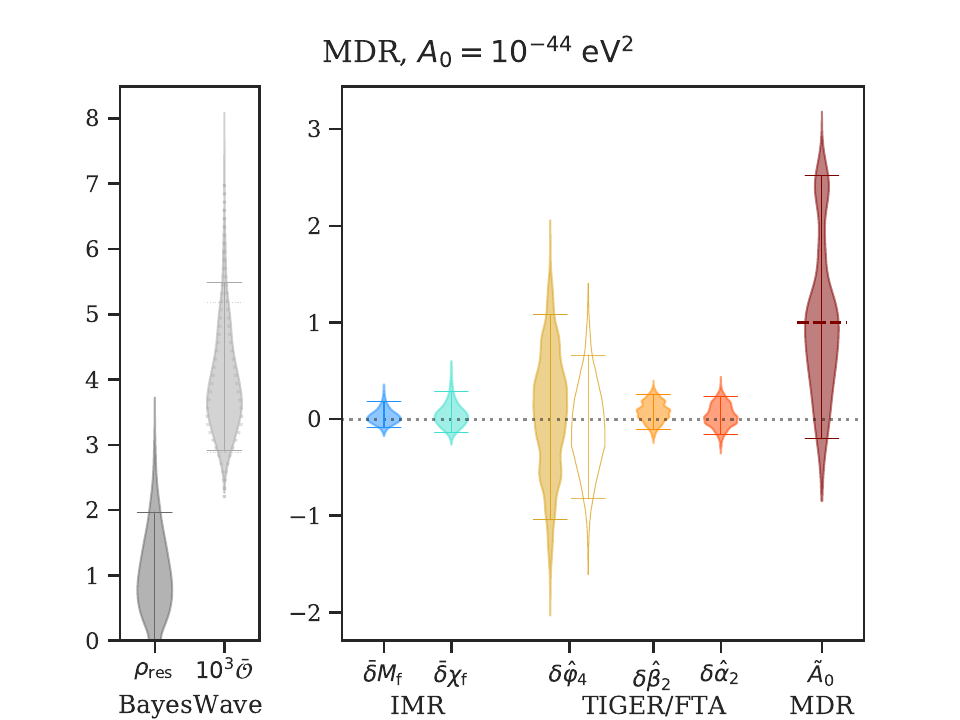}
}\\
\subfloat{
\includegraphics[width=0.42\textwidth]{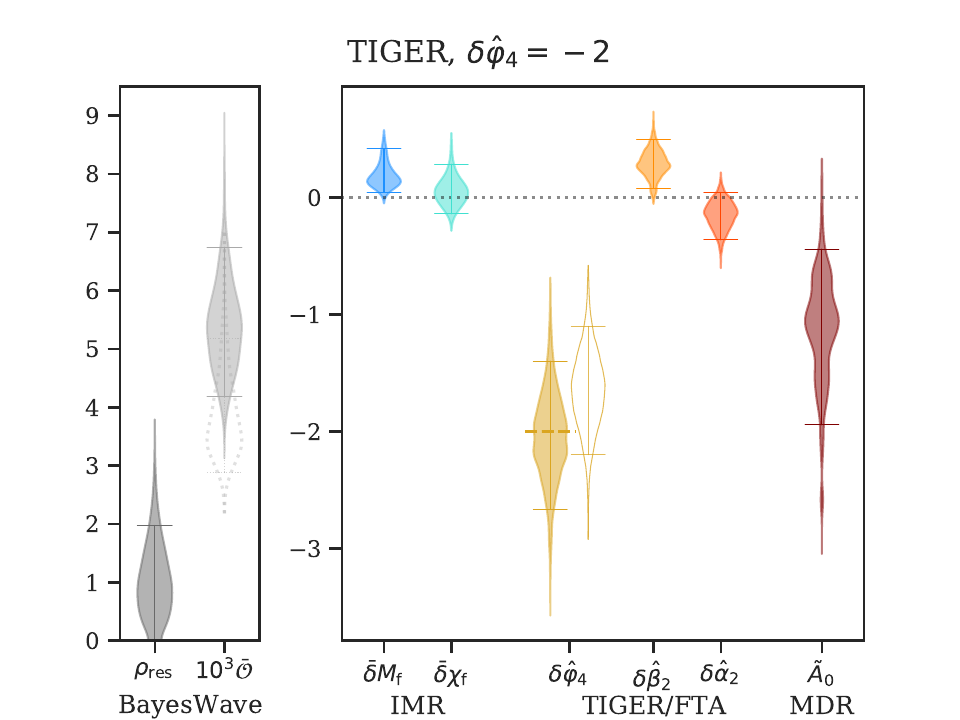}
}
\quad
\subfloat{
\includegraphics[width=0.42\textwidth]{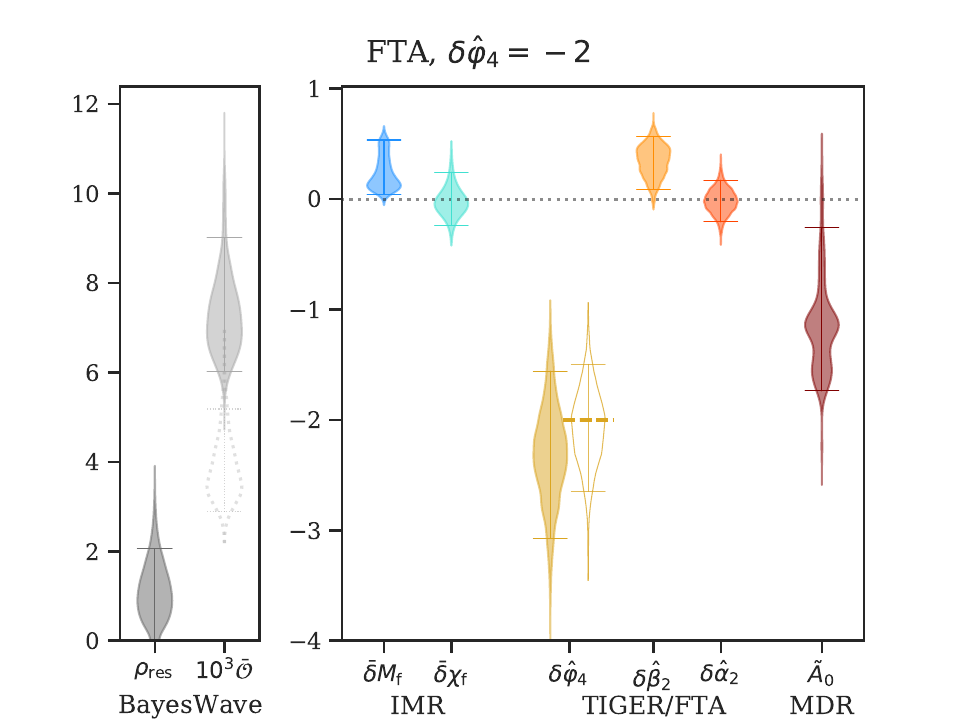}
}
\caption{\label{fig:violin_plots_GW150914-like_smaller} Like Fig.~\ref{fig:violin_plots_GW150914-like_larger}, except for the GW150914-like simulated observations with the smaller GR deviation. Additionally, we are able to scale $\bar{\mathcal{O}}$ the same way for all cases and also show the corresponding GR distribution in all cases.}
\end{figure*}

\begin{figure*}[htb]
\centering
\subfloat{
\includegraphics[width=0.42\textwidth]{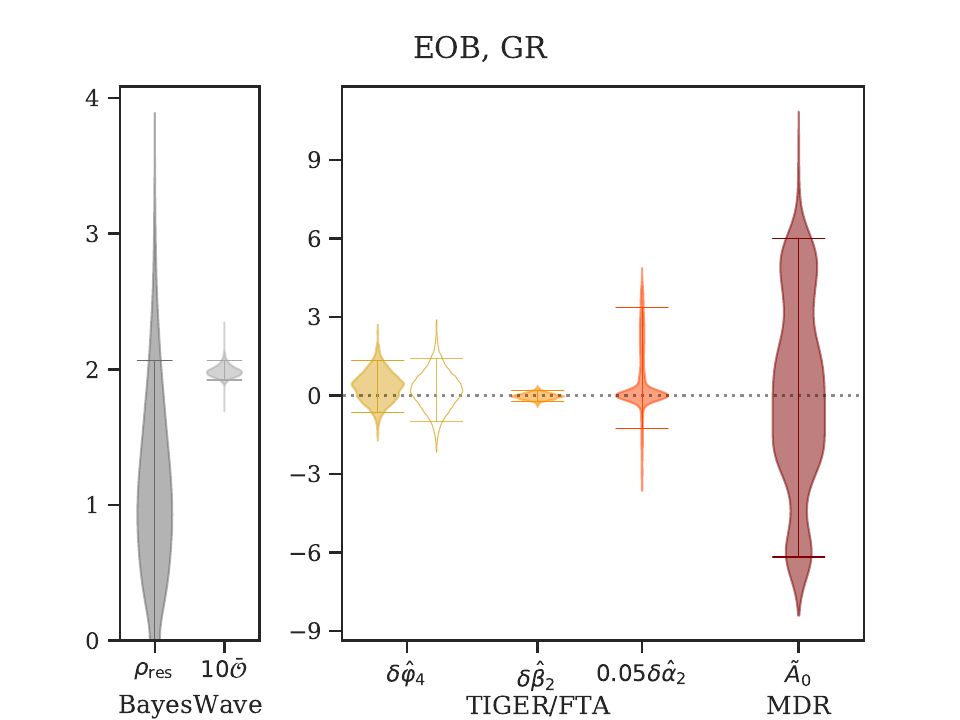}
}
\quad
\subfloat{
\includegraphics[width=0.42\textwidth]{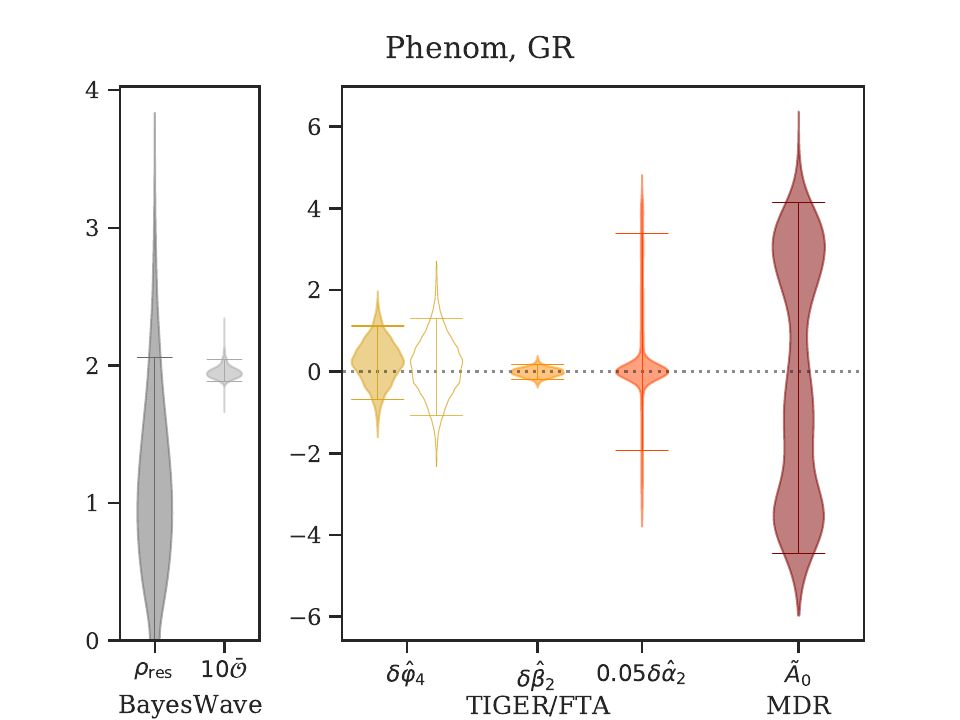}
}\\
\subfloat{
\includegraphics[width=0.42\textwidth]{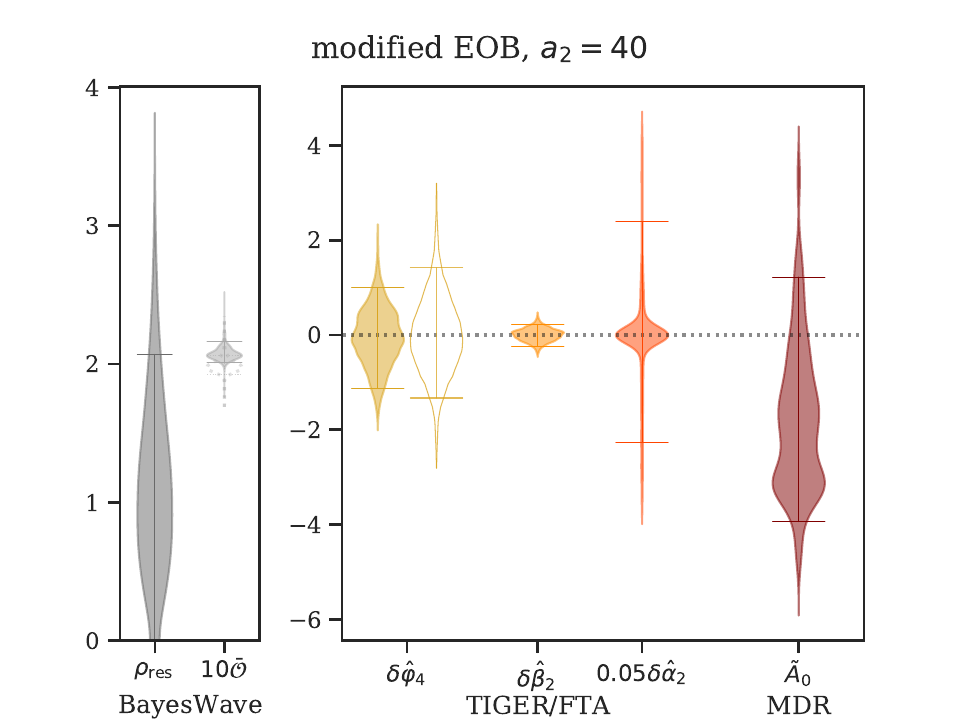}
}
\quad
\subfloat{
\includegraphics[width=0.42\textwidth]{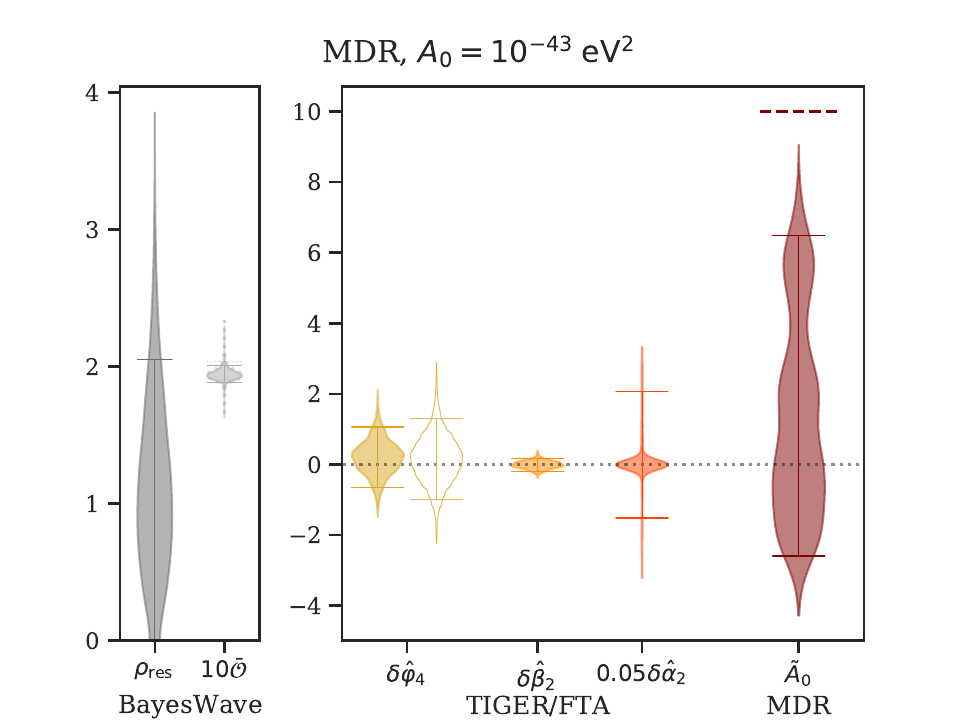}
}\\
\subfloat{
\includegraphics[width=0.42\textwidth]{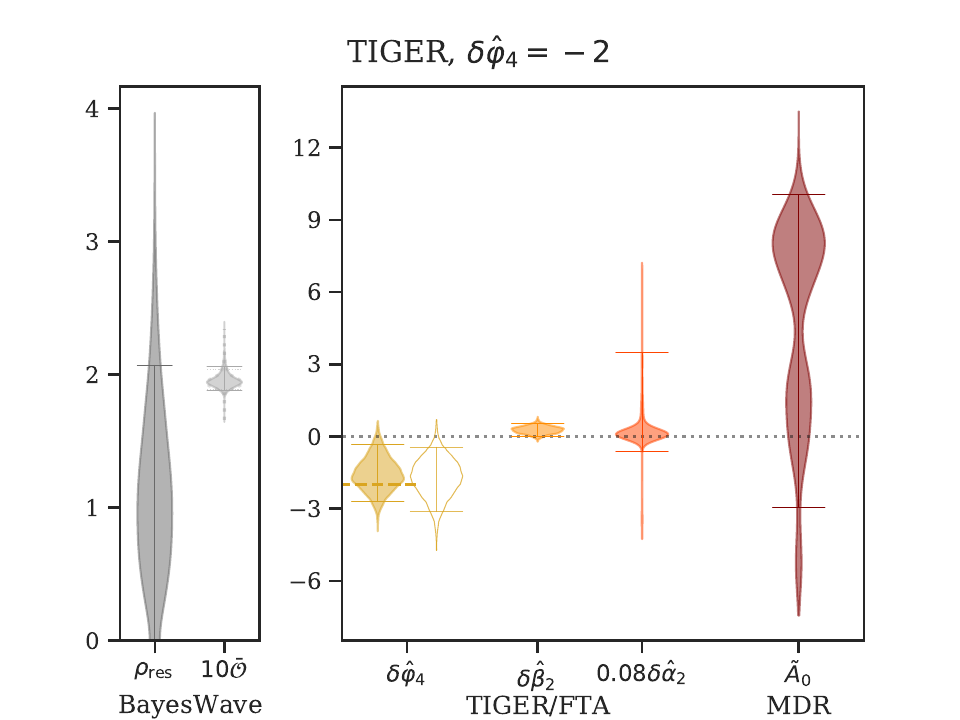}
}
\quad
\subfloat{
\includegraphics[width=0.42\textwidth]{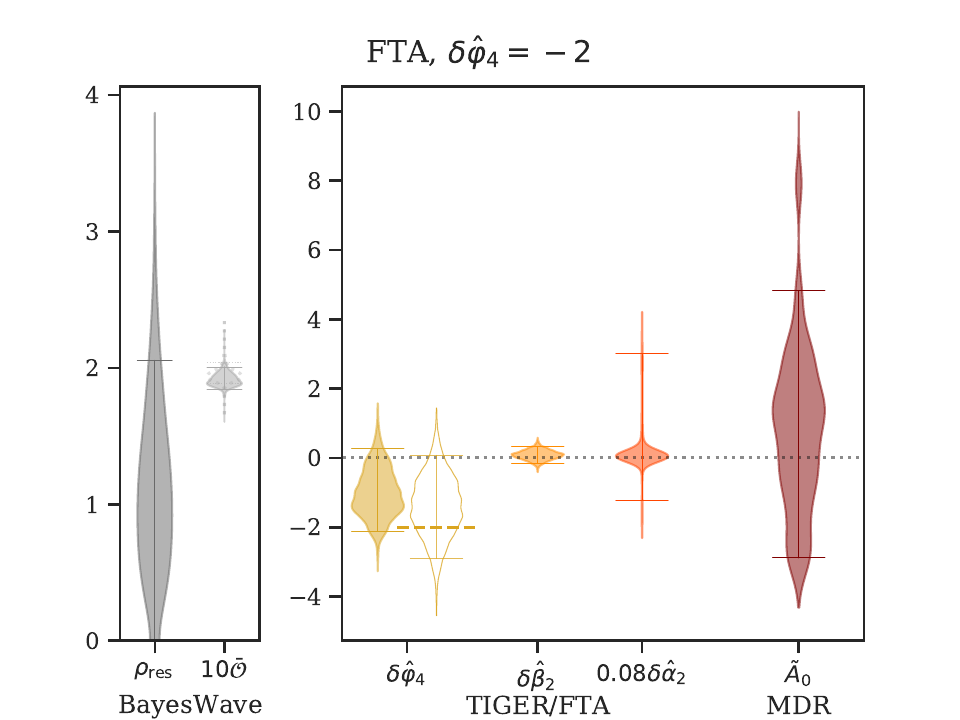}
}
\caption{\label{fig:violin_plots_GW170608} Like Fig.~\ref{fig:violin_plots_GW150914-like_larger}, except for the GW170608-like simulated observations (for which the IMR consistency test is not applicable). Here we scale the broad $\delta\hat{\alpha}_2$ posteriors to make the plot easier to read and scale $\bar{\mathcal{O}}$ by $10$ instead of the scalings of $10^2$ or $10^3$ used in the previous two plots. We also show the corresponding GR distribution of $\bar{\mathcal{O}}$ in all non-GR cases, though there is almost complete overlap except in the modified EOB case.}
\end{figure*}

\begin{table*}
\caption{\label{tab:summary}
Summary statistics for the test parameters for all tests on all simulated observations (simul.\ obs.)\ considered. (The IMR consistency test is not applicable to the GW170608-like cases.) For most tests, these are the GR quantiles as well as the median and surrounding $90\%$ credible interval. For the residual SNR, we give $90\%$ upper bounds instead, except for the TIGER and FTA simulated observations with larger GR deviations, where the probability distribution peaks well away from zero and we give the median and $90\%$ credible interval. The two BayesWave analyses do not provide GR quantiles. The GW150914-like and GW170608-like simulated observations are denoted by their abbreviations $\text{M}_{72}$ and $\text{M}_{20}$, respectively. For the GW150914-like simulated observations, $>$ denotes the case with the larger GR deviation and $<$ the one with the smaller deviation. The GR quantiles are one-dimensional (denoted $Q_\text{GR}$), so values around $50\%$ indicate good agreement with GR, in all cases except for the IMR consistency test, where they are two-dimensional, denoted $\cQ_\text{GR}$, and values around $0$ indicate good agreement with GR. In all cases they are rounded to the nearest percent. We discuss the strength to which GR is excluded when the GR quantile is in the tails of the distribution in the text. We bold the GR quantiles where GR is excluded at the $90\%$ credible level, so where the GR quantile is outside the $90\%$ credible interval around the median ($[5,95]\%$) for the one-dimensional quantiles.
}
\scalebox{0.86}{
\begin{tabular}{*{25}{c}}
\hline\hline
\multicolumn{2}{c}{\multirow{4}{*}{Simul.\ Obs.\ }} & & Res.\ & & Reconst.\ & & \multicolumn{3}{c}{IMR consistency} & & \multicolumn{11}{c}{TIGER/FTA} & & \multicolumn{2}{c}{MDR}\\
\cline{4-4}
\cline{6-6}
\cline{8-10}
\cline{12-22}
\cline{24-25}
& & & \multirow{3}{*}{$\rho_\text{res}$} & & \multirow{3}{*}{$\underset{\displaystyle [10^{-3}]}{1 - \mathcal{O}_\mathrm{B,L}}$} & & \multirow{3}{*}{$\underset{\displaystyle [\%]}{\cQ_\text{GR}}$} & \multirow{3}{*}{$\displaystyle \frac{\Delta M_\text{f}}{\bar{M}_\text{f}}$} & \multirow{3}{*}{$\displaystyle \frac{\Delta \chi_\text{f}}{\bar{\chi}_\text{f}}$} & & \multicolumn{2}{c}{$\varphi_4$, TIGER} & & \multicolumn{2}{c}{$\varphi_4$, FTA} & & \multicolumn{2}{c}{$\beta_2$} & & \multicolumn{2}{c}{$\alpha_2$} & & \multirow{3}{*}{$\underset{\displaystyle [\%]}{Q_\text{GR}}$} & \multirow{3}{*}{$\tilde{A}_0$}\\
\cline{12-13}
\cline{15-16}
\cline{18-19}
\cline{21-22}
& & & & & & & & & & & $Q_\text{GR}$ & \multirow{2}{*}{$\delta\hat{\varphi}_4$} & & $Q_\text{GR}$ & \multirow{2}{*}{$\delta\hat{\varphi}_4$} & & $Q_\text{GR}$ & \multirow{2}{*}{$\delta\hat{\beta}_2$} & & $Q_\text{GR}$ & \multirow{2}{*}{$\delta\hat{\alpha}_2$} & & \\
& & & & & & & & & & & $[\%]$ & & & $[\%]$ & & & $[\%]$ & & & $[\%]$ & & & &\\
\hline
\multirow{12}{*}{$\text{M}_{72}$} & EOB, GR & & $<2.1$ & & $3.8_{-1.0}^{+1.5}$ & & $4$ & $0.0_{-0.1}^{+0.2}$ & $\hphantom{-}0.0_{-0.2}^{+0.3}$ & & $50$ & $\hphantom{-}0.0_{-1.0}^{+0.8}$ & & $45$ & $\hphantom{-}0.1_{-0.8}^{+0.7}$ & & $62$ & $\hphantom{-}0.0_{-0.1}^{+0.1}$ & & $85$ & $-0.1_{-0.2}^{+0.2}$ & & $69$ & $-0.3_{-1.1}^{+0.9}$\\[1pt]
& Phenom, GR & & $<2.1$ & & $3.7_{-0.8}^{+1.5}$ & & $3$ & $0.0_{-0.1}^{+0.2}$ & $\hphantom{-}0.0_{-0.2}^{+0.3}$ & & $54$ & $\hphantom{-}0.0_{-0.8}^{+0.7}$ & & $61$ & $-0.1_{-0.6}^{+0.6}$ & & $52$ & $\hphantom{-}0.0_{-0.1}^{+0.1}$ & & $60$ & $\hphantom{-}0.0_{-0.2}^{+0.1}$ & & $68$ & $-0.2_{-0.8}^{+0.9}$\\[2.5pt]
& modified EOB, $>$ & & $<2.1$ & & $14.6_{-2.0}^{+2.6}$ & & $\mathbf{100}$ & $0.4_{-0.2}^{+0.2}$ & $\hphantom{-}0.2_{-0.7}^{+0.9}$ & & $\mathbf{99}$ & $-1.3_{-0.8}^{+0.8}$ & & $\mathbf{100}$ & $-2.0_{-0.6}^{+0.7}$ & & $64$ & $\hphantom{-}0.0_{-0.3}^{+0.1}$ & & $\mathbf{98}$ & $-0.3_{-0.3}^{+0.2}$ & & $\mathbf{100}$ & $-2.9_{-1.1}^{+0.8}$\\[1pt]
& modified EOB, $<$ & & $<2.0$ & & $6.0_{-1.0}^{+1.7}$ & & $14$ & $0.1_{-0.1}^{+0.1}$ & $\hphantom{-}0.0_{-0.2}^{+0.4}$ & & $57$ & $-0.1_{-0.7}^{+0.8}$ & & $50$ & $\hphantom{-}0.0_{-0.7}^{+0.7}$ & & $79$ & $-0.1_{-0.1}^{+0.2}$ & & $94$ & $-0.2_{-0.2}^{+0.2}$ & & $\mathbf{95}$ & $-0.8_{-0.7}^{+0.8}$\\[2.5pt]
& MDR, $>$ & & $<2.1$ & & $5.1_{-1.0}^{+1.5}$ & & $\mathbf{90}$ & $0.1_{-0.2}^{+0.2}$ & $\hphantom{-}0.3_{-0.2}^{+0.3}$ & & $93$ & $-0.8_{-0.9}^{+0.9}$ & & $91$ & $-0.6_{-0.7}^{+0.8}$ & & $\mathbf{1}$ & $\hphantom{-}0.2_{-0.1}^{+0.1}$ & & $85$ & $-0.1_{-0.2}^{+0.2}$ & & $\mathbf{0}$ & $\hphantom{-}4.9_{-1.7}^{+2.2}$\\[1pt]
& MDR, $<$ & & $<2.0$ & & $3.9_{-0.9}^{+1.6}$ & & $15$ & $0.0_{-0.1}^{+0.2}$ & $\hphantom{-}0.0_{-0.2}^{+0.3}$ & & $45$ & $\hphantom{-}0.1_{-1.1}^{+1.0}$ & & $59$ & $-0.1_{-0.7}^{+0.8}$ & & $25$ & $\hphantom{-}0.1_{-0.2}^{+0.2}$ & & $41$ & $\hphantom{-}0.0_{-0.2}^{+0.2}$ & & $10$ & $\hphantom{-}0.9_{-1.1}^{+1.6}$\\[2.5pt]
& TIGER, $>$ & & $23.5_{-1.7}^{+1.7}$ & & $114.4_{-2.4}^{+2.7}$ & & $\mathbf{100}$ & $0.6_{-0.1}^{+0.1}$ & $-1.7_{-0.1}^{+0.1}$ & & $\mathbf{100}$ & $-13.0_{-0.4}^{+0.4}$ & & $\mathbf{100}$ & $-11.3_{-0.3}^{+0.3}$ & & $\mathbf{0}$ & $\hphantom{-}0.8_{-0.1}^{+0.2}$ & & $7$ & $\hphantom{-}0.3_{-0.3}^{+0.3}$ & & $71$ & $-0.6_{-1.2}^{+1.6}$\\[1pt]
& TIGER, $<$ & & $<2.0$ & & $5.3_{-1.2}^{+1.4}$ & & $\mathbf{98}$ & $0.2_{-0.1}^{+0.2}$ & $\hphantom{-}0.0_{-0.2}^{+0.2}$ & & $\mathbf{100}$ & $-2.0_{-0.7}^{+0.6}$ & & $\mathbf{100}$ & $-1.6_{-0.6}^{+0.5}$ & & $\mathbf{1}$ & $\hphantom{-}0.3_{-0.2}^{+0.2}$ & & $89$ & $-0.1_{-0.3}^{+0.1}$ & & $\mathbf{99}$ & $-1.1_{-0.9}^{+0.6}$\\[2.5pt]
& FTA, $>$ & & $26.8_{-1.7}^{+1.7}$ & & $148.3_{-1.4}^{+1.7}$ & & $\mathbf{100}$ & $0.8_{-0.1}^{+0.1}$ & $-1.8_{-0.1}^{+0.1}$ & & $\mathbf{100}$ & $-13.9_{-0.4}^{+0.5}$ & & $\mathbf{100}$ & $-13.0_{-0.3}^{+0.3}$ & & $\mathbf{100}$ & $-2.6_{-0.1}^{+0.1}$ & & $\mathbf{0}$ & $\hphantom{-}0.9_{-0.3}^{+0.4}$ & & $\mathbf{100}$ & $-4.4_{-0.9}^{+1.2}$\\[1pt]
& FTA, $<$ & & $<2.1$ & & $7.2_{-1.2}^{+1.8}$ & & $\mathbf{97}$ & $0.2_{-0.2}^{+0.3}$ & $\hphantom{-}0.0_{-0.2}^{+0.3}$ & & $\mathbf{100}$ & $-2.3_{-0.8}^{+0.7}$ & & $\mathbf{100}$ & $-2.1_{-0.5}^{+0.6}$ & & $\mathbf{1}$ & $\hphantom{-}0.4_{-0.3}^{+0.2}$ & & $56$ & $\hphantom{-}0.0_{-0.2}^{+0.2}$ & & $\mathbf{97}$ & $-1.2_{-0.6}^{+0.9}$\\[1pt]
\hline
\multirow{7}{*}{$\text{M}_{20}$} & EOB, GR & & $<2.1$ & & $198_{-6}^{+8}$ & & -- & -- & -- & & $30$ & $\hphantom{-}0.3_{-0.9}^{+1.0}$ & & $38$ & $\hphantom{-}0.2_{-1.2}^{+1.2}$ & & $63$ & $\hphantom{-}0.0_{-0.2}^{+0.2}$ & & $38$ & $\hphantom{-}2_{-27}^{+65}$ & & $49$ & $\hphantom{-}0.1_{-6.3}^{+5.9}$\\[1pt]
& Phenom, GR & & $<2.1$ & & $195_{-6}^{+10}$ & & -- & -- & -- & & $32$ & $\hphantom{-}0.2_{-0.9}^{+0.9}$ & & $43$ & $\hphantom{-}0.1_{-1.2}^{+1.2}$ & & $56$ & $\hphantom{-}0.0_{-0.2}^{+0.2}$ & & $44$ & $\hphantom{-}1_{-40}^{+67}$ & & $53$ & $-0.4_{-4.1}^{+4.5}$\\[1pt]
& modified EOB & & $<2.1$ & & $207_{-6}^{+10}$ & & -- & -- & -- & & $53$ & $-0.1_{-1.0}^{+1.1}$ & & $52$ & $\hphantom{-}0.0_{-1.3}^{+1.4}$ & & $49$ & $\hphantom{-}0.0_{-0.2}^{+0.2}$ & & $49$ & $\hphantom{-}0_{-45}^{+48}$ & & $86$ & $-2.0_{-2.0}^{+3.2}$\\[1pt]
& MDR & & $<2.1$ & & $194_{-5}^{+7}$ & & -- & -- & -- & & $34$ & $\hphantom{-}0.2_{-0.9}^{+0.9}$ & & $41$ & $\hphantom{-}0.1_{-1.1}^{+1.2}$ & & $54$ & $\hphantom{-}0.0_{-0.2}^{+0.2}$ & & $49$ & $\hphantom{-}0_{-30}^{+41}$ & & $39$ & $\hphantom{-}0.9_{-3.5}^{+5.6}$\\[1pt]
& TIGER & & $<2.1$ & & $195_{-7}^{+12}$ & & -- & -- & -- & & $\mathbf{98}$ & $-1.6_{-1.1}^{+1.3}$ & & $\mathbf{99}$ & $-1.7_{-1.4}^{+1.3}$ & & $\mathbf{5}$ & $\hphantom{-}0.3_{-0.3}^{+0.2}$ & & $33$ & $\hphantom{-}2_{-10}^{+42}$ & & $15$ & $\hphantom{-}6.4_{-9.3}^{+3.6}$\\[1pt]
& FTA & & $<2.1$ & & $191_{-6}^{+10}$ & & -- & -- & -- & & $90$ & $-1.1_{-1.0}^{+1.4}$ & & $94$ & $-1.4_{-1.5}^{+1.5}$ & & $28$ & $\hphantom{-}0.1_{-0.3}^{+0.2}$ & & $35$ & $\hphantom{-}1_{-16}^{+37}$ & & $37$ & $\hphantom{-}0.9_{-3.7}^{+4.0}$\\[1pt]
\hline\hline
\end{tabular}
}
\end{table*}

\begin{figure*}[htb]
\centering
\subfloat{
\includegraphics[width=0.42\textwidth]{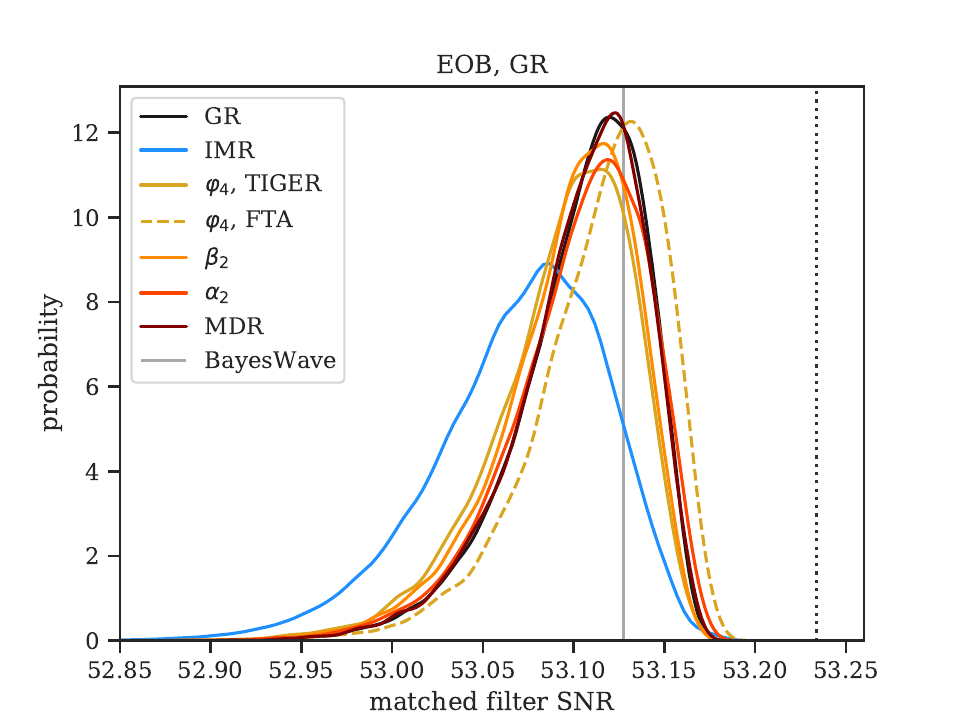}
}
\quad
\subfloat{
\includegraphics[width=0.42\textwidth]{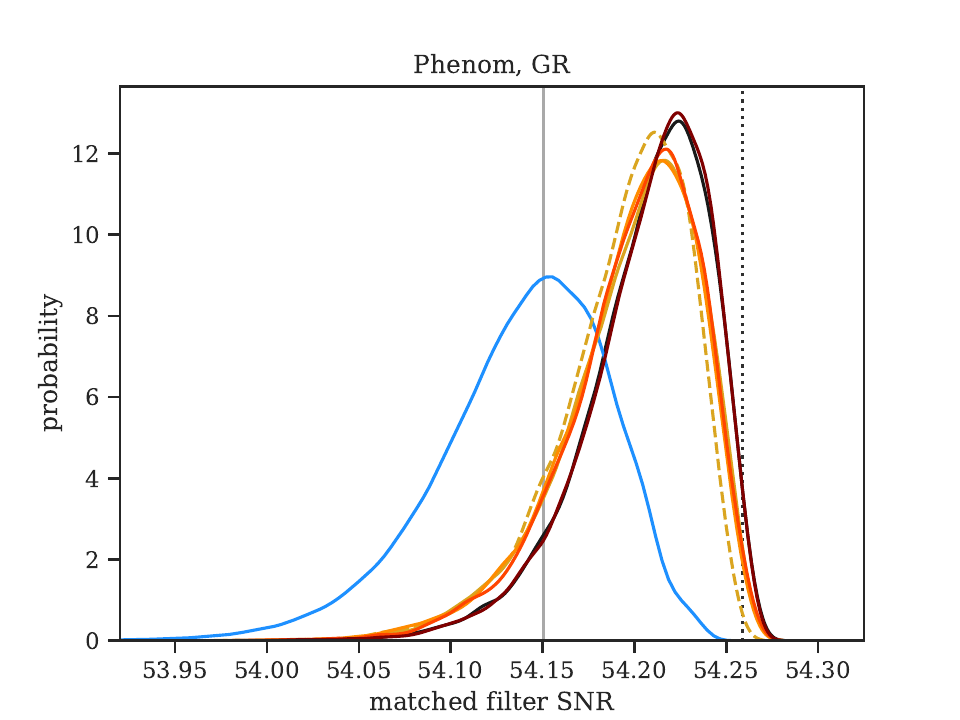}
}\\
\subfloat{
\includegraphics[width=0.42\textwidth]{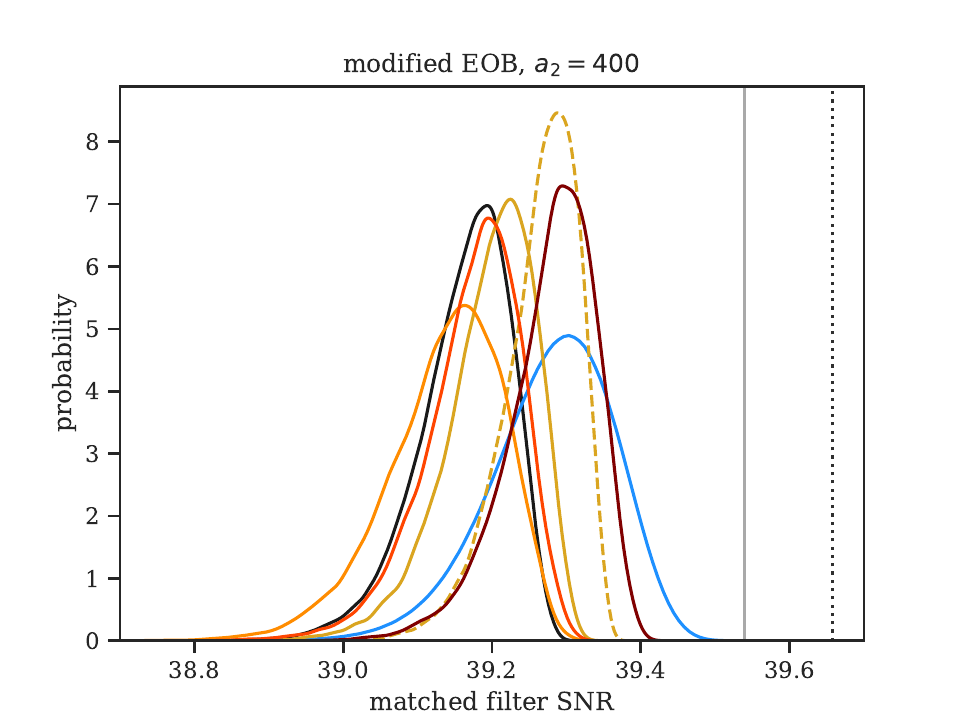}
}
\quad
\subfloat{
\includegraphics[width=0.42\textwidth]{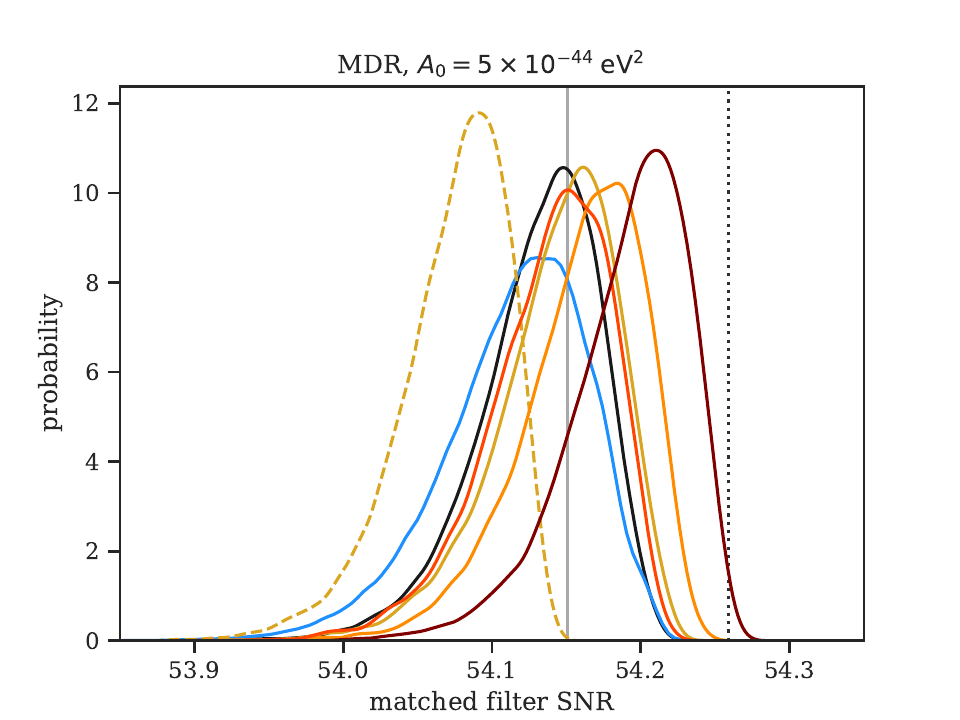}
}\\
\subfloat{
\includegraphics[width=0.42\textwidth]{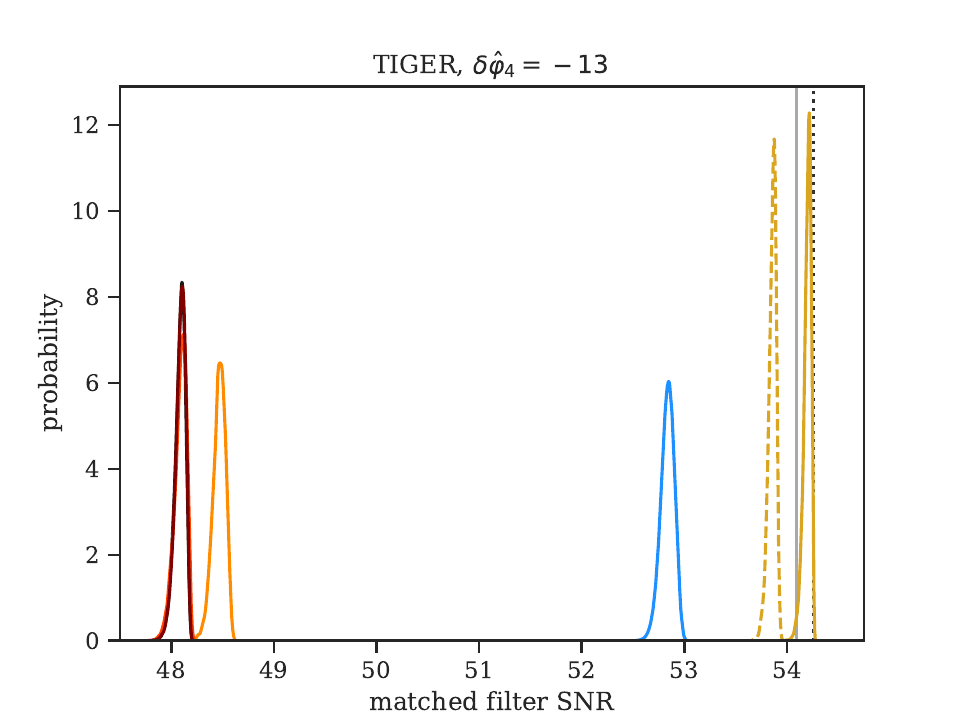}
}
\quad
\subfloat{
\includegraphics[width=0.42\textwidth]{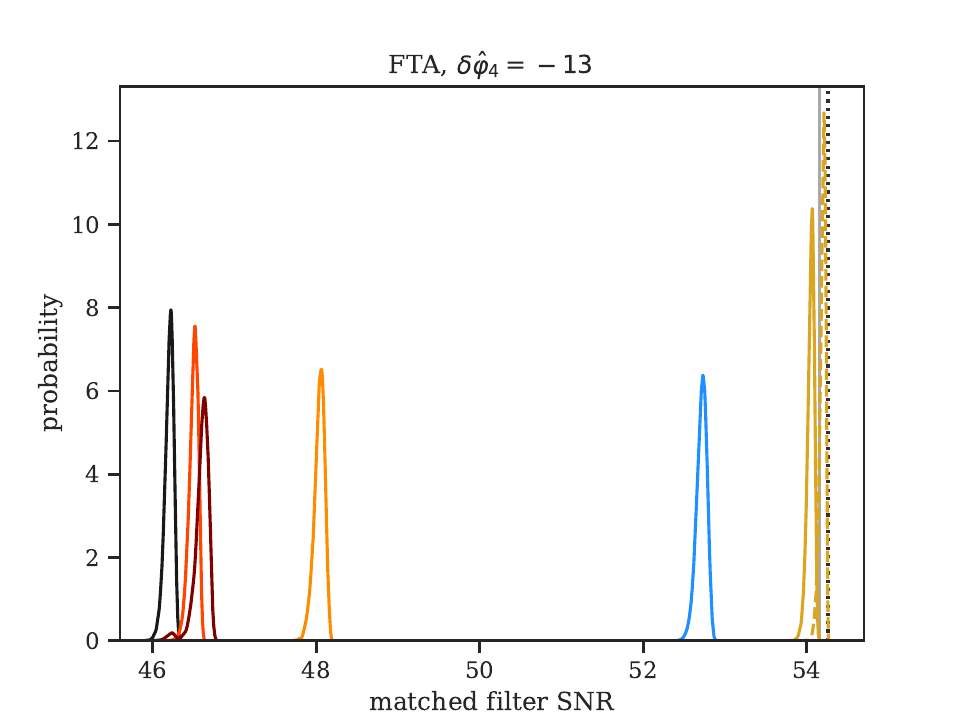}
}
\caption{\label{fig:SNR_plots_GW150914-like_larger} The posterior distributions of the recovered matched filter SNRs for the various tests applied to the GW150914-like GR simulated observations and those with the larger GR deviation, as well as the matched filter SNR recovered by the median BayesWave reconstruction for these cases. We also show the optimal SNR of the simulated observation, plotted as a vertical dotted line. The IMR results combine together the inspiral and postinspiral posteriors to give a posterior on the SNR for the full frequency range, while the MDR results combine together the positive and negative $A_0$ results.
}
\end{figure*}

\begin{figure*}[htb]
\centering
\subfloat{
\includegraphics[width=0.42\textwidth]{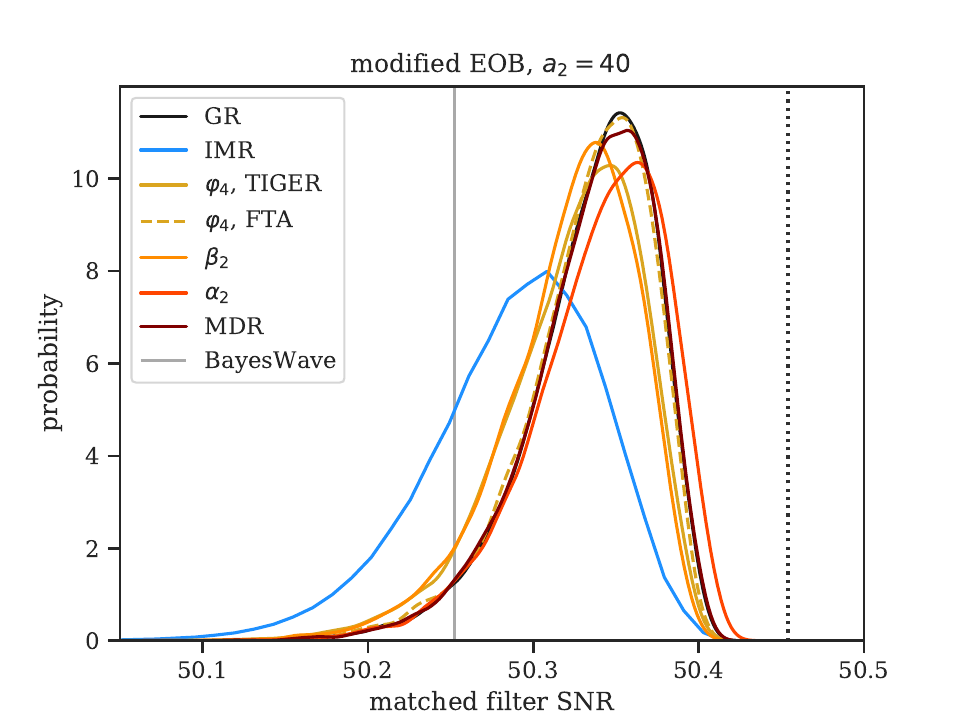}
}
\quad
\subfloat{
\includegraphics[width=0.42\textwidth]{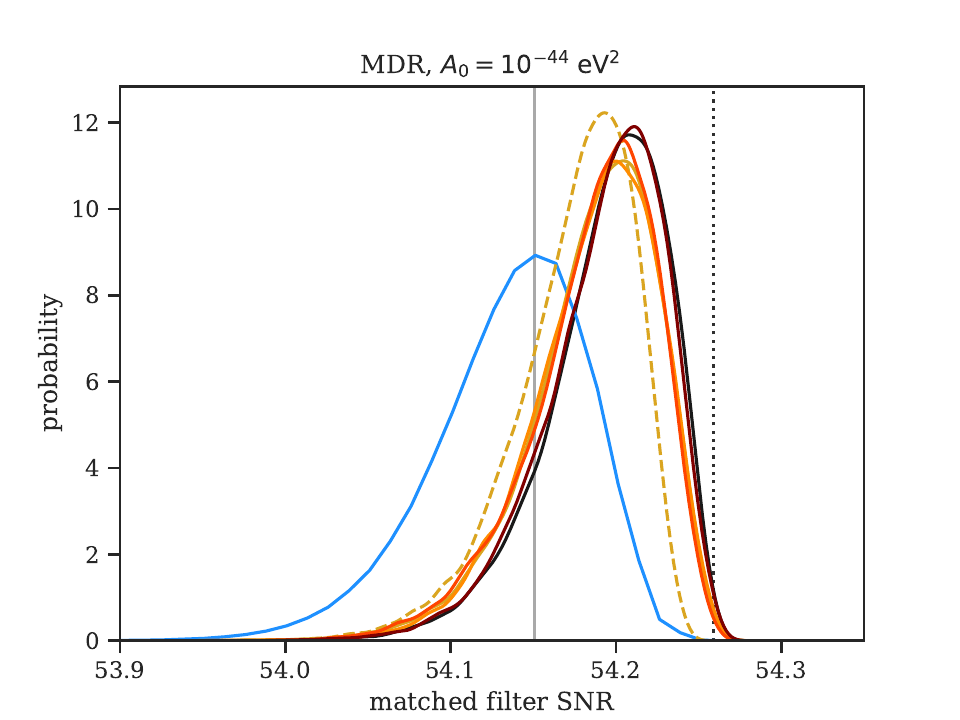}
}\\
\subfloat{
\includegraphics[width=0.42\textwidth]{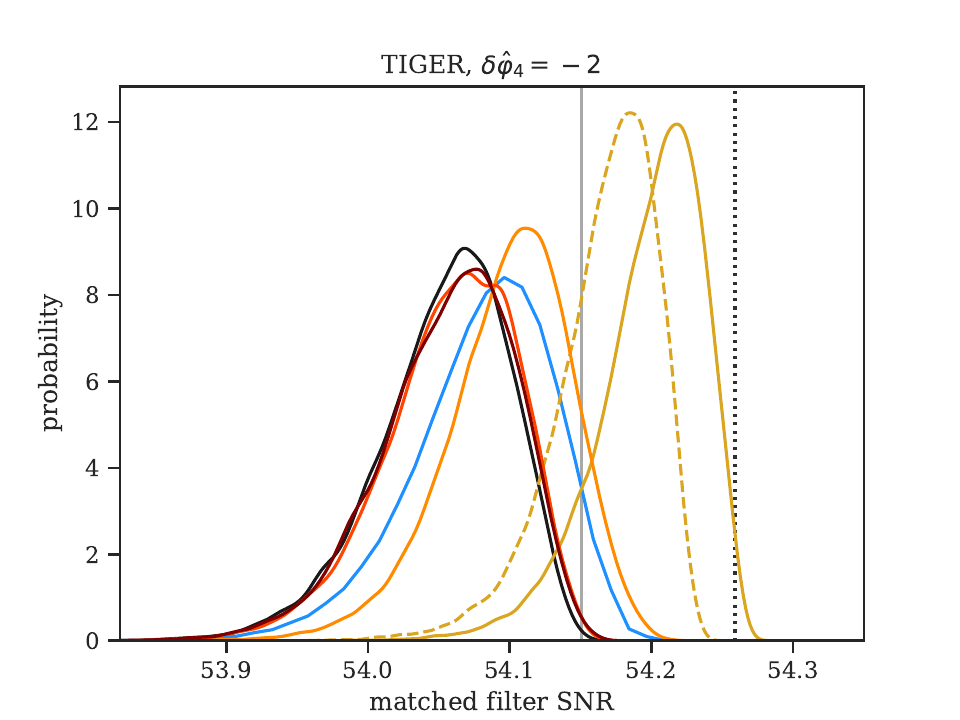}
}
\quad
\subfloat{
\includegraphics[width=0.42\textwidth]{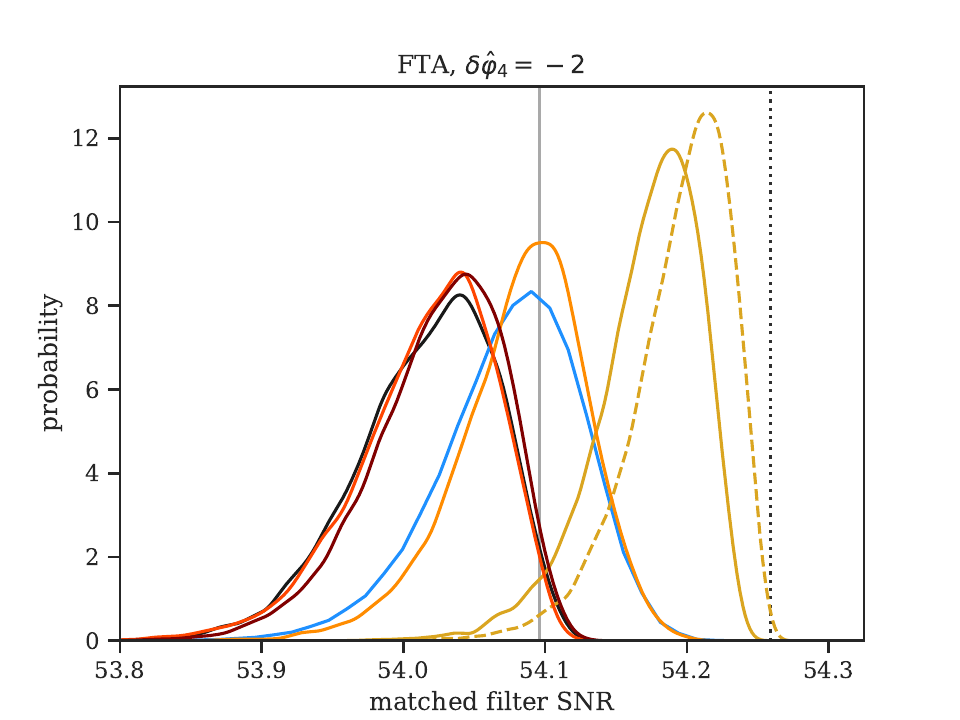}
}
\caption{\label{fig:SNR_plots_GW150914-like_smaller} The analog of Fig.~\ref{fig:SNR_plots_GW150914-like_larger} for the GW150914-like simulated observations with the smaller GR deviation.}
\end{figure*}

\begin{figure*}[htb]
\centering
\subfloat{
\includegraphics[width=0.42\textwidth]{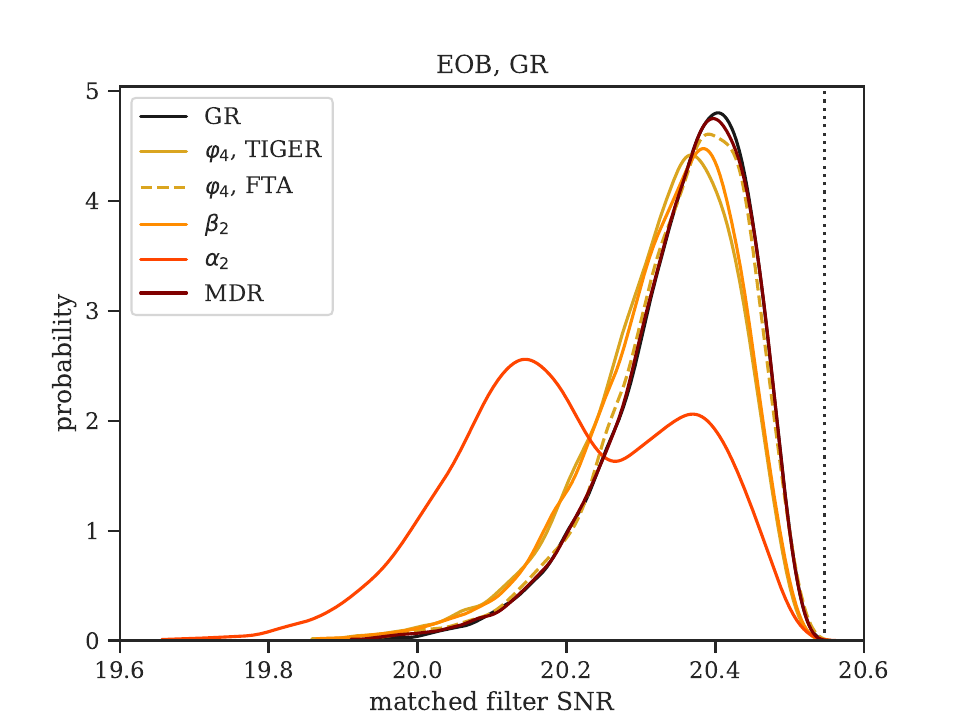}
}
\quad
\subfloat{
\includegraphics[width=0.42\textwidth]{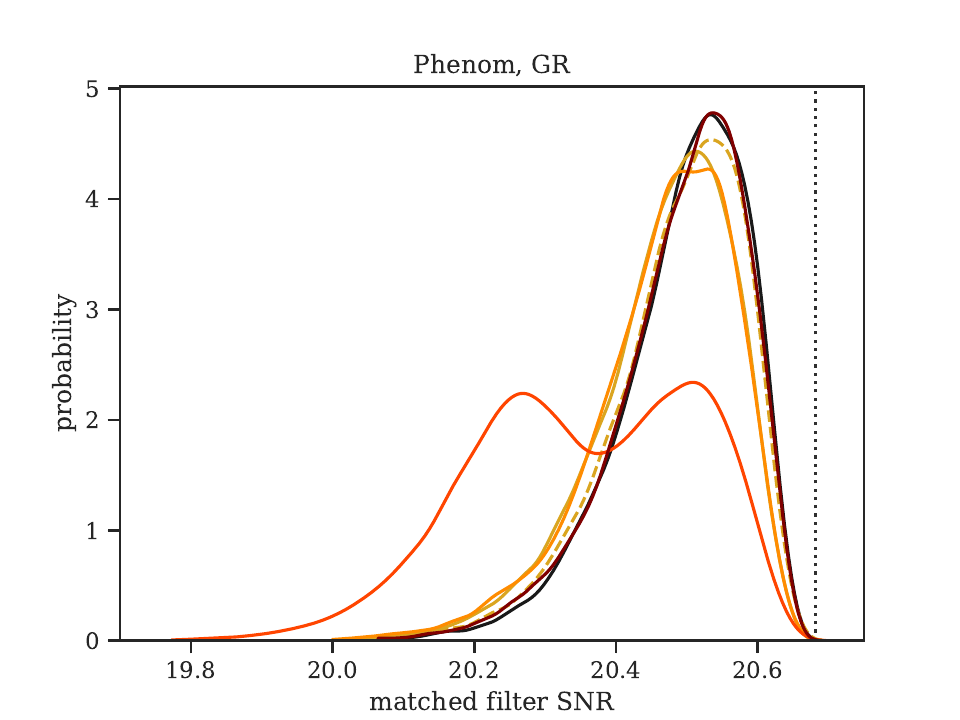}
}\\
\subfloat{
\includegraphics[width=0.42\textwidth]{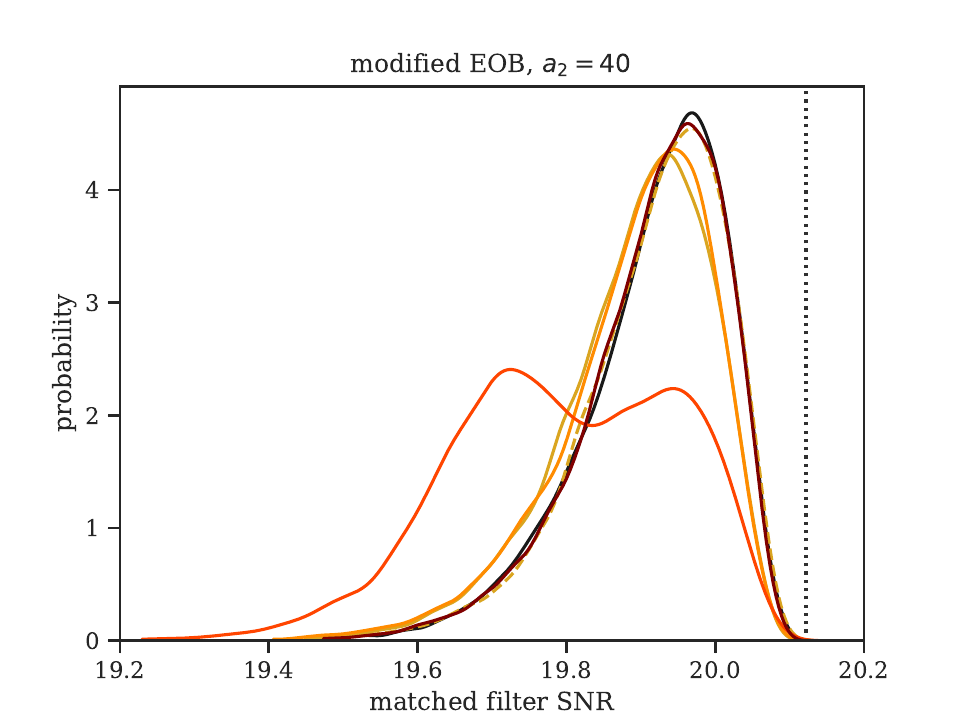}
}
\quad
\subfloat{
\includegraphics[width=0.42\textwidth]{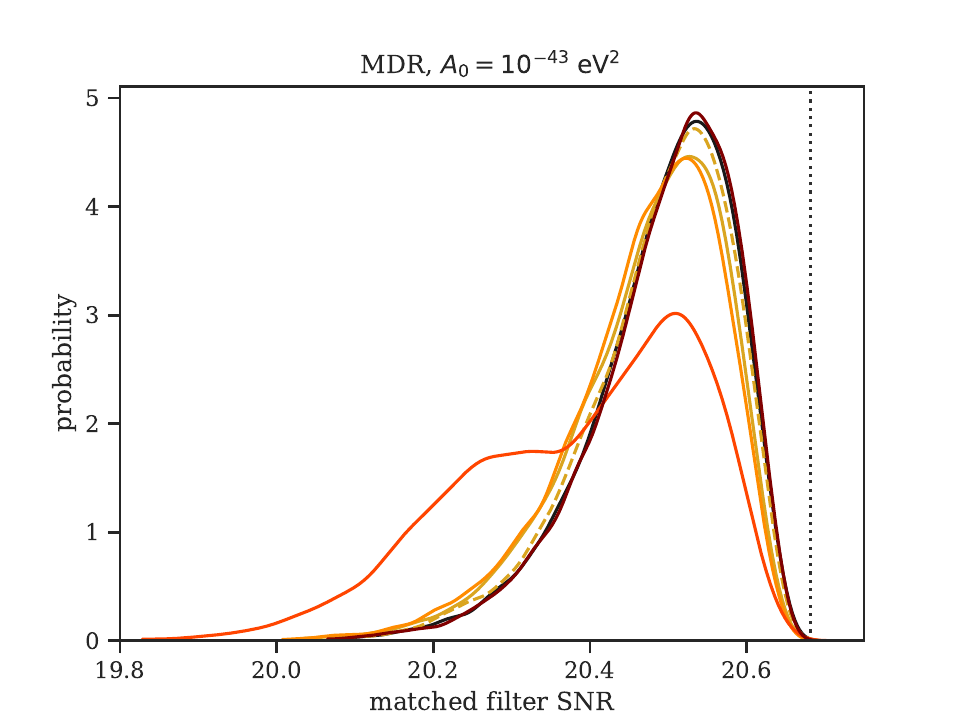}
}\\
\subfloat{
\includegraphics[width=0.42\textwidth]{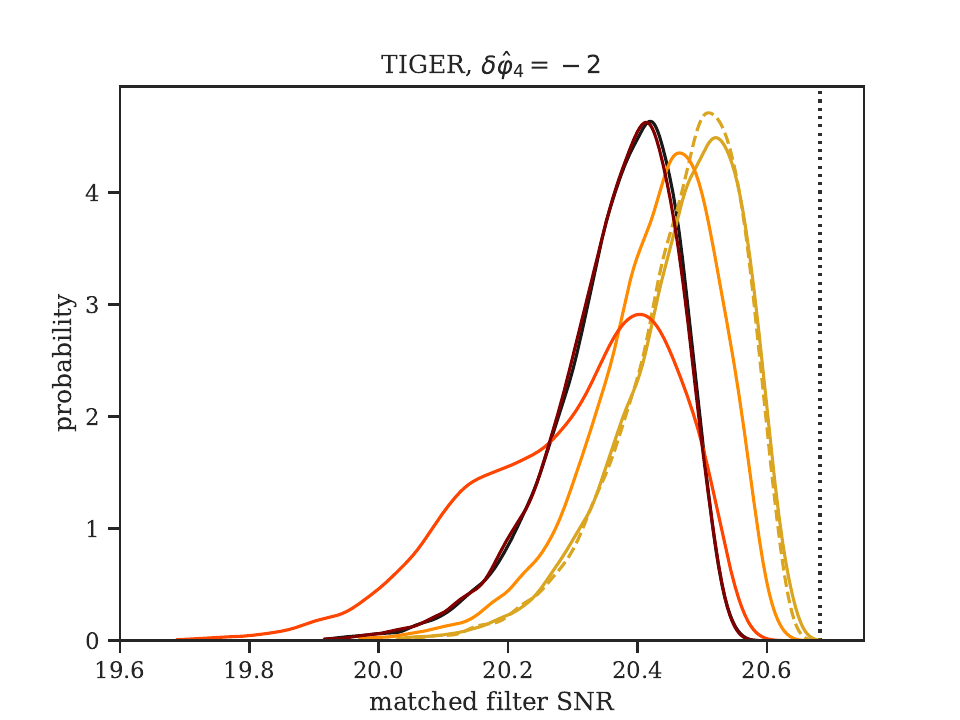}
}
\quad
\subfloat{
\includegraphics[width=0.42\textwidth]{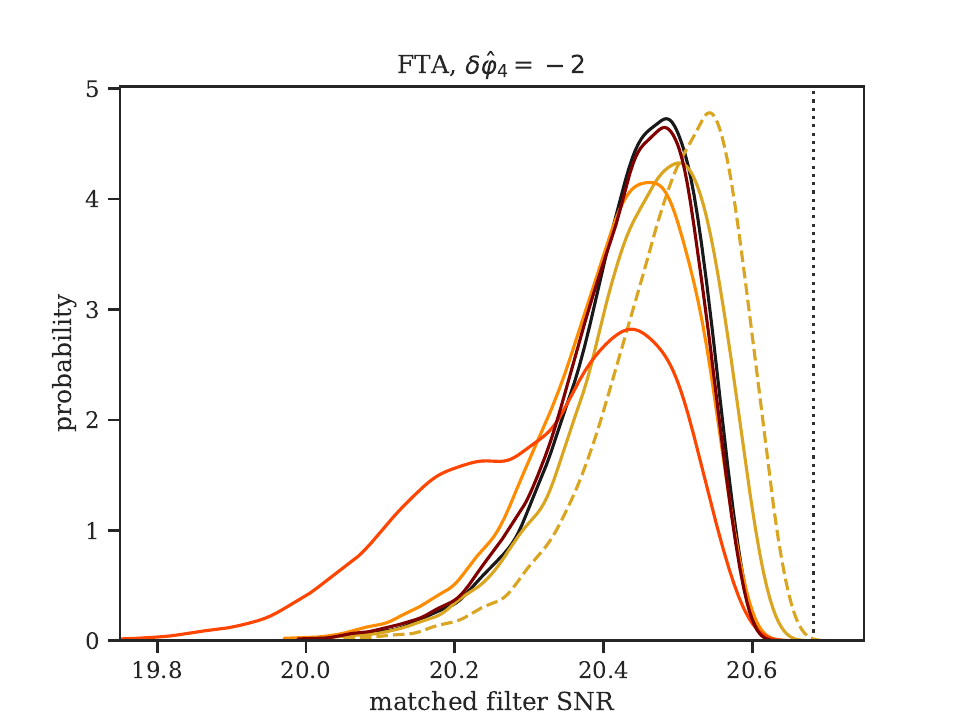}
}
\caption{\label{fig:SNR_plots_GW170608-like} The analog of Fig.~\ref{fig:SNR_plots_GW150914-like_larger} for the GW170608-like simulated observations. Here we do not show the SNR recovered by the median BayesWave reconstruction since it is much smaller than the SNRs plotted for these spread-out signals, as can be seen from the overlaps plotted in Fig.~\ref{fig:violin_plots_GW170608} and given in Table~\ref{tab:summary}.}
\end{figure*}

\begin{figure*}[htb]
\centering
\subfloat{
\includegraphics[width=0.42\textwidth]{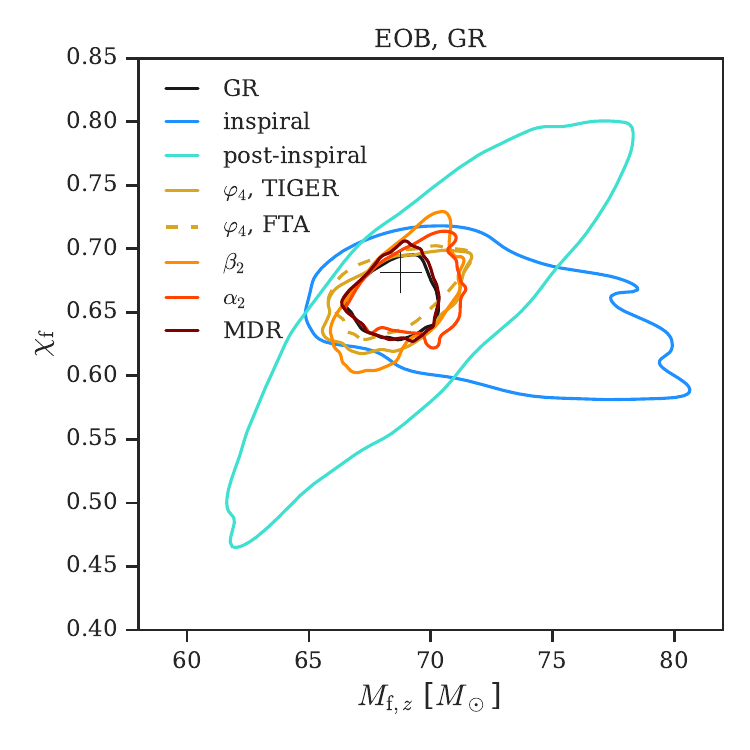}
}
\quad
\subfloat{
\includegraphics[width=0.42\textwidth]{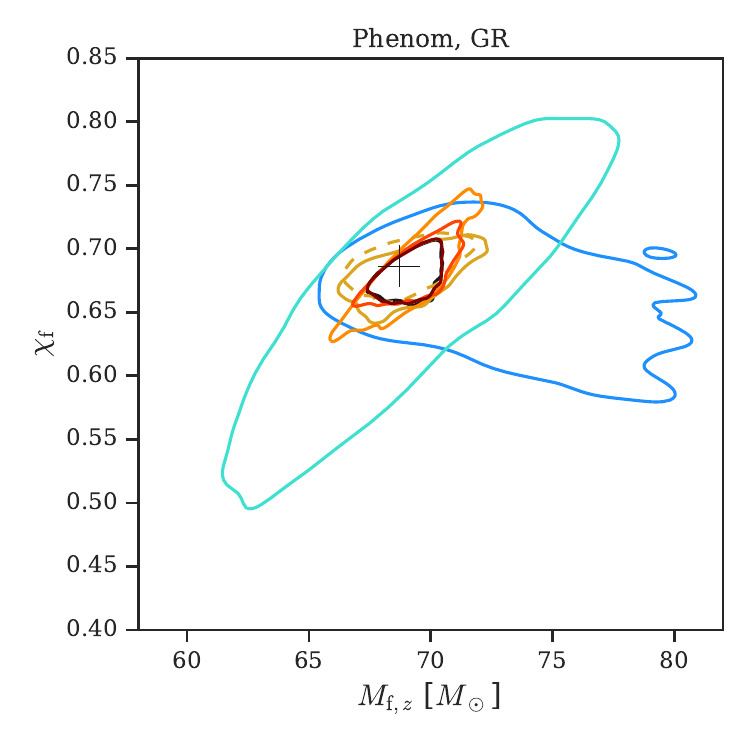}
}\\
\subfloat{
\includegraphics[width=0.42\textwidth]{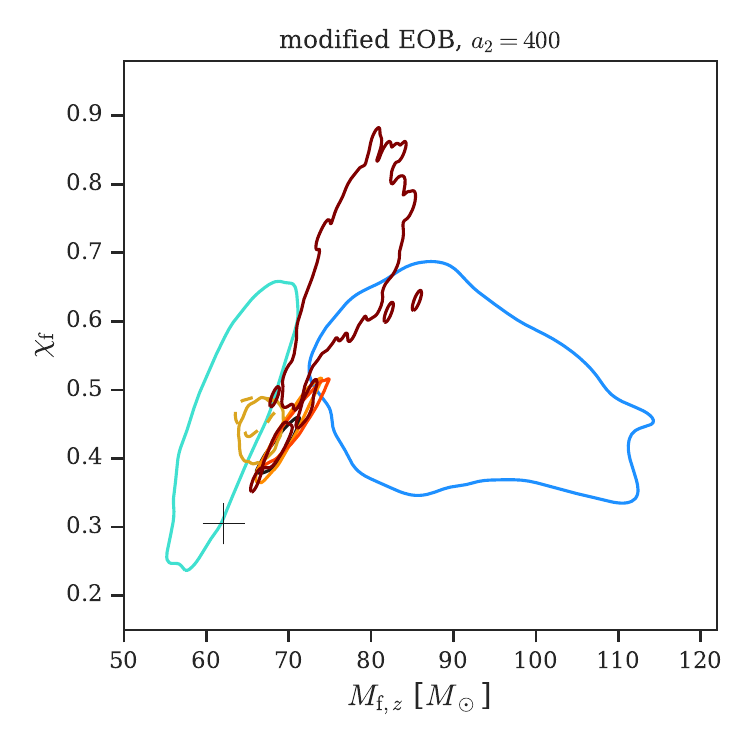}
}
\quad
\subfloat{
\includegraphics[width=0.42\textwidth]{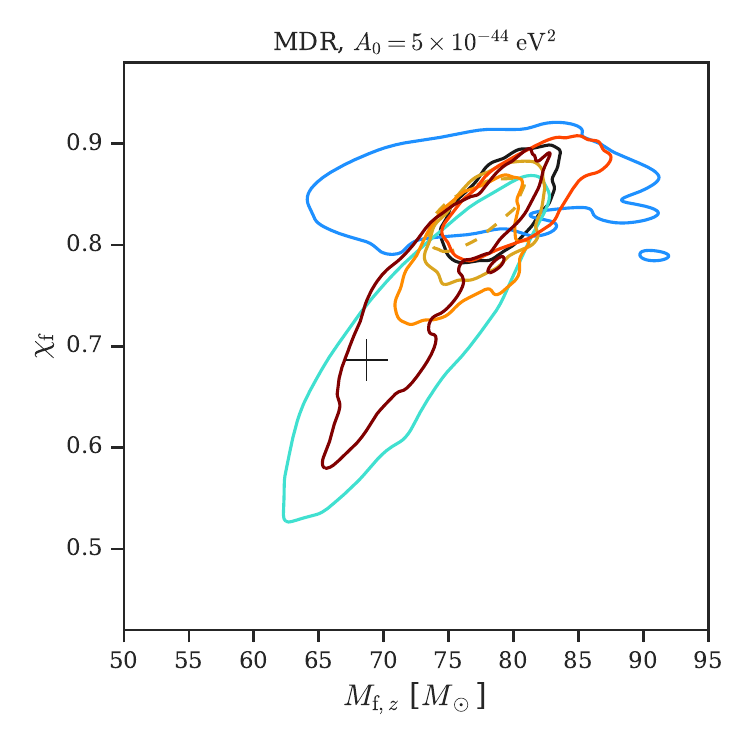}
}\\
\subfloat{
\includegraphics[width=0.42\textwidth]{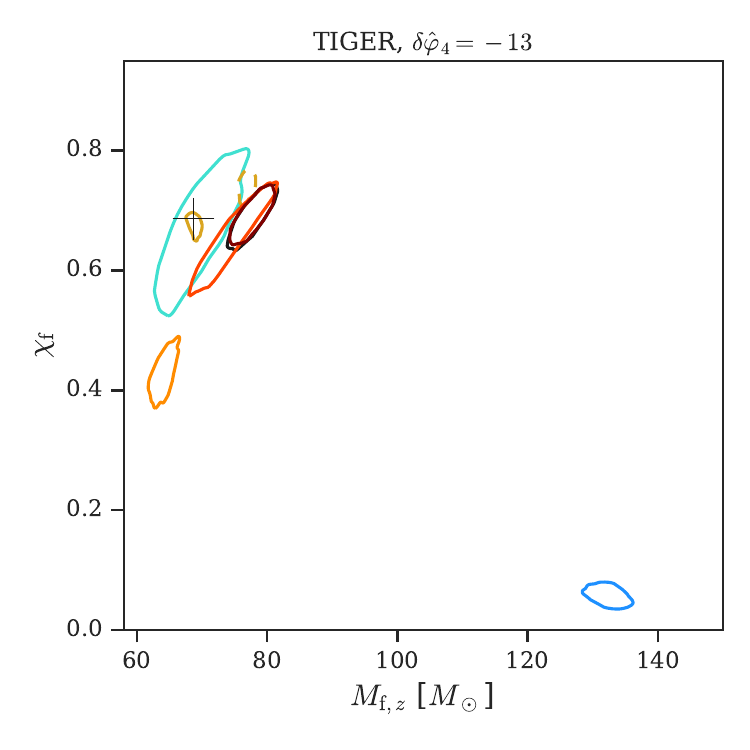}
}
\quad
\subfloat{
\includegraphics[width=0.42\textwidth]{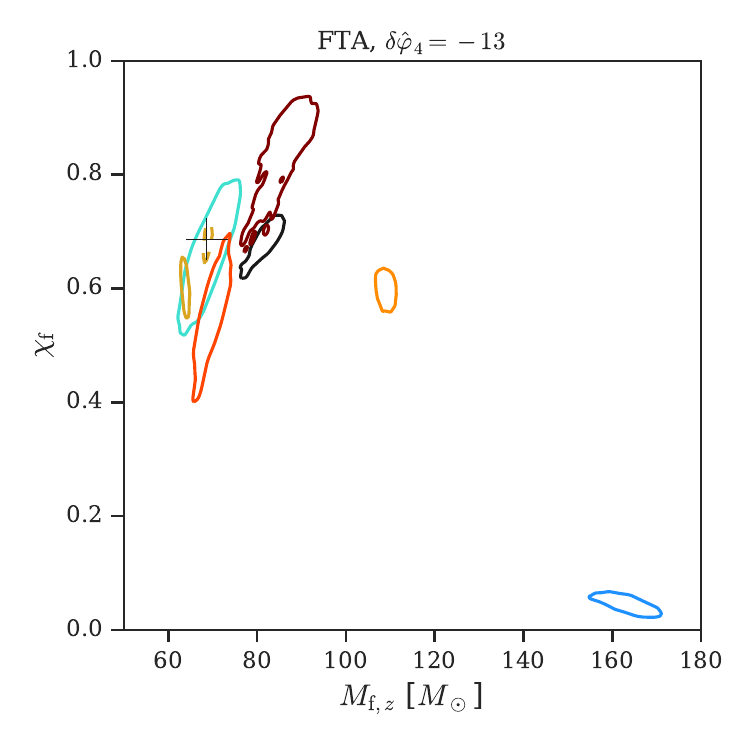}
}
\caption{\label{fig:Mf_af_plots_GW150914-like_larger} The $90\%$ credible regions of the joint posterior distributions of the recovered (redshifted) final mass and spin for the GW150914-like GR and larger GR deviation simulated observations, along with the values of the simulated observations, plotted as plus signs.}
\end{figure*}

\begin{figure*}[htb]
\centering
\subfloat{
\includegraphics[width=0.42\textwidth]{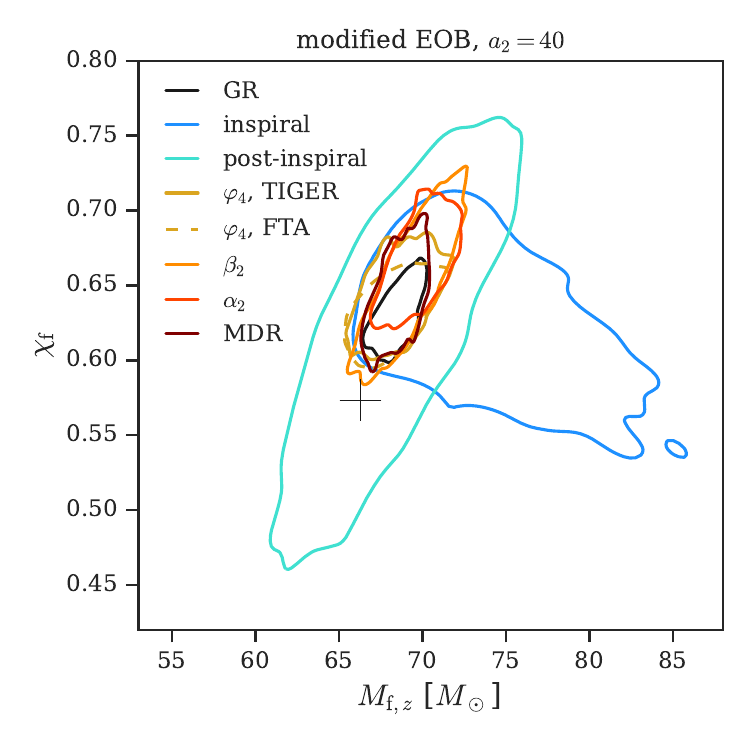}
}
\quad
\subfloat{
\includegraphics[width=0.42\textwidth]{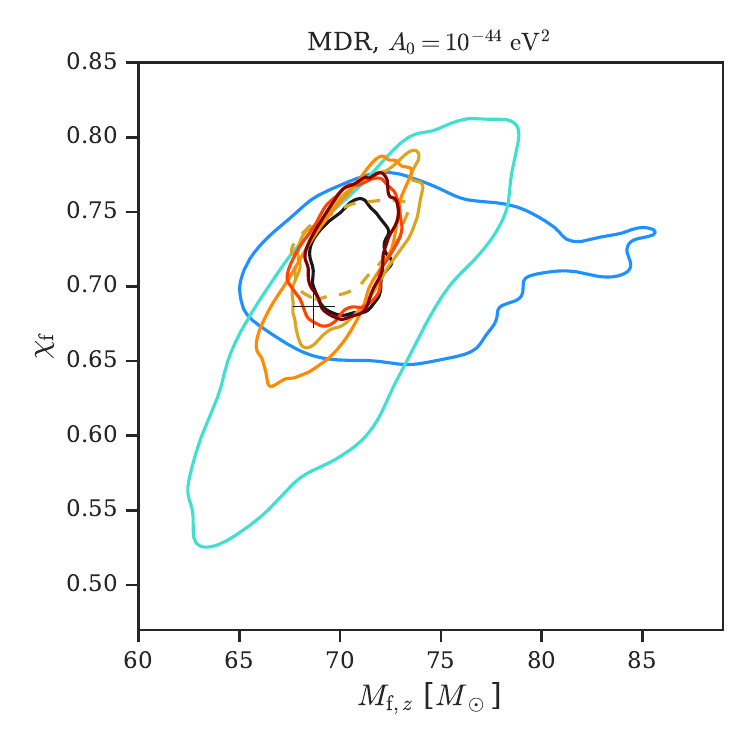}
}\\
\subfloat{
\includegraphics[width=0.42\textwidth]{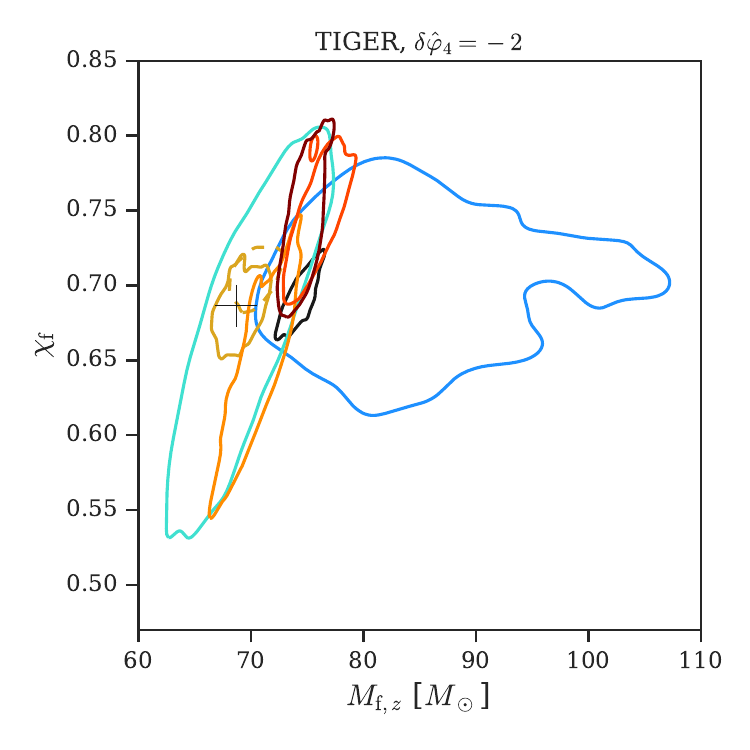}
}
\quad
\subfloat{
\includegraphics[width=0.42\textwidth]{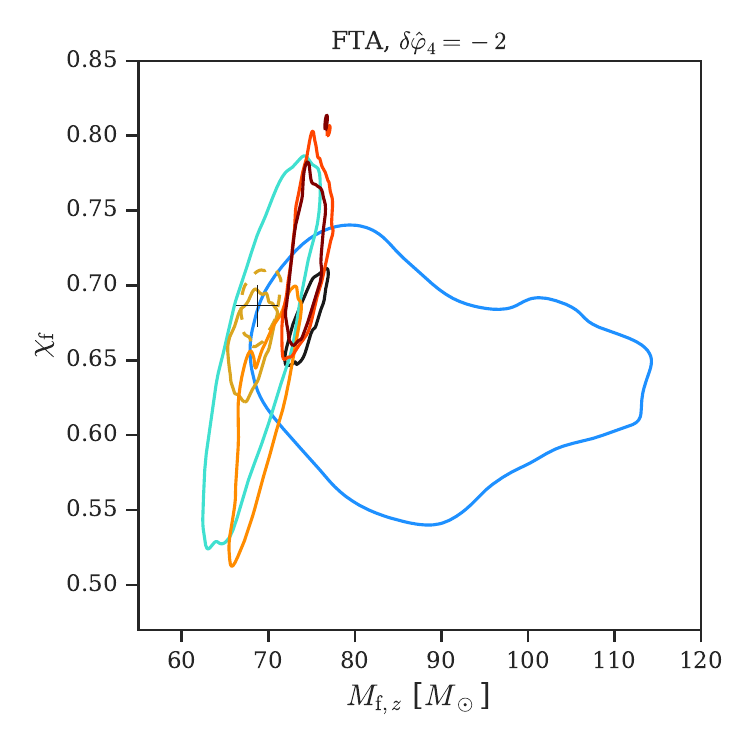}
}
\caption{\label{fig:Mf_af_plots_GW150914-like_smaller} The analog of Fig.~\ref{fig:Mf_af_plots_GW150914-like_larger} for the GW150914-like smaller GR deviation simulated observations.}
\end{figure*}

\begin{figure*}[htb]
\centering
\subfloat{
\includegraphics[width=0.42\textwidth]{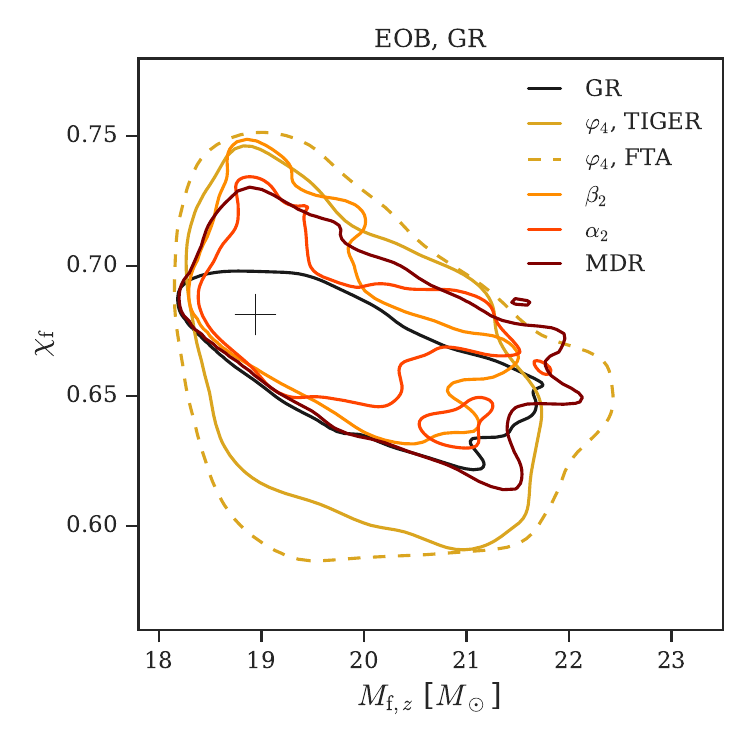}
}
\quad
\subfloat{
\includegraphics[width=0.42\textwidth]{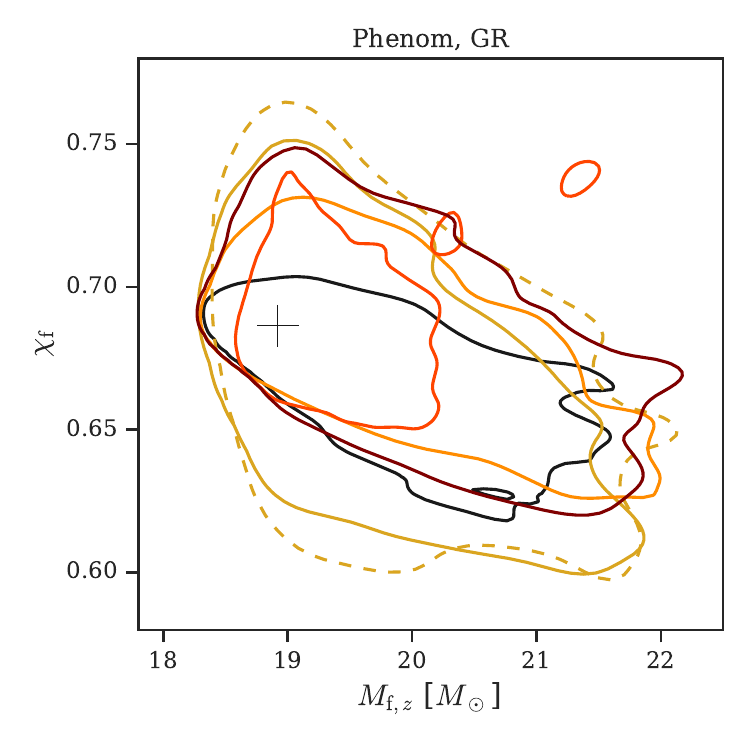}
}\\
\subfloat{
\includegraphics[width=0.42\textwidth]{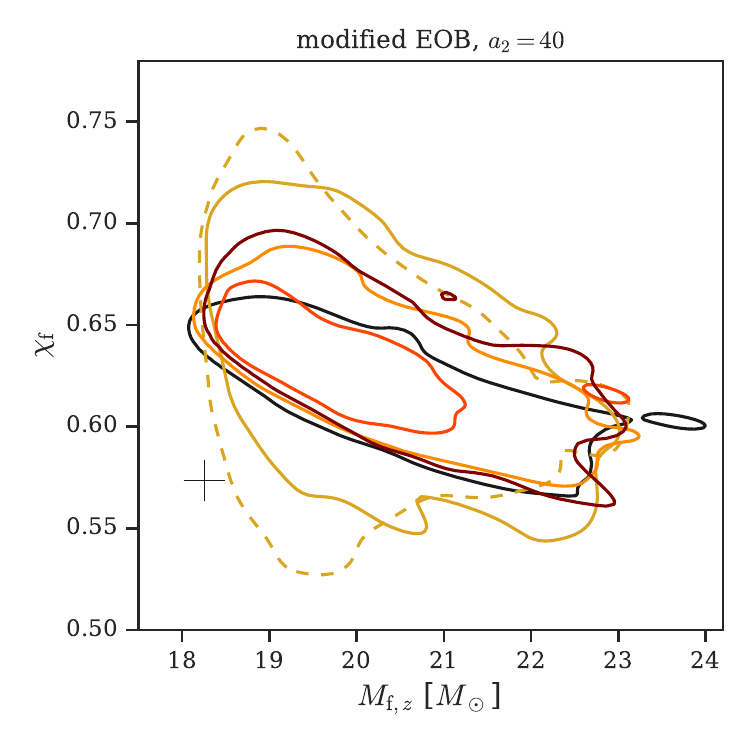}
}
\quad
\subfloat{
\includegraphics[width=0.42\textwidth]{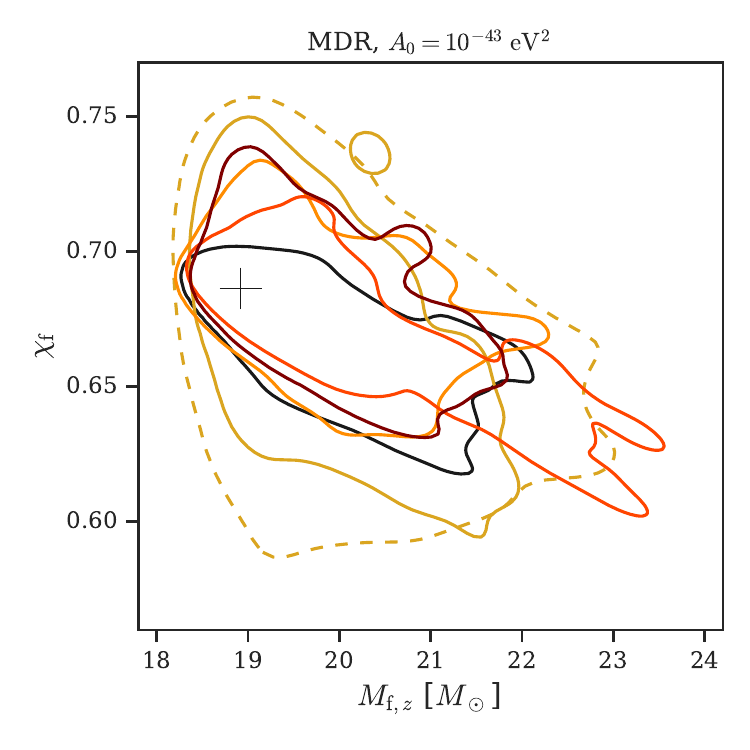}
}\\
\subfloat{
\includegraphics[width=0.42\textwidth]{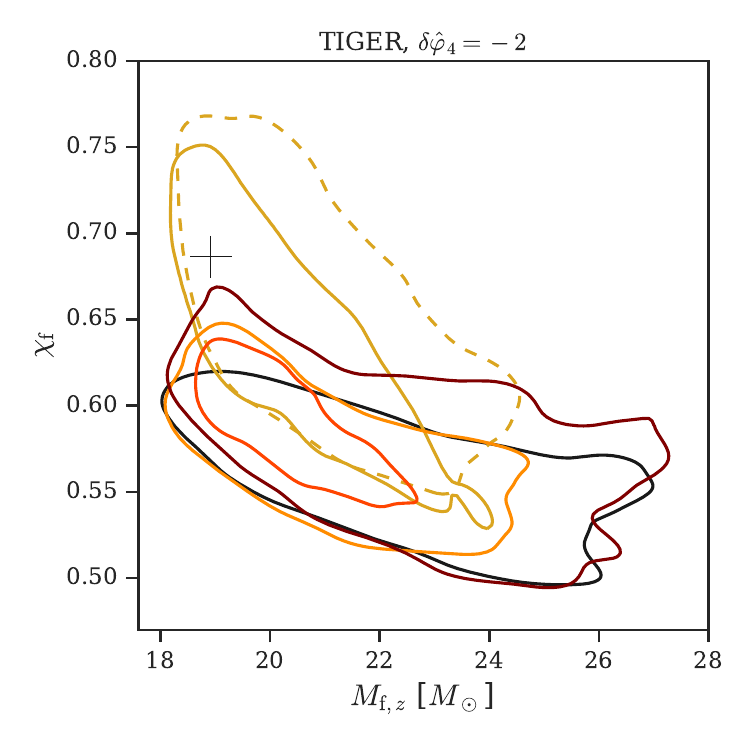}
}
\quad
\subfloat{
\includegraphics[width=0.42\textwidth]{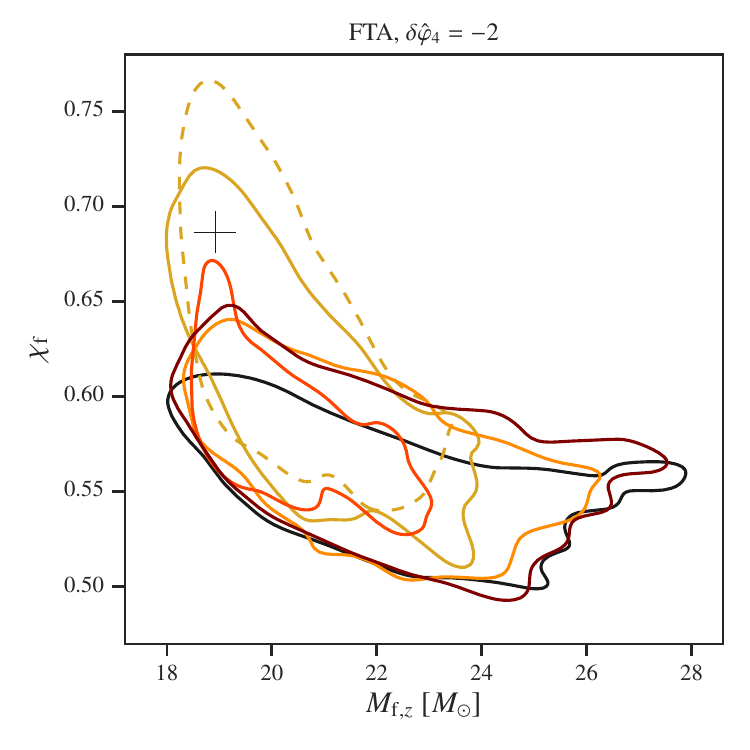}
}
\caption{\label{fig:Mf_af_plots_GW170608-like} The analog of Fig.~\ref{fig:Mf_af_plots_GW150914-like_larger} for the GW170608-like simulated observations.}
\end{figure*}

We give a summary of the results in Figs.~\ref{fig:violin_plots_GW150914-like_larger}, \ref{fig:violin_plots_GW150914-like_smaller}, and \ref{fig:violin_plots_GW170608} and Table~\ref{tab:summary}. The GR quantiles in the table are the quantile at which the GR value of the test is recovered. They are two-dimensional for the IMR consistency test, where smaller values indicate better consistency with GR, and one-dimensional for all other tests, where values around $50\%$ indicate better consistency with GR. We see that the tests all find that the GR simulated observations are consistent with GR within the $90\%$ credible level and discuss the results on the non-GR simulated observations in detail in the following. We show the network matched-filter SNRs recovered by the various analyses in Figs.~\ref{fig:SNR_plots_GW150914-like_larger}, \ref{fig:SNR_plots_GW150914-like_smaller}, and \ref{fig:SNR_plots_GW170608-like}. In the notation of Sec.~\ref{ssec:rec}, the network matched-filter SNR of data $\mathbf{d}$ with waveform model $\mathbf{h}$ is $\rho_\text{MF} := \langle\mathbf{d}|\mathbf{h}\rangle/\sqrt{\langle\mathbf{h}|\mathbf{h}\rangle}$. Finally, we show the recovery of the final mass and spin (for the non-BayesWave tests) in Figs.~\ref{fig:Mf_af_plots_GW150914-like_larger}, \ref{fig:Mf_af_plots_GW150914-like_smaller}, and \ref{fig:Mf_af_plots_GW170608-like}. These are computed the same way as for the IMR consistency test, as described in Sec.~\ref{ssec:imr}.

\subsection{GW150914-like cases}

\subsubsection{Larger GR deviations}

We find that all of the GW150914-like simulated observations with the larger GR deviation are identified as not consistent with GR at the $90\%$ credible level by at least three tests (see Table~\ref{tab:summary}). For the TIGER and FTA simulated observations, with their very large GR deviations, all the tests pick up the deviations, except surprisingly for the MDR test for the TIGER simulated observation, though the $\alpha_2$ TIGER case recovers GR just inside the $90\%$ credible interval. For the modified EOB simulated observation, all the tests recover a strong GR deviation except for the residuals and TIGER $\beta_2$ tests. The TIGER and FTA $\varphi_4$ tests exclude GR at the $99\%$ credible level or higher, but recover a value of $\delta\hat{\varphi}_4$ that is much smaller than the true value of $\sim -14$. This is not surprising: These tests are only varying a single PN coefficient, while the simulated observation also has all $3$PN and higher coefficients modified and additionally modifies the merger and ringdown. This illustrates that these tests are not designed to measure the true PN coefficients, just to detect deviations from GR. The massive graviton (MDR) simulated observation is identified as a GR violation at the $90\%$ credible level only with the IMR consistency test, TIGER $\beta_2$, and MDR analyses.

\begin{figure}[htb]
\centering
\includegraphics[width=0.45\textwidth]{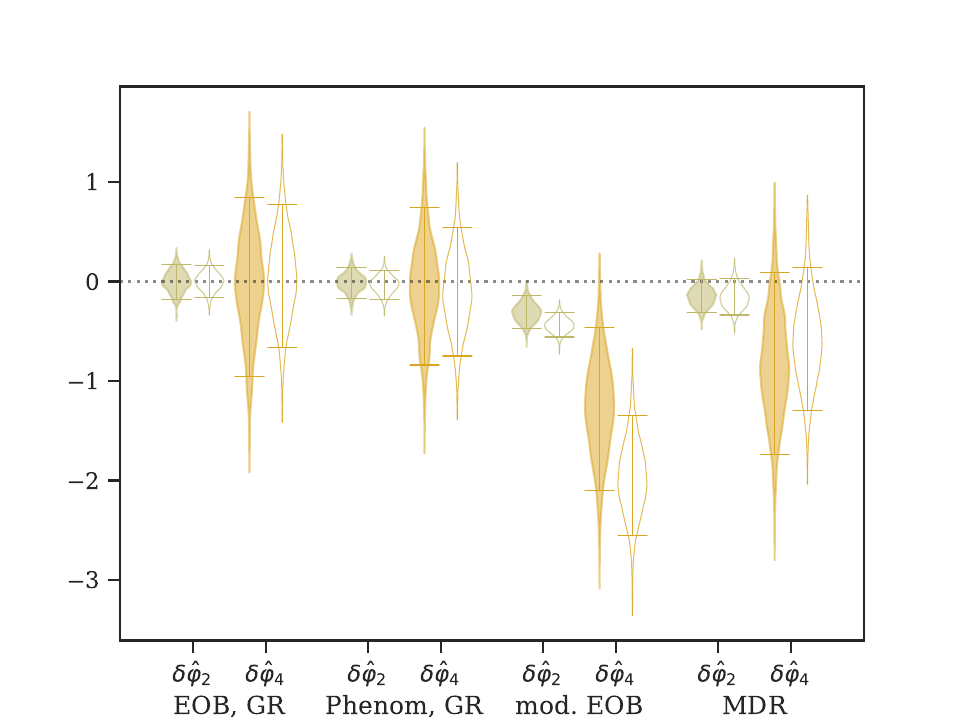}
\caption{\label{fig:dchi2_violins} Comparing the results of the TIGER and FTA $\varphi_2$ and $\varphi_4$ analyses on the GW150914-like simulated observations with GR waveforms and the modified EOB and massive graviton (MDR) waveforms with the larger GR violations. The FTA results are shown as unfilled violins.}
\end{figure}

\begin{figure}[htb]
\centering
\includegraphics[width=0.45\textwidth]{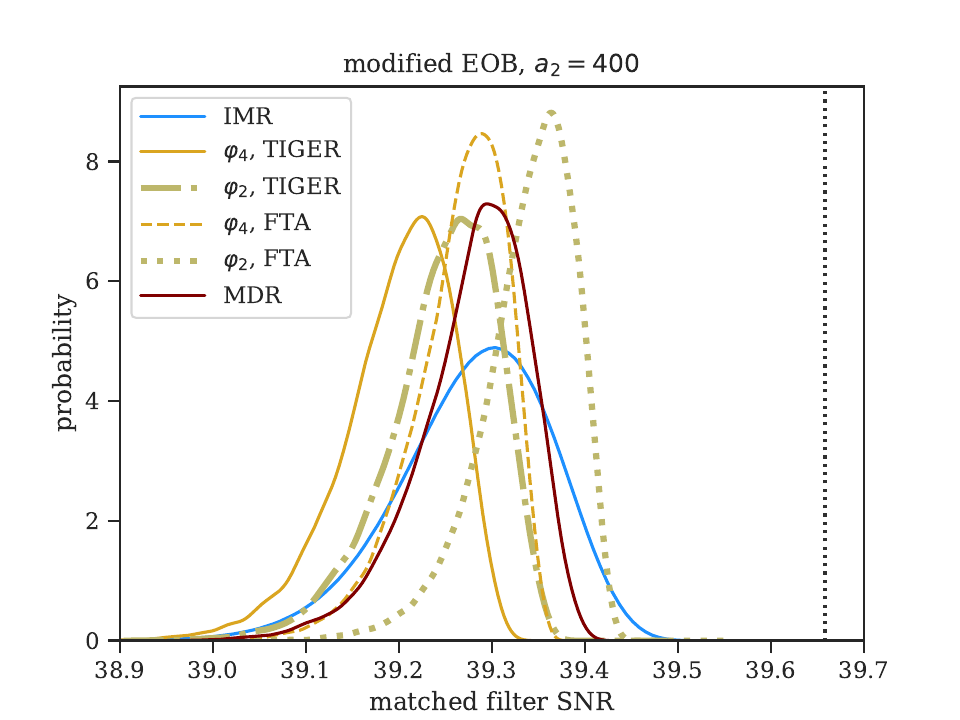}
\caption{\label{fig:SNR_plots_dchi2} The analog of Fig.~\ref{fig:SNR_plots_GW150914-like_larger} for the TIGER and FTA $\varphi_2$ analyses of the GW150914-like modified EOB simulated observation with the larger GR violations. We only show the IMR consistency, TIGER and FTA inspiral, and MDR results to give a cleaner plot.}
\end{figure}

\begin{table}
\caption{\label{tab:dchi2}
The analog of Table~\ref{tab:summary} for the TIGER and FTA $\varphi_2$ analyses of the GW150914-like simulated observations (simul.\ obs.)\ with the GR waveforms as well as the modified EOB and massive graviton (MDR) waveforms with the larger GR deviation.
}
\begin{tabular}{*{8}{c}}
\hline\hline
\multirow{3}{*}{Simul.\ Obs.} & & \multicolumn{2}{c}{$\varphi_2$, TIGER} & & \multicolumn{2}{c}{$\varphi_2$, FTA}\\
\cline{3-4}
\cline{6-7}
& & $Q_\text{GR}$ & \multirow{2}{*}{$\delta\hat{\varphi}_2$} & & $Q_\text{GR}$ & \multirow{2}{*}{$\delta\hat{\varphi}_2$}\\
& & $[\%]$ & & & $[\%]$ &\\
\hline
EOB, GR & & $50$ & $\hphantom{-}0.0_{-0.2}^{+0.2}$ & & $49$ & $\hphantom{-}0.0_{-0.2}^{+0.2}$\\[1pt]
Phenom, GR & & $54$ & $\hphantom{-}0.0_{-0.2}^{+0.1}$ & & $64$ & $\hphantom{-}0.0_{-0.2}^{+0.1}$\\[1pt]
modified EOB, $>$ & & $\mathbf{100}$ & $-0.3_{-0.2}^{+0.2}$ & & $\mathbf{100}$ & $-0.4_{-0.2}^{+0.1}$\\[1pt]
MDR, $>$ & & $93$ & $-0.1_{-0.2}^{+0.1}$ & & $92$ & $-0.2_{-0.1}^{+0.2}$\\[1pt]
\hline\hline
\end{tabular}
\end{table}

Since we find that the MDR test finds a strong deviation from GR for the modified EOB simulated observation, we also apply the TIGER and FTA $\varphi_2$ analyses to this simulated observation and the massive graviton simulated observation, since this PN coefficient matches the $\alpha = 0$ MDR dephasing in the inspiral. We also ran the TIGER and FTA $\varphi_2$ analyses on the GR simulated observations, for comparison. We compare with the TIGER and FTA $\varphi_4$ analyses in Fig.~\ref{fig:dchi2_violins} and give the analog of Table~\ref{tab:summary} in Table~\ref{tab:dchi2}. We find that the $\varphi_2$ analyses indeed pick up these GR deviations somewhat more strongly than the $\varphi_4$ analyses, except for the TIGER analysis of the massive graviton simulated observation, where the GR quantile is the same as the $\varphi_4$ analysis. The TIGER and FTA $\varphi_2$ analyses of the modified EOB simulated observation find increased SNR, illustrated in Fig.~\ref{fig:SNR_plots_dchi2}. The SNR distributions for the $\varphi_2$ analyses of the massive graviton and GR simulated observations are almost identical to those for the $\varphi_4$ analyses. 
For the massive graviton case, the TIGER and FTA $\varphi_2$ analyses recover a posterior that excludes the true value of $\delta\hat{\varphi}_2 = -0.66$, with a $90\%$ lower bound of about half the true value, similar to (though less dramatic than) their significant underestimate of the true value of the testing parameter in the modified EOB case.

\begin{figure*}[htb]
\centering
\includegraphics[width=\textwidth]{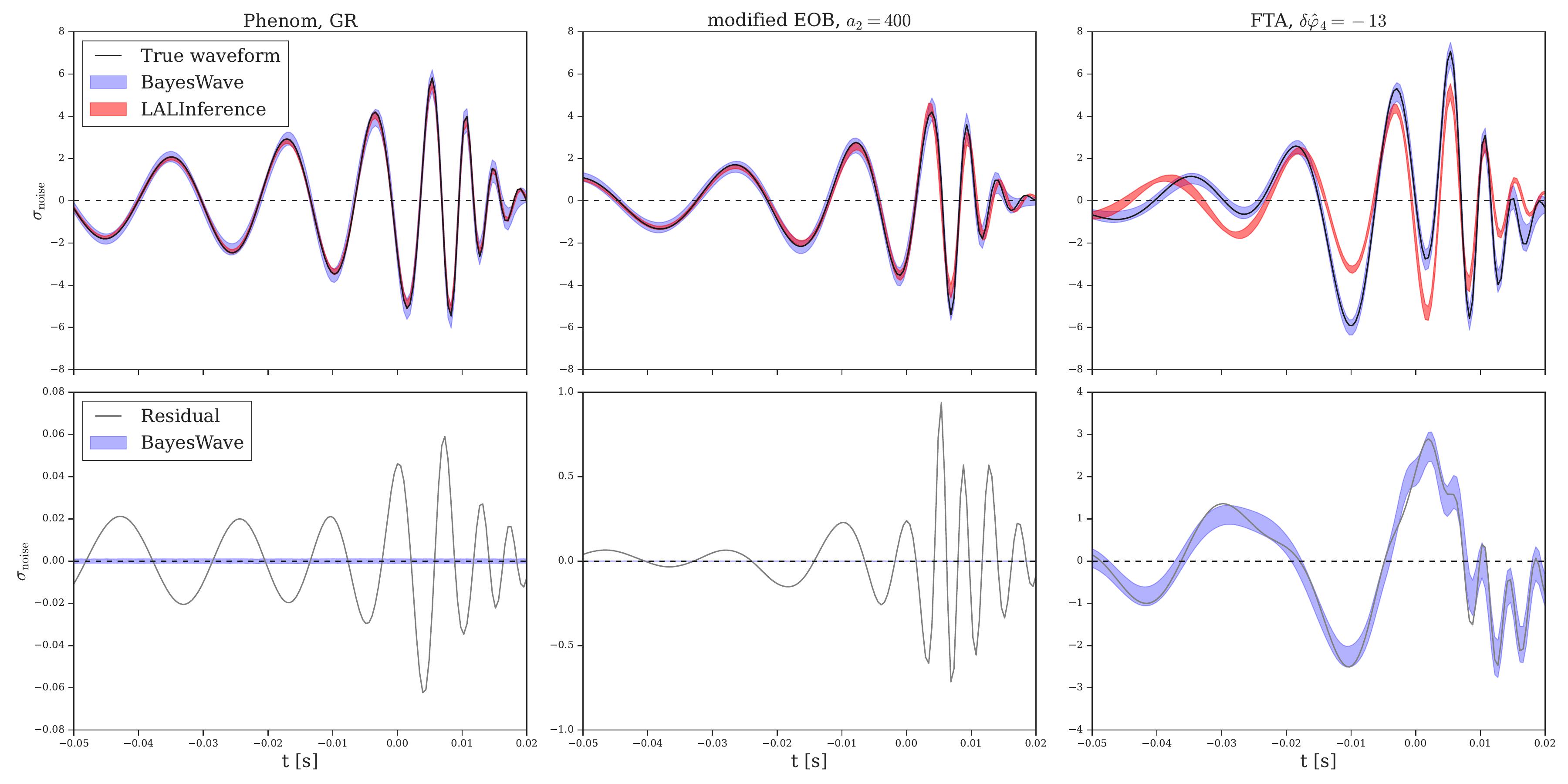}
\caption{\label{fig:rec_and_res} The waveform reconstruction and residuals analysis results on three cases of the GW150914-like simulated observations: the Phenom GR case and the modified EOB and FTA cases with larger GR deviations. All quantities are shown here as they would appear in the LIGO Livingston detector. The top panels show the true waveform, BayesWave $90\%$ credible intervals (CI), and the LALInference $90\%$ CI. The bottom panels show the residual data obtained by subtracting the maximum likelihood waveform obtained by the LALInference GR analysis, and the 90\% CI obtained by analyzing the residual data using BayesWave. The horizontal axis gives the time from the peak of the waveform, and vertical axis gives the strain amplitude whitened using a filter given by the inverse amplitude spectral density of the detector noise. The whitened strain is measured in units of the standard deviation of the noise, $\sigma_\mathrm{noise}$. Note that the disagreement between the LALInference reconstruction and the simulated waveform increases from left to right consistent with the increase of the deviation of the waveform morphology from GR from left to right. This is seen clearly in the bottom panels where the average amplitude of the residual time series grows approximately by an order of magnitude in each plot.}
\end{figure*}

We find that the residuals test is only able to identify the TIGER and FTA simulated observations, with their very large GR violations, as deviations from GR. Even though the modified EOB and massive graviton simulated observations also have significant GR violations that are easily picked up by some of the other tests, the distribution of residual SNRs is almost identical to that for the GR simulated observations. This is in agreement with the results in Fig.~\ref{fig:SNR_plots_GW150914-like_larger}, which show that the GR analysis is able to recover most of the SNR in those cases. We illustrate the residuals and their BayesWave recovery in a few cases in the bottom panels of Fig.~\ref{fig:rec_and_res}, which show the residual detector data in the LIGO Livingston detector and the recovered 90\% credible interval with BayesWave. This illustrates that while BayesWave is able to recover the residual signal very well when it is relatively significant, as for the FTA case, it does not find any coherent signal in the residual for the GR and modified EOB cases with their quite small and relatively small residuals, respectively.

The comparison of reconstructions is more sensitive, finding a distribution of overlaps that is disjoint from the one for the GR simulated observations for the modified EOB case, and a clear shift to larger mismatches for the massive graviton case, as well as clearly picking up the very large TIGER and FTA GR violations. This can be seen qualitatively in the top panels of Fig.~\ref{fig:rec_and_res}, where the difference between the BayesWave and LALInference reconstructions increases from left to right with increasing size of the GR deviation, and quantitatively through the overlaps given in Fig.~\ref{fig:violin_plots_GW150914-like_larger} and Table~\ref{tab:summary}. The reconstruction is also able to recover the most SNR of any analysis for the modified EOB case, and is second only to the TIGER (FTA) analysis for the TIGER (FTA) simulated observation, as shown in Fig.~\ref{fig:SNR_plots_GW150914-like_larger}. However, it only recovers about as much SNR as the GR analysis for the massive graviton simulated observation, likely because the dispersion spreads out the waveform, and the reconstructions do better at recovering short waveforms---the modified EOB, TIGER, and FTA waveforms are all significantly shorter than their GR counterparts, as shown in Fig.~\ref{fig:inj_TD_GW150914-like}.

For the cases where the GR quantile rounds to $0$ or $100\%$ in Tables~\ref{tab:summary} and~\ref{tab:dchi2}, it is interesting to consider how strongly GR is excluded. Here we use the scale of Gaussian standard deviations $\sigma$, and quote a lower bound of $7\sigma$ for cases where the GR quantile is even closer to $0$ or $100\%$. We impose such a lower bound because we have neglected the uncertainties in determining such high credible levels with a finite number of posterior samples, given that these are rather extreme scenarios, so these results should just be taken as roughly indicative of the constraining power of the tests in such cases. For the modified EOB simulated observation, we find that GR is excluded at greater than $7\sigma$ by the IMR consistency test, slightly greater than $3\sigma$ by the TIGER $\varphi_2$ test, by slightly greater than $4.5\sigma$ and $4\sigma$ in the FTA $\varphi_4$ and $\varphi_2$ tests, and by slightly greater than $5\sigma$ by the MDR test. For the massive graviton simulated observation, GR is excluded at slightly greater than $4\sigma$ by the MDR test. For the TIGER and FTA simulated observations, GR is excluded at greater than $7\sigma$ by the IMR consistency test as well as the TIGER and FTA $\varphi_4$ tests and the TIGER $\beta_2$ test. For the FTA simulated observation, GR is also excluded at greater than $7\sigma$ by the TIGER $\alpha_2$ test and at slightly greater than $4\sigma$ by the MDR test.

We now consider the recovery of the GR parameters. We start by considering the TIGER simulated observation, where the MDR analysis does not find a GR deviation, recovering an unequal-mass (mass ratio of $0.66_{-0.08}^{+0.08}$, giving the median and surrounding $90\%$ credible interval) precessing system with a nearly edge-on inclination and a highly spinning primary (the primary spin posterior rails strongly against the high-spin prior bound) with most of the primary spin in the orbital plane. Both signs of $A_0$ give very similar posteriors, which are also very similar to those from the GR recovery. For instance, the mass ratio median and $90\%$ credible interval is the same for the MDR recovery with both signs of $A_0$ and the GR recovery to the precision quoted above. There is also a much larger difference in the recovered SNRs for the different tests for the TIGER and FTA simulated observations than for the others---see Fig.~\ref{fig:SNR_plots_GW150914-like_larger}. This figure illustrates that the MDR recovery of the TIGER simulated observation finds a matched filter SNR very similar to the GR recovery, as would be expected, given the similarity of the posteriors for other parameters.

Now considering the modified EOB simulated observation, we find that all the tests except the IMR consistency test inspiral and MDR $A_0 < 0$ cases find close to equal masses, with an inclination angle and distance close to the true ones, but favor large antialigned spins to give a large negative effective spin\footnote{The effective spin is defined by $\chi_\text{eff} := (m_1\chi_1^\parallel + m_2\chi_2^\parallel)/(m_1 + m_2)$, where $m_A$ and $\mathbf{\chi}^\parallel_A$ ($A\in\{1,2\}$) are the holes' masses and components of the dimensionless spins parallel to the (Newtonian) orbital angular momentum, respectively. We consider this quantity here since it is a simple, well-measured combination of the spins that is closely related to the dominant spin-orbit coupling (see, e.g.,~\cite{Racine:2008qv,Santamaria:2010yb}).} and thus reduce the length of the signal and the final spin from their nonspinning GR values. The IMR consistency test inspiral recovery favors an unequal-mass system (mass ratio posterior peaking around $0.3$) with a small spin on the larger black hole and the smaller black hole's spin unconstrained. The MDR $A_0 < 0$ case finds large in-plane spins with an effective spin posterior that peaks close to zero. In both cases, they favor a slightly smaller inclination angle and larger distance than the true values. These two tests also recover slightly more matched filter SNR than the other tests---see Fig.~\ref{fig:SNR_plots_GW150914-like_larger}.

We also consider how well the (redshifted) final mass and spin are recovered by the non-BayesWave analyses, plotting the values of the simulated observations and the joint posterior distributions in Fig.~\ref{fig:Mf_af_plots_GW150914-like_larger}. We find that the MDR, TIGER, and FTA analyses always recover the true value in the $90\%$ credible region for their associated simulated observations, as does the IMR consistency test postinspiral analysis in all cases (albeit just barely for the modified EOB simulated observation). Additionally, the posteriors from many of the different tests are disjoint for all of the GR violating cases except for the massive graviton simulated observation.

For the MDR simulated observation, the IMR consistency postinspiral and MDR analyses are the only ones to recover the true values of the final mass and spin in the $90\%$ credible region. These and the IMR consistency inspiral analyses are also the only ones to recover the true values of the individual (redshifted) masses in the $90\%$ credible region, though most analyses recover the true value of the redshifted chirp mass ($M_z\eta^{3/5}$) in the $90\%$ credible region, except for the TIGER $\beta_2$ case, which recovers it at the $96\%$ credible level. All the analyses besides the IMR consistency and MDR tests recover unequal masses, with a mass ratio $\lesssim 0.5$, and a positive effective spin, generally $\chi_\text{eff} \gtrsim 0.4$, except for the TIGER $\beta_2$ analysis, for which $\chi_\text{eff} \gtrsim 0.2$. The IMR consistency inspiral analysis also recovers $\chi_\text{eff} \gtrsim 0.4$ and prefers unequal masses, even though there is support for equal masses. This preference for unequal masses and positive effective spins is not surprising, since both of these act to extend the inspiral, and the signal is stretched out in time by the propagation effect, as illustrated in Fig.~\ref{fig:inj_TD_GW150914-like}.

\subsubsection{Smaller GR deviations}

For the GW150914-like simulated observations with smaller GR deviations, we find that none of the tests recover the GR deviations in the modified EOB and massive graviton simulated observations above the $90\%$ credible level (when rounded), though the TIGER $\alpha_2$ and MDR analyses in the modified EOB case find that GR is excluded at the $88\%$ and $90\%$ credible level, respectively. The massive graviton simulated observations only have GR excluded at most at the $80\%$ credible level, with the MDR analysis. For the TIGER and FTA simulated observation, several tests find GR to be excluded at the $90\%$ credible level or higher, even at slightly greater than $4\sigma$ up to slightly greater than $4.5\sigma$ for the TIGER and FTA $\varphi_4$ tests. This is not surprising, given that the frequency-domain dephasings are significantly larger for the TIGER and FTA cases than for the modified EOB and massive graviton cases, as illustrated in Fig.~\ref{fig:inj_FD}. What is interesting is that the MDR analysis finds GR to be excluded at the $99\%$ credible level for the TIGER simulated observation, even though it only excluded GR at the $42\%$ credible level for the TIGER simulated observation with the larger deviation from GR. This is likely due to the larger GR deviation leading to a very short signal that is easier to fit with a GR template.

For the modified EOB simulated observation, the TIGER and FTA $\varphi_4$ analyses not only do not find a deviation from GR but have no support at the true value of $\delta\hat{\varphi}_4 \simeq -2$, showing again that the constraints one obtains from such analyses cannot be straightforwardly interpreted as constraints on PN parameters.

We find the residuals test to be insensitive to all these smaller modifications of GR, returning SNR distributions that are almost identical to those for the GR simulated observations, likely because the GR analysis recovers almost all of the SNR, as seen in Fig.~\ref{fig:SNR_plots_GW150914-like_smaller}. The reconstruction comparison finds shifts to larger mismatches for the modified EOB, TIGER, and FTA cases, compared to the GR cases, though not for the massive graviton case. However, there is still considerable overlap of the posteriors in the TIGER and FTA cases. The largest shift is seen for the modified EOB case, and even there the posteriors have some overlap, unlike the disjoint posteriors found in the case with the larger GR deviation. Interestingly, the reconstruction recovers less SNR in the modified EOB case than the median from the GR analysis, as seen in Fig.~\ref{fig:SNR_plots_GW150914-like_smaller}. This is presumably because the signal is more spread out with this smaller GR deviation than in the case of the large GR deviation, where the reconstruction found more SNR than the GR analysis did (see Fig.~\ref{fig:SNR_plots_GW150914-like_larger}).

The final mass and spin recovery is shown in Fig.~\ref{fig:Mf_af_plots_GW150914-like_smaller}. We find the same general pattern as before, with the true values for these quantities lying inside the $90\%$ credible regions for the IMR consistency test postinspiral analysis and for the test associated with the waveform in the TIGER and FTA cases. The true values fall just outside of the $90\%$ credible region in the massive graviton case with the MDR analysis, discussed further below. The same patterns in the recovery of the mass ratio and effective spin for the modified EOB and massive graviton simulated observations noted above for the larger GR violations are still present here, just with reduced amplitude. That is, the recoveries of the modified EOB simulated observation prefer close to equal masses and negative effective spins to give a shorter waveform and smaller final spin, while the recoveries of the massive graviton simulated observation prefer unequal masses and positive effective spins, to give a longer waveform. In fact, this preference is even seen in the MDR $A_0 > 0$ recovery of the massive graviton simulated observation and likely explains the bias seen in the final mass and spin noted above.

\subsection{GW170608-like cases}

For the GW170608-like cases, we do not consider the IMR consistency test, since it is not applicable to these low-mass, moderate-SNR systems, and is thus not applied to GW170608 in~\cite{O2_TGR, O3a_TGR, O3b_TGR}. The BayesWave analyses are also not as well suited to these more spread-out signals as to the shorter GW150914-like signals considered previously, but we show their results anyway, for comparison, since these analyses are applied to GW170608 itself in~\cite{O2_TGR, GWTC-1_paper}.

In the GW170608-like cases, we find that the tests only identify the GR deviations at or above the $90\%$ credible level in the TIGER case, and even there this is only for the TIGER and FTA $\varphi_4$ tests and the TIGER $\beta_2$ test. However, in the FTA case the FTA analysis finds a GR deviation at the $88\%$ credible level. The most significant GR deviation for the modified EOB case is again found by the MDR analysis, though this time only at the $72\%$ credible level. The TIGER and FTA analyses also find that the true value of the deviation parameter ($\delta\hat{\varphi}_4 \simeq -2$) is outside the $90\%$ credible interval, though it is closer here than in the GW150914-like cases. The reconstructions analysis finds a distribution of mismatches for the modified EOB case that is shifted to larger values than for the GR cases (and the other non-GR cases), though the distributions still overlap. 

In the massive graviton case, not even the MDR analysis finds a significant deviation from GR. In fact, the true value of $\tilde{A}_0 = 10$ is well outside of the $90\%$ credible interval. This is due largely to a bias in the recovery of the distance, since the inferred distance determines how one converts the observed dephasing into a bound on $A_0$. This bias on the distance comes from the distance-inclination degeneracy, where the distance and inclination angle both peak at significantly larger values than the true ones. See, e.g., Fig.~9 in~\cite{Veitch:2014wba} for an example of this bias for a simulated binary black hole observation in Gaussian noise and Fig.~1 in~\cite{Rodriguez:2013oaa} for an example for a simulated binary neutron star observation with zero noise. However, the bias we find is a bit more extreme than in those cases, with a median and $90\%$ credible interval for the distance of $591^{+82}_{-168}$~Mpc for the GR analysis of the Phenom GR simulated observation, compared to the true value of $364$~Mpc. We find this bias in all of our analyses of the GW170608-like cases. In particular, the MDR recovery with both signs of $A_0$ gives very similar results for the distance median and $90\%$ credible interval to the analysis of the Phenom GR simulated observation.

If one uses the true values of the distance and redshift to obtain the posterior on $A_0$ from the posterior on $\lambda_{A,\text{eff}}$, which is the parameter that directly enters the phase and is sampled on [see, e.g., Eq.~(2) in~\cite{O2_TGR}], and scales $\lambda_{A,\text{eff}}$ by $(D_L/D_L^\text{true})^2$ so that the dephasing is unchanged, then one obtains a median and $90\%$ credible interval for $\tilde{A}_0$ of $1.0^{+8.8}_{-3.9}$, so the true value of $\tilde{A}_0 = 10$ is much closer to being included. There are no noticeable biases in the other parameters, though all the analyses of the massive graviton case favor unequal masses and a positive effective spin, as was found for the other massive graviton cases. This preference may explain the remaining bias in the recovery of $A_0$.

As illustrated in Fig.~\ref{fig:SNR_plots_GW170608-like}, the recovered SNR is not significantly different between the different tests except in a few cases. Two of these cases are the TIGER and FTA simulated observations, with their somewhat larger GR deviations. The other cases are the TIGER $\alpha_2$ test for all simulated observations, with its very broad posteriors on the testing parameter. These cases all have a broader posterior on the SNR, as well, extending to lower values. The recovery of the final mass and spin is shown in Fig.~\ref{fig:Mf_af_plots_GW170608-like}. None of the tests find the true values for the modified EOB case in their $90\%$ credible regions and all the analyses prefer unequal masses and a negative effective spin, as we found for the other modified EOB cases. The final mass and spin are recovered in all the $90\%$ credible regions in the massive graviton case, but just for both the TIGER and FTA tests in the TIGER and FTA cases.

\section{Summary and conclusions}
\label{sec:concl}

We have studied how a selection of standard tests of GR that are regularly applied to LIGO-Virgo observations of binary black holes respond to a variety of phenomenological deviations from GR. Specifically, we considered the residuals test, IMR consistency test, TIGER and FTA parameterized tests, and the MDR test. We also considered how well the unmodeled reconstructions of the waveforms agree with the GR waveforms that are found to describe the signal well. The non-GR waveforms we considered are the ones with phenomenological deviations in post-Newtonian coefficients used in the TIGER and FTA tests, as well as the propagation effects from a massive graviton, and a self-consistent modification of the binary's energy flux in the EOB framework. For all of these waveforms, we considered a GW150914-like system with larger and smaller GR deviations and a GW170608-like system with smaller GR deviations. We also considered the GR analogs of the non-GR waveform models considered.

For the GW150914-like case with larger deviations of GR, we found that the deviations from GR are detected at a high credible level by most of the tests considered. However, even for these large deviations, some tests find consistency with GR at the $90\%$ credible level. In particular, in the massive graviton case with the larger graviton mass, only the IMR consistency test, TIGER $\beta_2$, and MDR analyses exclude GR at the $90\%$ credible level (and just barely for the IMR consistency test). However, all other cases with large GR deviations are identified as deviations from GR at the $90\%$ credible level or greater by at least five tests. Indeed, many of the larger GR deviations are identified as such at very high credible levels, greater than a Gaussian $5\sigma$. (These very high credible levels are likely because our simulated observations do not contain noise.) For the GW150914-like smaller GR deviations, the number of tests that find a significant GR deviation decreases considerably. Most notably, none of the tests identify the massive graviton case as a GR violation above the $80\%$ credible level and the modified EOB case is only (just) identified as a GR violation at the $90\%$ credible level by the MDR analysis. However, the TIGER and FTA modifications are identified as GR violations at the $90\%$ credible level or greater by all but two of the tests considered.

For the GW170608-like case, with its smaller SNR, we found that only the TIGER case is identified as a GR deviation at the $90\%$ credible level (by the TIGER and FTA tests), though the FTA analysis almost identifies the FTA case as a GR deviation at the $90\%$ credible level and the MDR analysis identifies the modified EOB waveform as a GR violation at the $72\%$ credible level. These are the only cases that are identified as GR violations at such high credible levels.

One does not always find that the tests one expects to detect a given GR violation strongly are actually effective in doing so. Conversely, one finds that tests that one might not expect to be effective in detecting a given GR violation detect it strongly. The most striking example of both of these is likely the GW150914-like modified EOB waveform with the smaller GR deviation. Here one might expect that the TIGER and FTA tests that look for deviations in the $2$PN phase coefficient would find significant deviations from GR, since the leading deviation in the inspiral phase in the modified EOB waveform is at $2$PN. However, this is not the case: Both of these tests find excellent consistency with GR in this case, while the MDR analysis recovers a deviation from GR at the $90\%$ credible level (albeit just barely). For another case where the TIGER and FTA tests of inspiral PN coefficients do not recover the deviation from GR as strongly as one might expect, in the GW150914-like massive graviton case with the larger GR deviation, where the leading order of the deviation in the inspiral is at $1$PN, the TIGER $\beta_2$ intermediate coefficient test finds a GR deviation at the $98\%$ credible level, while the TIGER and FTA $1$PN analyses only find a deviation at about the $85\%$ credible level.

In fact, for the modified EOB waveform, the $2$PN TIGER and FTA analyses do not even recover the true value of the deviation parameter within the $90\%$ credible interval. This is particularly true for the GW150914-like cases, though in the case with the larger GR deviation, the TIGER and FTA analyses do find a strong deviation from GR, even though they underestimate the size of the deviation parameter by almost an order of magnitude. In the GW170608-like case, the true value of the deviation parameter is slightly closer to the boundary of the $90\%$ credible region than in the GW150914-like case with the smaller GR deviation, but still outside it. The $1$PN TIGER and FTA analyses of the GW150914-like massive graviton case with the larger massive graviton mass (the only massive graviton case we analyze with the $1$PN analyses) also recover significantly smaller deviations than the true value.

The fact that the TIGER and FTA analyses do not recover the true value of the modified PN parameter is not surprising. These analyses are designed to detect deviations from GR, not to measure individual PN coefficients: They only modify one PN coefficient at a time and include the post-inspiral part of the signal in the analysis without attempting to account for the expected modifications to this part of the waveform in modified theories. This means that analyses that interpret the TIGER and FTA results as constraints on PN parameters, e.g.,~\cite{Nair:2019iur,Wang:2021yll}, may be obtaining apparent constraints on modified theories that are significantly more stringent than actually allowed by the data. Additionally, this suggests that it would be a good idea to apply similar checks to the method for modifying the PN coefficients in~\cite{Perkins:2021mhb}, where the frequency domain dephasing is applied to the entire signal. Given the results here, it seems likely that this method will also underestimate the size of a potential GR deviation, invalidating the constraints on alternative theories presented there. Developing a method to constrain deviations from PN coefficients accurately in as generic a situation as possible would be a very worthwhile endeavor. The method in~\cite{Wang:2021yll} that restricts to the low-frequency portion of the signal is a possible way to proceed, though it would still need to be validated with these sorts of tests. In particular, it seems unlikely that the current setting of the IMRPhenomD value for the end of the inspiral for the high-frequency cutoff is the optimal choice. We provide the frame files for our simulated observations~\cite{frame_files_Zenodo} so they can be used to perform such checks.

One also finds that the residuals test is not very sensitive to most of the deviations from GR considered here. It only excludes GR for the extreme deviations from GR in the GW150914-like TIGER and FTA cases with the larger GR violations, where the waveforms do not look at all like those from binary black hole coalescences in GR (see Fig.~\ref{fig:inj_TD_GW150914-like}). Thus, while the residuals test seems like a promising way to identify deviations from GR (or more generally from the quasicircular binary black hole hypothesis) without making assumptions about the exact nature of the deviations, it is likely only effective in detecting extreme deviations from GR, at least for the relatively moderate SNRs that one expects for most detections by current and near-future detectors.

The comparison of unmodeled and GR reconstructions appears to be more effective at identifying deviations from GR than the residuals test: The distribution of mismatches between the two reconstructions is well separated from the distribution for the GR waveform in several cases where the residuals test does not identify any deviation from GR, notably for the GW150914-like modified EOB waveform with the larger GR deviation. However, in our analysis we are only comparing the mismatches between reconstructions to a single GR case and with no noise. It is likely that the expected distribution of mismatches in the GR case would broaden considerably when considering a larger range of GR waveforms and detector noise, considerably weakening these results. Nevertheless, it is likely worth pursuing the reconstruction comparison as a test of GR for high-mass binary black hole signals. It will, however, not be applicable to low-mass signals, like the GW170608-like cases we consider, where the power is spread out over about a second or more, making it difficult for the unmodeled reconstructions to recover the waveform accurately.

Finally, we found that the final mass and spin distributions recovered by the different tests have disjoint $90\%$ credible regions for many of the tests with larger GR deviations. This suggests that it might be worthwhile to develop  a ``meta IMR consistency test'' by comparing the recovery of the final mass and spin (or other parameters) between different tests.

Of course, this is still quite a preliminary study, and there is much more to do to assess the relation between different tests of GR on gravitational wave data. For instance, it is important to consider the effects on tests of GR of missing physics in the waveform models, e.g., higher modes, for which there are initial studies in~\cite{Pang:2018hjb,Islam:2021pbd}, as well as eccentricity (for which there are fairly well developed numerical relativity calculations for binary black holes and some waveform models that reproduce these results reasonably well in certain portions of the parameter space, e.g.,~\cite{Hinder:2017sxy,Huerta:2019oxn,Ramos-Buades:2019uvh,Liu:2019jpg,Chiaramello:2020ehz,Chen:2020lzc,Gayathri:2020coq,Setyawati:2021gom,Islam:2021mha,Liu:2021pkr,Yun:2021jnh,Nagar:2021xnh,Placidi:2021rkh,Ramos-Buades:2021adz}). Other important physical effects to consider are those from gravitational lensing (e.g., the effects calculated in~\cite{Ezquiaga:2020gdt} in the geometrical optics regime as well as wave optics effects~\cite{Takahashi:2003ix,Pagano:2020rwj}), and the presence of a third body (e.g., the calculations in~\cite{Meiron:2016ipr,DOrazio:2019fbq,Gupta:2019unn,Toubiana:2020drf,Cardoso:2021vjq,Chandramouli:2021kts,Yu:2021dqx,Gondan:2021fpr}) and other environmental effects (e.g., from gas or dark matter)~\cite{Barausse:2014tra,Toubiana:2020drf}.

Similarly, one should consider waveforms from binaries of black hole mimickers (see~\cite{Johnson-McDaniel:2018uvs, Islam:2019dmk} for some simple checks using rescaled binary neutron star and black hole--neutron star waveforms, respectively, and~\cite{Toubiana:2020lzd} for a toy model for such waveforms). Finally, one needs to assess the effects of systematic errors in the baseline GR waveform models, which could plausibly start to affect current combined constraints, as discussed in~\cite{Moore:2021eok}, and will definitely be important even for loud individual events in future detectors, as discussed in, e.g.,~\cite{Purrer:2019jcp,Ferguson:2020xnm}.

One will also want to include more tests in future studies and use waveforms from various alternative theories, once they are computed with sufficient accuracy (there are also constructions of self-consistent waveforms based on analytical knowledge of modified theories~\cite{Carson:2020cqb}, also used in~\cite{Carson:2020ter}, that could be useful in these sorts of studies before full numerical waveforms are available). It is also important to consider the effects of detector noise and calibration as well as systematics in the GR waveform models; see~\cite{Kwok:2021zny} for a study of the effects of transient non-Gaussian noise features (``glitches'') and their removal on TIGER. However, the most important study will likely be considering populations of signals to determine how well the tests perform when combining together multiple observations to potentially detect smaller deviations from GR, e.g., using the method in~\cite{Isi:2019asy}. Here it will be particularly important to include the effects of spins and higher modes in the simulated observations, which were not included in this initial study.

Nevertheless, this study already indicates that one will require quite high SNRs, above the SNRs of $\sim 50$ we considered here in the GW150914-like case, to be able to detect some moderate deviations from GR in individual events with the tests we consider here. This strongly motivates the need for improvements in gravitational wave detectors, particularly third generation ground-based detectors~\cite{Hild:2010id,Reitze:2019iox,Hall:2020dps}, to provide the much larger SNRs that will allow one to distinguish relatively small deviations from GR. Additionally, improvements in the design of tests of GR and methods for combining together multiple observations will also be necessary to fully exploit current and future gravitational wave detector data to test GR.

\acknowledgments

We wish to thank all the LIGO-Virgo-KAGRA testing GR group members who implemented these tests in publicly available code. Additionally, we thank Anuradha Samajdar for assistance with the MDR test, initial work on this project, and a careful reading of the paper, Archisman Ghosh for the code used to create the frame files to analyze, Noah Sennett and Michalis Agathos for the FTA reweighting script, Parameswaran Ajith for initial discussions, and Chris Van Den Broeck and B.~S.~Sathyaprakash for useful comments.

N.~K.~J.-M.\ acknowledges support from STFC Consolidator Grant No.~ST/L000636/1. S.~G.\ gratefully acknowledges support from National Science Foundation (NSF) grant PHY-1809572. M.~S.\ acknowledges support from the Infosys Foundation, the Swarnajayanti fellowship grant DST/SJF/PSA-01/2017-18, and NSF grants PHY-1806630, PHY-2010970, and PHY-2110238.
N.~V.~K.\ acknowledges support from the Max Planck Society's Independent Research Group Grant. J.~A.~C.\ acknowledges support from NSF grants PHY-1700765 and PHY-1764464.

The authors are grateful for computational resources provided by the LIGO Laboratory and supported by NSF Grants PHY-0757058 and PHY-0823459 as well as by the Open Science Grid~\cite{Pordes:2007zzb,osg09}, which is supported by the NSF award \#2030508.
Additional computations were performed on the clusters Alice at the International Centre for Theoretical Sciences, Tata Institute of Fundamental Research and Hypatia at the Max Planck Institute for Gravitational Physics (Albert Einstein Institute), Potsdam-Golm.

This material is based upon work supported by NSF's LIGO Laboratory which is a major facility fully funded by the NSF. This research has made use of data obtained from the Gravitational Wave Open Science Center (www.gw-openscience.org), a service of LIGO Laboratory, the LIGO Scientific Collaboration and the Virgo Collaboration. LIGO is funded by the US NSF. Virgo is funded by the French Centre National de Recherche Scientifique (CNRS), the Italian Istituto Nazionale della Fisica Nucleare (INFN) and the Dutch Nikhef, with contributions by Polish and Hungarian institutes.

We used the following software in this study: BayesWave~\cite{Cornish:2014kda,Cornish:2020dwh}, LALSuite~\cite{LALSuite}, matplotlib~\cite{Hunter:2007ouj}, numpy~\cite{Harris:2020xlr}, PESummary~\cite{Hoy:2020vys}, PyCBC~\cite{PyCBC}, scipy~\cite{Virtanen:2019joe}, and seaborn~\cite{waskom2020seaborn}.

This is LIGO document P2100322.

\appendix*

\section{Two-dimensional IMR consistency plots}
\label{app:imr}

\begin{table}
\caption{\label{tab:fcut}
Cutoff frequencies $f_\text{cut}$ inferred from the GR analysis of each of the GW150914-like simulated observations (``simul.\ obs.''), rounded to the nearest Hz. As in Table~\ref{tab:summary}, $>$ denotes the case with the larger GR deviation and $<$ the one with the smaller deviation. The results are given in two groups, first the EOB GR result and the results for the modified EOB waveforms and then the Phenom GR result and the results for the Phenom-based non-GR waveforms.
}
\begin{tabular}{ccc}
\hline\hline
Simul.\ Obs.\ & \multicolumn{2}{c}{$f_\text{cut}$ [Hz]}\\
\cline{2-3}
& $>$ & $<$\\
\hline
EOB, GR & \multicolumn{2}{c}{$129$}\\
mod.\ EOB & $92$ & $122$\\
\hline
Phenom, GR & \multicolumn{2}{c}{$131$}\\
MDR & $164$ & $137$\\
TIGER & $129$ & $125$\\
FTA & $117$ & $121$\\
\hline\hline
\end{tabular}
\end{table}

\begin{figure*}[htb]
\centering
\subfloat{
\includegraphics[width=0.42\textwidth]{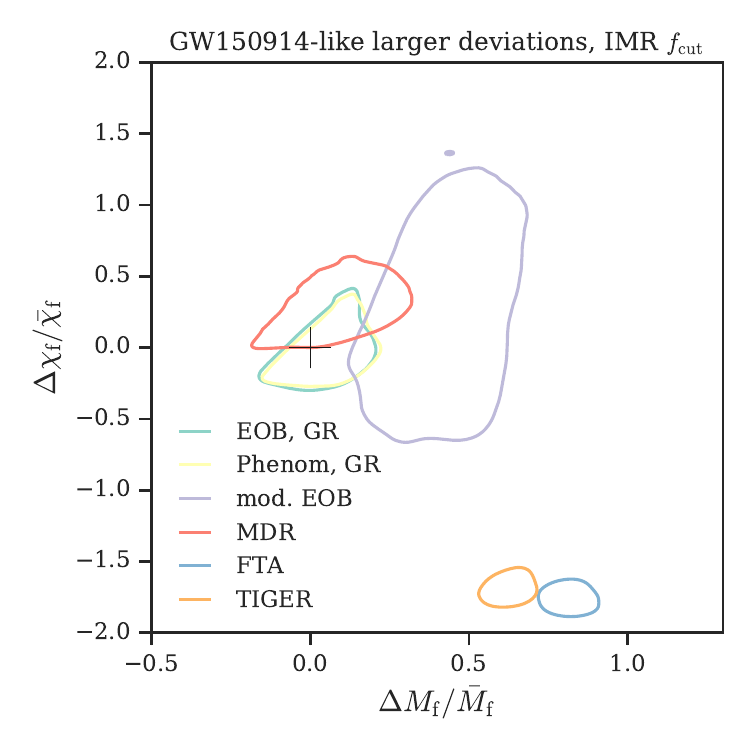}
}
\quad
\subfloat{
\includegraphics[width=0.42\textwidth]{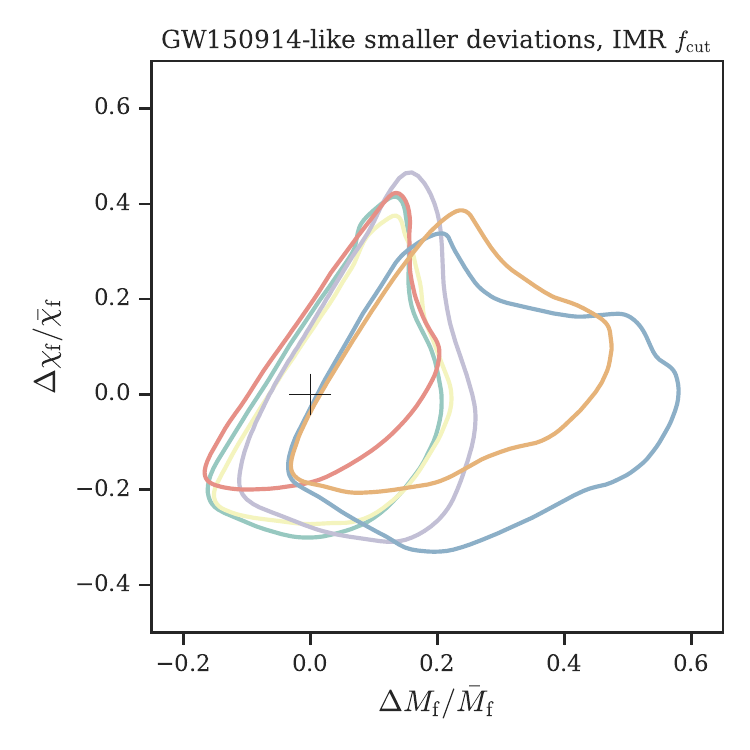}
}\\
\subfloat{
\includegraphics[width=0.42\textwidth]{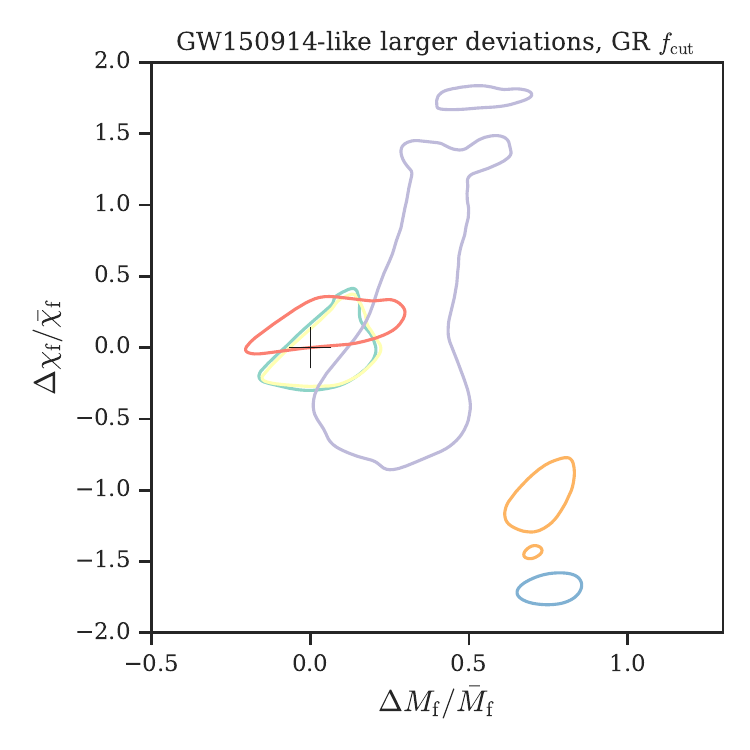}
}
\quad
\subfloat{
\includegraphics[width=0.42\textwidth]{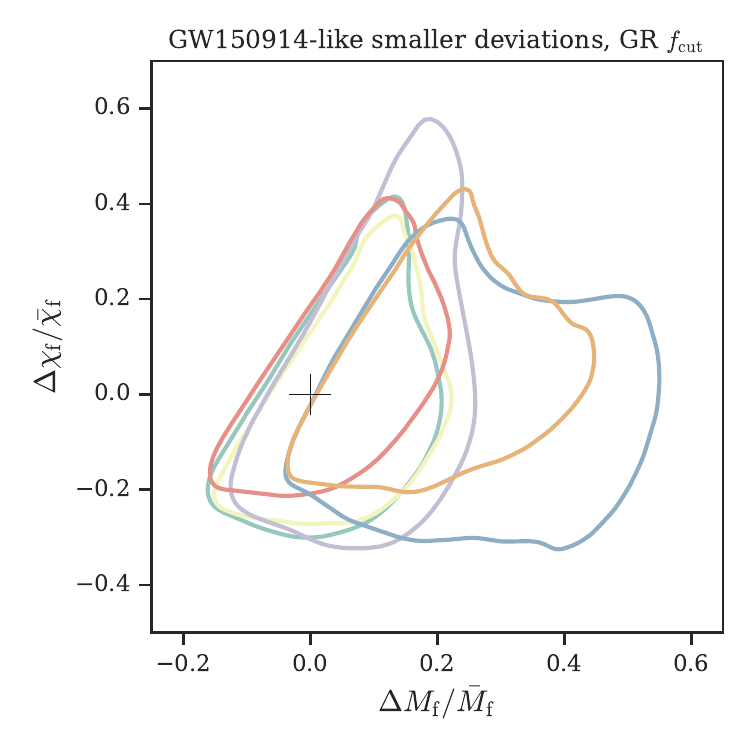}
}
\caption{\label{fig:IMR_2d} The $90\%$ credible regions of the joint probability distributions of $\Delta M_\mathrm{f}/\bar{M}_\mathrm{f}$, $\Delta \chi_\mathrm{f}/\bar{\chi}_\mathrm{f}$ for the IMR consistency test applied to the GW150914-like simulated observations. The left (right) column shows the results with larger (smaller) GR deviations. The top row shows the results with the test's cutoff frequency obtained using the GR analysis of the simulated observation being analyzed (given in Table~\ref{tab:fcut}), while the bottom row fixes these to the values from each simulated observation's corresponding GR case. The results for the GR simulated observations are the same in all four panels.}
\end{figure*}

Here we give the two-dimensional (2d) $\Delta M_\mathrm{f}/\bar{M}_\mathrm{f}$, $\Delta \chi_\mathrm{f}/\bar{\chi}_\mathrm{f}$ joint probability distributions for the IMR consistency test, for comparison with analogous plots shown for analyses of gravitational wave detections in~\cite{O3a_TGR,O3b_TGR}. (The 2d plots in the methods papers~\cite{Ghosh:2016qgn,Ghosh:2017gfp} and earlier LIGO-Virgo papers~\cite{GW150914_TGR,GW170104,O2_TGR,GW190412} are not exactly comparable, since they do not use flat priors in $\Delta M_\mathrm{f}/\bar{M}_\mathrm{f}$ and $\Delta \chi_\mathrm{f}/\bar{\chi}_\mathrm{f}$, and \cite{Ghosh:2016qgn,GW150914_TGR} also use a different normalization.) We show two sets of results in Fig.~\ref{fig:IMR_2d}. We first show the results corresponding to the  results shown in Sec.~\ref{sec:results}, which infer $f_\text{cut}$ from the full IMR GR analysis of each of the simulated observations, as discussed in Sec.~\ref{ssec:imr}. Then, for comparison, we show the results obtained with the same $f_\text{cut}$ as the corresponding GR simulated observation (i.e., $f_\text{cut} = 129$~Hz for the modified EOB observations and $f_\text{cut} = 131$~Hz for all the others).

As expected, the difference between the results with the IMR $f_\text{cut}$ and GR simulated observation $f_\text{cut}$ is largest for the modified EOB case with the larger GR deviation, since this is the case with the largest difference between the two cutoff frequencies. Surprisingly, the $2$~Hz difference in the TIGER case with the larger GR deviation leads to disjoint probability distributions, due to the extreme GR deviation in this case. In all other cases, the differences with different $f_\text{cut}$ values are not so significant, with substantial overlap of the probability distributions, even in the MDR case with the larger GR deviation, which has almost as large a difference in $f_\text{cut}$ as the modified EOB case with the larger GR deviation, though in the opposite direction.

\bibliography{TGR_relation}

\end{document}